\documentclass{aa}
\usepackage{xcolor}
\usepackage{pdflscape}
\usepackage{txfonts}
\def\l{$\lambda$}
\def\mbh{$M_{\rm BH}$\/}
\def\nh{$n_{\mathrm{H}}$\/}
\def\lledd{$L/L_{\rm Edd}$}

\def\msol{M$_\odot$\/}

\def\chm{$c(\frac{1}{2})$\/}

\def\ltsima{$\; \buildrel < \over \sim \;$}
\def\ltsim{\lower.5ex\hbox{\ltsima}}  
\def\gtsima{$\; \buildrel > \over \sim \;$}

\def\gtsim{\lower.5ex\hbox{\gtsima}} 

\def\lya{{ Ly}$\alpha$}
\def\civ{{\sc{Civ}}$\lambda$1549\/}

\def\cm3{cm$^{-3}$\/}
\def\hb{{\sc{H}}$\beta$\/}

\def\mgii{{Mg\sc{ii}}$\lambda$2800\/}

\def\niv{{\sc{Niv]}}$\lambda$1486\/}
\def\ciii{{\sc{Ciii]}}$\lambda$1909\/}
\def\oiiiopt{{\sc{[Oiii]}}$\lambda\lambda$4959,5007\/}

\def\oiiiuv{{\sc{Oiii]}}$\lambda$1663\/}
\def\niii{{\sc{Niii]}}$\lambda$1750\/}
\def\siiii{Si{\sc iii]}$\lambda$1892\/}
\def\aliii{Al{\sc iii}$\lambda$1860\/}
\def\heiiuv{He{\sc{ii}}$\lambda$1640}
\def\nv{{N\sc{v}}$\lambda$1240}

\def\feii{{Fe\sc{ii}}\/}
\def\siii{{Si\sc{ii}}$\lambda$1814\/}
\def\feiii{{Fe\sc{iii}}\/}
\def\fe{{\sc{Fe}}\/}

\def\fe76087{{\sc [Fe vii]}$\lambda$6087\/}

\def\kms{km~s$^{-1}$}

\def\ergss{erg s$^{-1}$\/}
\def\hii{H{\sc ii}\/}

\def\siiv{Si{\sc iv}$\lambda$1397\/}
\def\oiv{O{\sc iv]}$\lambda$1402\/}

\usepackage{enumerate}

\begin{document}
\authorrunning{Sulentic et al.}
\titlerunning{Spectra of  Quasars at $z \approx$ 2.3}
\title{GTC Spectra of $z \approx$ 2.3 Quasars: Comparison with Local Luminosity Analogues}
\author{Jack W. Sulentic\inst{1} \and Paola Marziani\inst{2}\and  Ascensi\'on del Olmo\inst{1} \and  
Deborah Dultzin\inst{3} \and  Jaime Perea\inst{1} \and  C. Alenka Negrete\inst{4}}
\institute{Instituto de Astrofis\'{\i}ca de Andaluc\'{\i}a, IAA-CSIC, Glorieta 
de la Astronomia s/n 18008 Granada, Spain. \email{sulentic@iaa.es,chony@iaa.es,jaime@iaa.es} 
\and  
INAF, Osservatorio Astronomico di Padova, vicolo dell' Osservatorio 5, IT 35122, Padova, Italy. 
\email{paola.marziani@oapd.inaf.it}
\and
Instituto de Astronom\'{\i}a, Universidad Nacional Aut\'onoma de M\'exico, Mexico D.F. 04510, 
Mexico \email{deborah@astro.unam.mx}
\and
Instituto Nacional de Astrof\'{\i}sica, \'Optica y Electr\'onica, Tonantzintla, Puebla, 
Mexico. \email{cnegrete@inaoep.mx}
}
\date{}
\abstract
{The advent of 8-10m class telescopes makes possible for the first time detailed 
comparison of quasars with similar luminosity and very different  redshifts.}
{A search for $z$-dependent gradients in line emission diagnostics and 
derived physical properties by comparing, in a narrow bolometric luminosity 
range ($\log L \sim 46.1  \pm$ 0.4 [\ergss]), some of the most luminous local $z < 0.6$ 
quasars with some of the lowest luminosity sources yet found at redshift $z = 2.1 - 2.5$.}
{Moderate S/N spectra for 22 high redshift sources were obtained with the 
10.4m Gran Telescopio Canarias (GTC) while the HST (largely Faint Object Spectrograph) 
archive provides a low redshift control sample. Comparison is 
made in the context of the 4D Eigenvector 1 formalism meaning that 
we divide both source samples into high accreting Population A and low
 accreting Population B sources.}
{\civ, the strongest and most reliable diagnostic line, shows very 
similar properties at both redshifts confirming at high redshift the
\civ\ profile differences between Pop. A and B that are well established 
in local quasars. The \civ\ blueshift that appears quasi-ubiquitous in 
higher $L$\ sources is found in only half (Population A) of quasars observed
 in both of our samples. A \civ\ evolutionary Baldwin effect is 
certainly disfavored. We find evidence for lower metallicity 
in the GTC sample that may point toward a gradient with $z$. No
 evidence for a gradient in \mbh\ or \lledd\ is found.}
{Spectroscopic differences established at low $z$\ are also present in 
much higher redshift quasars. Our results on the \civ\ blueshift suggest that it depends 
on both source luminosity and \lledd. Given that our samples involve sources with very 
similar luminosity the evidence for a systematic $Z$\ decrease, if real, points toward 
an evolutionary effect. Our samples are not large enough to effectively constrain 
possible changes of \mbh\ or \lledd\ with redshift. Both samples appear representative 
of a slow evolving quasar population likely present at all redshifts.}  
\keywords{
                quasars: emission lines --
                quasars: supermassive black holes --
                ISM: abundances --
                line: profiles  --
                cosmology: observations
               }
\maketitle
\section{Introduction}

It has been known for decades that quasars show evolution with redshift
\citep{schmidt68,schmidtgreen83}. The brightest quasars at redshifts
z$\gg$1.0  are much more luminous than any local quasars (often called
Seyfert 1s if they show a host galaxy). Locally ($z \lesssim $0.5) we
find quasars with absolute magnitudes in the range $M_\mathrm{B}
= $  --23 to --25 (we ignore here AGN below --23 which was
a kind of historical boundary between type-1 quasars and Seyfert 1
galaxies).  At redshift $z \approx$ 2.3 (the quasar number density peaks
near there: {\citealt{schmidtetal95,boyleetal00}}) we find sources
in the range $M_\mathrm{B} =$  --25 to --29 while  sources in this  luminosity
range are almost nonexistant locally.  At redshift $z \approx 4$\ we
observe  $M_\mathrm{B} =$  --27 to --30 superluminous  quasars
that are 2dex more luminous than any local sources. 

\begin{figure*}
\begin{center}
\includegraphics*[width=13.25cm,angle=0]{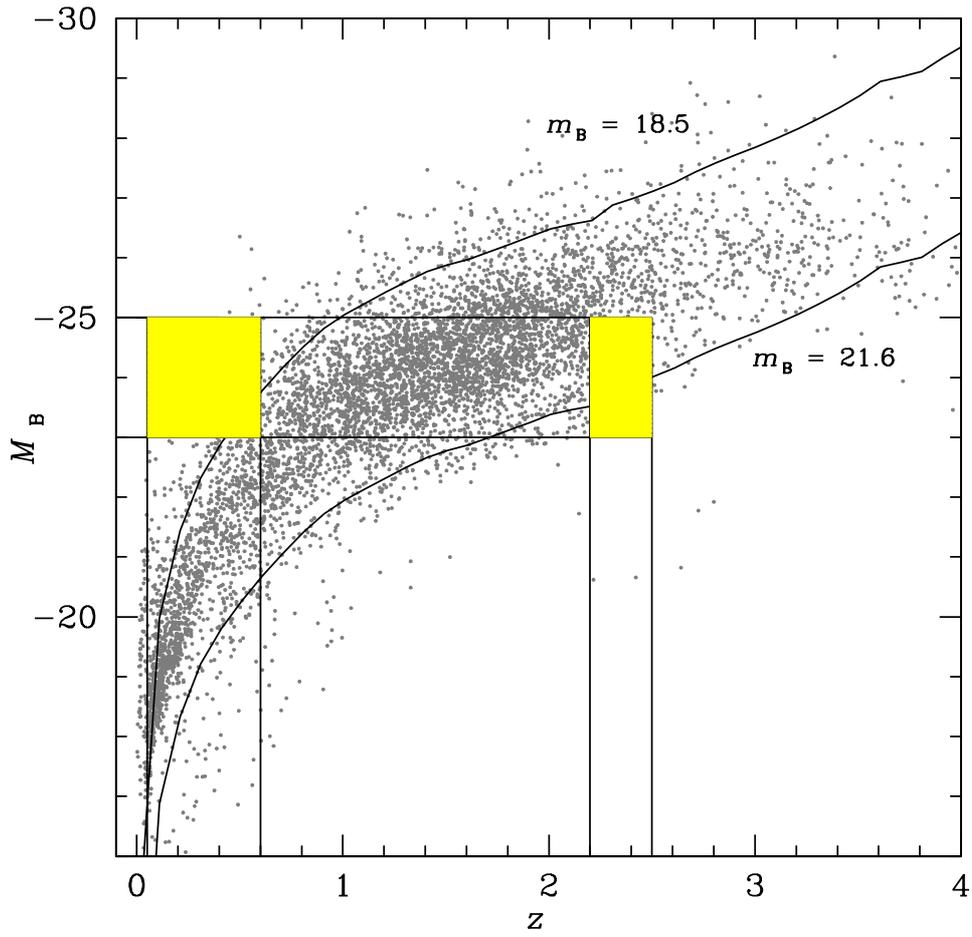}
\end{center}
\caption{Absolute magnitude of quasars vs. redshift for the SDSS-based catalog of \cite{schneideretal10}. 
Curves are computed for two limiting (K-corrected) apparent magnitudes 
$m_\mathrm{B} = 18.5$  and $m_\mathrm{B} = 21.6$. The shaded boxes identify the loci in the 
$z -  M_\mathrm{B} $\ planes of two volume limited sample within the same luminosity limits: 
one  between $2.2 \le z \le 2.5$\ as for the GTC sample of this paper, and a control sample at 
$0.05 \le z \le 0.6$. \label{fig:mabsz}}
\end{figure*}

We are forced to conclude that the most luminous quasars show
strong evolution in their space density. In fact they no longer exist.  
Does this mean that all quasars show evolution or is it purely
luminosity driven? Figure \ref{fig:mabsz} illustrates the observational
situation  by plotting the distribution of a representative  SDSS quasar
subsample \citep{schneideretal10} in the $z$ -- $B$\ absolute magnitude
plane. Note that only sources listed as quasars in the
\citet{schneideretal10} catalogues are shown so that a large population of
local AGN \citep{koehleretal97} and AGN with the narrowest 
(FWHM $<$ 1000 \kms) broad lines \citep{zhouetal06} are  not shown 
in the lower part of the plot. The low-$z$ yellow box identifies a range of 
absolute magnitudes where quasars are rare in the local Universe (they
would be easily detected, but they do not exist in large numbers) and
belong to the high luminosity tail of the optical luminosity function
\citep{chengetal85,grazianetal00,richardsetal05}.    If analogues  to
local  quasars exist at all  redshifts then the lower right quadrant of
the plot will be  well populated when they are found. They would be
generally fainter than $m_\mathrm{B} \approx$ 22 and until very recently 
were not easy to find; however, that situation is changing rapidly 
thanks to SDSS-III/BOSS \citep{ahnetal14}.

Historical attempts to compare quasars with similar luminosity at
significantly different redshifts were few and difficult.  We have more
often compared the brightest  quasars at different redshifts 
 because it has been almost impossible to obtain
good spectra for fainter sources (luminosity analogues of low $z$\ sources)
using 2-4m class telescopes. This means that we are
usually comparing quasars in very different optical luminosity  ranges.
This limitation is quite relevant if optical luminosity plays an
important role in quasar evolution and obscures other physical drivers. We
are unable to search  effectively for signatures of luminosity-independent
evolution (e.g. metallicity, BLR structure/geometry geometry, \mbh\
and accretion rate). Spectroscopy provides the most  powerful clues about
most of these things. Obtaining spectra for  $z \geq $ 2.0 -- 2.5
luminosity analogues of nearby quasars involves spectroscopy of faint
sources $m_\mathrm{B} =$  21 -- 23. It is one thing to obtain spectra in
order to confirm the existence of faint  quasars and quite another to
obtain spectra with s/n and resolution permitting more detailed studies.

There is a growing consensus that the leading parameter governing quasar
diversity involves the Eddington ratio
\citep[e.g.,][]{marzianietal01,baskinlaor04,dongetal09,brightmanetal13}.
Other factors likely play important  roles and have been discussed in
recent review papers \citep[e.g.,][]{marzianietal06,marzianisulentic13}.
We note that a  key complication affecting the  quasars (and not stars)
is that our measures are affected by line-of-sight orientation
\citep{sulenticetal03,collinetal06,boroson11,runnoeetal13}.  Metallicity
effects also significantly affect the observed spectra of quasars
\citep[e.g.,][]{ferlandetal96,shemmeretal04,netzertrakhtenbrot07}.

One might ask if high redshift analogues of local low luminosity quasars
even exist? Perhaps all quasars were luminous ``monsters'' at high
redshift and we are witnessing a systematic downgrading of activity with
increasing cosmic time. In this view the youngest quasars at any epoch
grow so fast that we observe very few in the process of rapid
accretion. The currently favored view is that rapidly accreting  ``monster''
quasars exist side-by-side with slower growing quasars (AGN) and that the
latter exist at all redshifts \citep{trakhtenbrotetal11}. The monsters
have gradually disappeared and are no longer seen below $z \approx 2.0$.
The slower evolving quasar population is assumed to be plentiful, perhaps
even  more numerous than at low redshift?  Is the peak in quasar density
near $z \approx 2.3$ \citep[e.g.,][]{fan06} driven largely or totally  by the
luminous population that has ceased to exist? Do high redshift
analogues of local quasars show the same space density at all redshifts as
well as the same emission line  properties, black holes masses and
Eddington ratios?  The few studies of low $L$\ quasars  at high $z$\  so
far have focussed on estimating the space density of these
sources \citep{glikmanetal11,ikedaetal12}. We focus instead on trying to
answer some of the questions raised above. The fainter, assumed
slower evolving, quasars of interest in this study are largely absent
above $z \sim 1$ in Figure \ref{fig:mabsz} because they were not spectroscopically sampled
by SDSS. Up to the present they have more often been
discovered in deep radio, X-ray or optical pencil beam  surveys. At
redshifts above $z \sim$ 1,  SDSS I/II surveyed quasars largely brighter than
$m_\mathrm{B} \approx $   20 -- 21.  If high redshift analogues of low
luminosity/redshift  quasars exist in abundance are they
spectroscopically different in any way? This question motivated our study.

One of the many technological changes that has taken place over the 50
years since the discovery of quasars involves the advent of 8-10m class
telescopes equipped with CCD imaging spectrographs. We are now entering
the era when we can effectively compare quasars of very similar luminosity
at very different redshifts. We report here on a pilot survey of $z
\approx $ 2.1 -- 2.4 quasars with luminosities in the range $M_\mathrm{B} =$ --23
to --25 which is the same range as low luminosity quasars found at low $z$\
(down to the most luminous local  Seyfert galaxies). Section \ref{sample} describes
the survey, source sample, low-$z$\ control sample and reduction procedures.
Section \ref{results} presents the high-$z$ source spectra and emission line parameters
derived from them.   Section \ref{compa} presents a comparison of mean/median
properties for the high and low-$z$ samples. This is followed by comparison
in the 4D Eigenvector (4DE1) context where we distinguish and
compare Population A and B sources separately.  Section \ref{discuss} 
discusses implication of our findings in terms of quasar evolution, focusing on 
observational parameters, gas metallicity, black hole mass and Eddington ratio.  
Finally we present in  Appendix \ref{red}  the analysis of a red quasar that
showed up in our sample.

\section{Spectroscopic Samples and Reductions}
\label{sample}

\subsection{Sample Selection Strategy}

The goal of the present study was to make a spectroscopic survey of
quasars in the range $z = $ 2.1 -- 2.5 with luminosities similar to
local sources. We required high enough s/n to allow detailed comparison
with properties of a low redshift control sample taken
from the HST archive. This was a pencil-beam survey in the sense that
the same spectral region (Ly$\alpha$--\ciii) was sampled over a very
narrow $z$, $M_\mathrm{B}$ range. The $z$\ range was chosen to be as
high as possible in order to sample quasars at a cosmologically
significant epoch. The $M_B$ range involved luminosities as low as
possible (just above the nominal Seyfert 1 -- type 1 quasar boundary at
$M_\mathrm{B} =  -23.0$). We would have preferred to go 1-2 magnitudes fainter but
the adopted range was dictated by available telescope time
and instrumentation. This translates to  sources in the apparent
magnitude range $m_\mathrm{B} \sim $  20.0 -- 22.0 which corresponds 
to $M_\mathrm{B} \approx$ --23 to --25 (and bolometric luminosity 
$\log L \sim 45.7 - 46.5$).

\begin{figure*}
\begin{center}
\includegraphics*[width=10.75cm,angle=0]{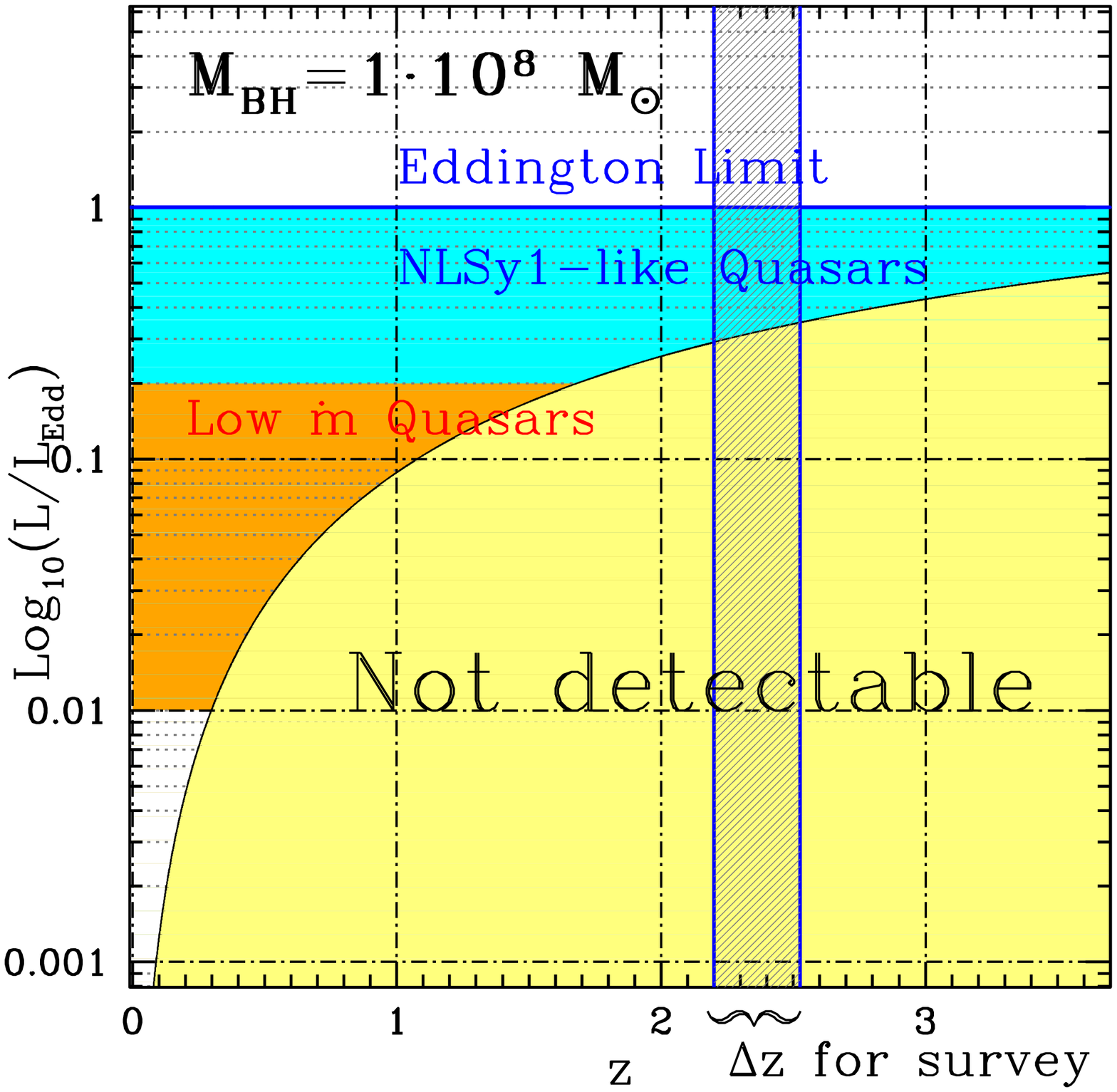}
\includegraphics*[width=10.75cm,angle=0]{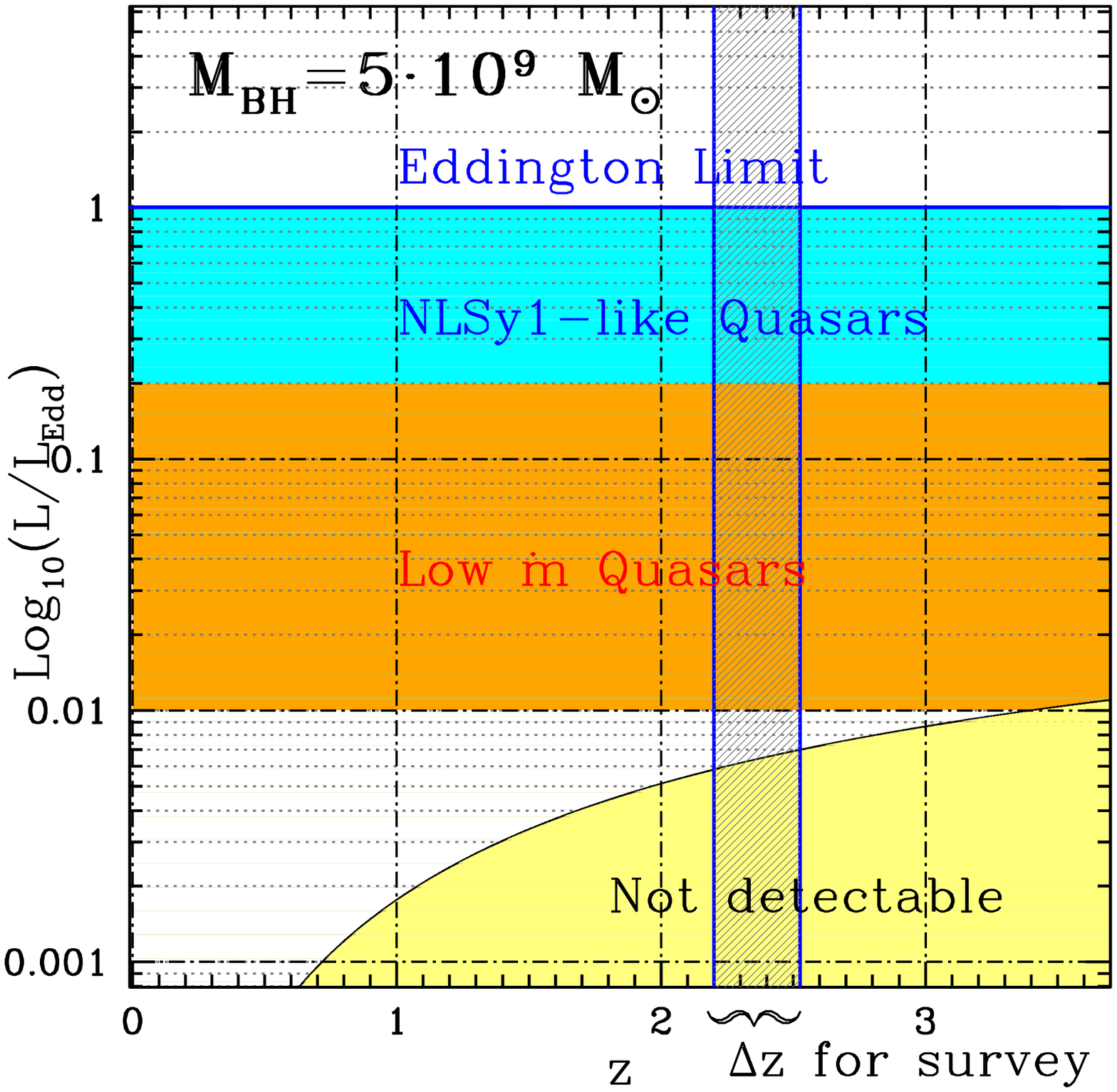}
\end{center}
\caption{Eddington ratio as a function of redshift, with the area of undetectable sources below a limiting 
magnitude $m_\mathrm{B} \approx 21.5$) colored in yellow. The redshift range of the  quasar survey 
is identified by the dashed strip. See text for further details. \label{fig:eb}}
\end{figure*}

Steps to find low luminosity quasars  at $z \approx 2.5$\ involved a
search for faint previously identified quasars in the V\'eron-Cetty \&
V\'eron catalog \citep[e.g.,][]{veroncettyveron10}: since  we did not
carry out  a new quasar survey. Many of our adopted targets
are SDSS sources, but lacked spectroscopic follow-up in either 
SDSS-I/II or SDSS-III/BOSS at the time of our survey.
Sources were selected in narrow redshift
($2.3 \lesssim z \lesssim 2.5$) and absolute magnitude ($-25 \lesssim
M_\mathrm{B} \lesssim -23$) ranges. The high and low $z$\ regions of interest are
marked in (Fig. \ref{fig:mabsz}).  More than 2700 candidate sources are
found in the 13th edition of \citet{veroncettyveron10}. Clearly the
number of low/intermediate luminosity quasars at high $z$\ is increasing
rapidly. Targets were selected randomly and are listed in Table
\ref{tab:id} which presents quasar identification, apparent magnitude,
redshift  $\pm 1 \sigma$ uncertainty (see \S \ref{anal}), absolute B
magnitude from \citet{veroncettyveron10}, discovery technique or detection
band, a reference to the discovery paper, an alternate name and some
additional information notably on radio flux for radio detected
sources.  As expected many chosen sources have names reflecting discovery
in X-ray surveys or deep optical searches (Table \ref{tab:id}, columns
7-8).  The flux limits associated with optical surveys
cause an ``Eddington ratio bias'' as shown in Figure \ref{fig:eb}.   Fig.
\ref{fig:eb} gives a graphical representation for the case of a typical
(10$^{8}$\ \msol) and  a large black hole mass   (10$^{9.5}$\ \msol). Bands
of different colors identify sources radiating in different Eddington
ranges. Sources radiating close to or somewhat below the Eddington limit
are indicated as NLSy1-like quasars (Pop. A in 4DE1 parlance,
\citealt{sulenticetal00b,netzer13} and \S \ref{discuss}). 
Below \lledd\ $\approx$ 0.2 we find no change in observational properties \citep[Pop.
B;][]{sulenticetal08,sulenticetal11}, while below   \lledd\
$\approx$ 0.01 accretion processes may become unstable and the quasar
state may be transient. The flux limit associated with a survey
introduces a bias in the discovery of quasars: in the example of Fig.
\ref{fig:eb} all quasars at $z \la$ 0.3 with \mbh =10$^{8}$ \msol\ can 
be detected while at  $z \approx$ 2.3 only quasars radiating close  to the 
Eddington limit will be detected within a limiting magnitude 
$m_\mathrm{B} \approx 21.5$. Only quasars with the largest masses will be 
detected at  $z \approx$ 2.3.  This mass-dependent loss of low Eddington ratio  
sources  will be a source of bias at high $z$.

\subsection{Observations and Data Reduction}

Using the 10m GTC equipped with the faint-object spectrograph {\sc osiris} it is possible to obtain  S/N
$\approx$ 20 spectra of quasars in the $m_\mathrm{B} \approx $ 21 -- 22 magnitude
range with exposures of about 40 minutes in good seeing. Under less
than optimal conditions we will obtain spectra with S/N $\approx$  7 --12. Within the 
observed wavelength range we obtain several lines (from \lya\ to \ciii) from 
which diagnostics intensity ratios can be computed. The GTC
spectra are similar to or  better than the HST archival UV spectra
for low redshift analogues \citep{bachevetal04,sulenticetal07}. Figure
\ref{fig:overview} shows a montage of the deredshifted GTC spectra
obtained for 22 quasars. One source (Wee 155)  turned out to have an
incorrect redshift ($z\approx$1.59) derived from the low S/N discovery
spectrum. An additional source is obviously a heavily reddened
(radio-loud if no correction for extinction is applied) quasar FIRST J15318+2423.
 (discussed in Appendix \ref {red}).
The remaining 20 sources represent our working sample of unreddened 
low luminosity quasars (2 radio-loud). Spectra show  S/N near 10 -- 20 
covering Ly$\alpha$ -- C{\sc iii}]$\lambda$1909 in the rest frame.
They can be  effectively compared with low $z$\ UV spectra from the
HST archive  already studied in the 4DE1 context
\citep{bachevetal04,sulenticetal07}. Our high $z$  sample clusters near
$\log L_\mathrm{bol} \sim $46.0 with redshift range from  $z = $
2.21 -- 2.40.

Long slit spectroscopic observations of the selected quasars were
carried out with OSIRIS (Optical System for Imaging and
low-Intermediate-Resolution Integrated Spectroscopy) located at the
Nasmyth-B focus of the 10.4m GTC telescope of the Observatorio del
Roque de los Muchachos (La Palma, Spain). The observations were
obtained in service mode in two approved observing programs during 2011
and 2013. In the first run 15 QSOs were observed while 7 more were
obtained during the spring 2013 run. The R1000B grism was used for all
observations  with $2 \times 2$\ binning yielding a wavelength range of 3650 -- 7750
\AA\ with a reciprocal dispersion of about 2.1 \AA/pixel  
($R = \lambda/\delta \lambda \approx$ 1000).
This coverage corresponds to the wavelength region of interest for our
study from \lya\ to \ciii\ and includes  \nv, \siiv\ \& \oiv, \civ,
\aliii\ or \siiii\ emission lines. The slit was oriented at  parallactic
angle to minimize effects of atmospheric differential refraction in the
spectra. Table \ref{tab:obs} gives a log of the observations  where  columns list in 
this order: QSO ID, date of observation, total exposure time in seconds, 
the number of split exposures,
the signal/noise measured in the combined spectrum for each
object (at  1450 \AA; blueward of the \civ\ line),   the slit
width in arcsec, and the seeing measured by fitting several stars
in the acquisition image of each quasar.

Data reduction was carried out in a standard way using the {\sc iraf} package.
The task {\sc ccdproc} was used to trim the spectra and make the overscan
correction. Bias subtraction was performed nightly by subtracting a
median zero-level bias image constructed  by combining all bias frames
provide by the telescope team. Also each spectrum was flat-field corrected
with the normalized flat-field obtained after median combination of the
flats obtained with the same instrumental setup.  2D wavelength
calibration was obtained using the combination of HgAr and Ne lamp
exposures  and {\sc iraf} routines {\sc identify}, {\sc reidentify}, {\sc
fitcoords}, and {\sc transform}. We checked wavelength calibration
for all the individual exposures of each objects before final combination. The {\sc
apall} task was used for object extraction and background subtraction.
Instrumental response and flux calibration were obtained 
from spectrophotometric standard star exposures with the same instrumental configuration,
provided for each night our sources were observed (Ross
640, Feige 34, GD 24, GRW+70.8247 and Feige 92 [for the
first run]; Feige92, Ross 640, G157-32 and GD 140 [for the
second run]).

\subsection{Data Analysis}
\label{anal}

Spectra were reduced to rest frame wavelength and specific flux scale.
An unbiased estimate of the redshift (i.e. source rest frame) is not
trivial because the GTC spectra only cover the UV domain where
narrow low ionization lines (LILs), usually the most reliable rest frame
indicators \citep{eracleoushalpern03,huetal08} are not available. We
therefore estimated redshifts from low-ionization \ion{O}{i}\l1304, 
\ion{Si}{ii}$\lambda$1264, \ion{C}{ii}$\lambda$1334 features as well as 
\ciii, including \civ\ and other high
ionization lines only if they showed values consistent with the LILs.
Uncertainties are reported in Table \ref{tab:id} with the rest frame estimated
to be accurate to $\pm$ 200 \kms\  for most sources ($\pm$ 300 \kms\ for
three sources). The LIL features were not measured in one source
([VCV96] 1721.4+3401) so the redshift rests on \civ, \lya, and \siiii. The first
two lines yield a significantly lower value than \siiii\ leading to
an estimated uncertainty  of $\pm $ 500 \kms.

One of the first steps after data reduction was an attempt to
separate sources into Population A and B following the 4DE1 formalism.
Both optical and UV (as well as X-ray) criteria are summarized
in \citet{sulenticetal07}. Criteria employed in the present paper
are also discussed in \citet{negreteetal14} and entail: (1) line widths
of \aliii\ and \siii\ are found to be correlated with \hb\ line
width \citep{negreteetal13a,negreteetal13b} making them useful Pop. A-B discriminators
and virial estimators; (2) EW \civ\ values for Pop. A1 and B sources are
a factor of two higher than for the rest of Pop. A sources and (3) blue
and red \civ\ profile asymmetries  appear to be signatures of
Pop. A and B sources, respectively. The criteria allow a reasonable
population A-B discrimination. Those near the intersection (bin A1)
are much less certain and the bin is likely a mix of A and B sources.
The fact that we find an even division between source populations (11
Pop. A and 9 Pop. B), as has been found at low redshift
\citep{zamfiretal10}, suggests that our sample
is a reasonably random selection of quasars in the 
context of 4DE1 (in \citep{zamfiretal10} the magnitude limited SDSS 
sample divides 45\% and 55\% for pop. A and B respectively).

Intensity and profile measures of emission lines was carried out
following the approach described in \citet{marzianietal10} and
\citet{negreteetal13a}. The salient assumption is that each line in
a Pop A source can be interpreted as a composite of: an unshifted
symmetric component plus a blueshifted, asymmetric
component (BLUE) related to systematic outflow motions. In the case of
Pop. B sources any blueshifted component is much weaker however a
very broad and redshifted component (VBC) is usually seen along with
the unshifted  component that is assumed to arise in a virialized medium
(the ``classical'' broad component; BC). These line components (BLUE, BC and VBC) 
were included when modelling the strongest emission lines \lya,
\nv, \civ, \heiiuv, \aliii, \siiii, and \ciii. They were
simultaneously fit along with continuum and \feii\ + \feiii\ emission
(the latter relevant only in the 1900 blend and for Pop. A only). In the case of the UV
\feii\ and \feiii\ emissions we considered the templates provided by
\citet{bruhweilerverner08} and \citet{vestergaardwilkes01}, respectively.
The fitting routine scaled and broadened the original templates to
reproduce the observed emission. \feii\ is always set to zero by the fitting 
routine reflecting its low intensity and the low S/N in the observed spectral 
range.   The S/N of some spectra allow only a marginal fit with 
the greatest uncertainty involving the 1900 blend and the blue component 
in Pop B sources.

The 1900  blend has become important for several reasons: 1)
FWHM \aliii\ and \siiii\ can be used as a a virial  estimator
\citep{negreteetal13a}, and their FWHM can also be used for distinguishing
Pop. A and B sources; 2) the relative strength of \aliii, \siiii, and
\ciii\ can be used as additional diagnostics for distinguishing between
Pop. A and B quasars \citep{bachevetal04,negreteetal13a}, 3) line ratios
involving \siiii/\civ\ and \aliii/\siiii\  are important for constraining
BLR physical conditions that can be used  to estimate the BLR radius
\citep{negreteetal13a}. The key is to use spectra with S/N high enough to
allow modelling and decomposition of the lines in
the blend. 

\begin{figure*}
\begin{center}
\includegraphics*[width=14.25cm,angle=0]{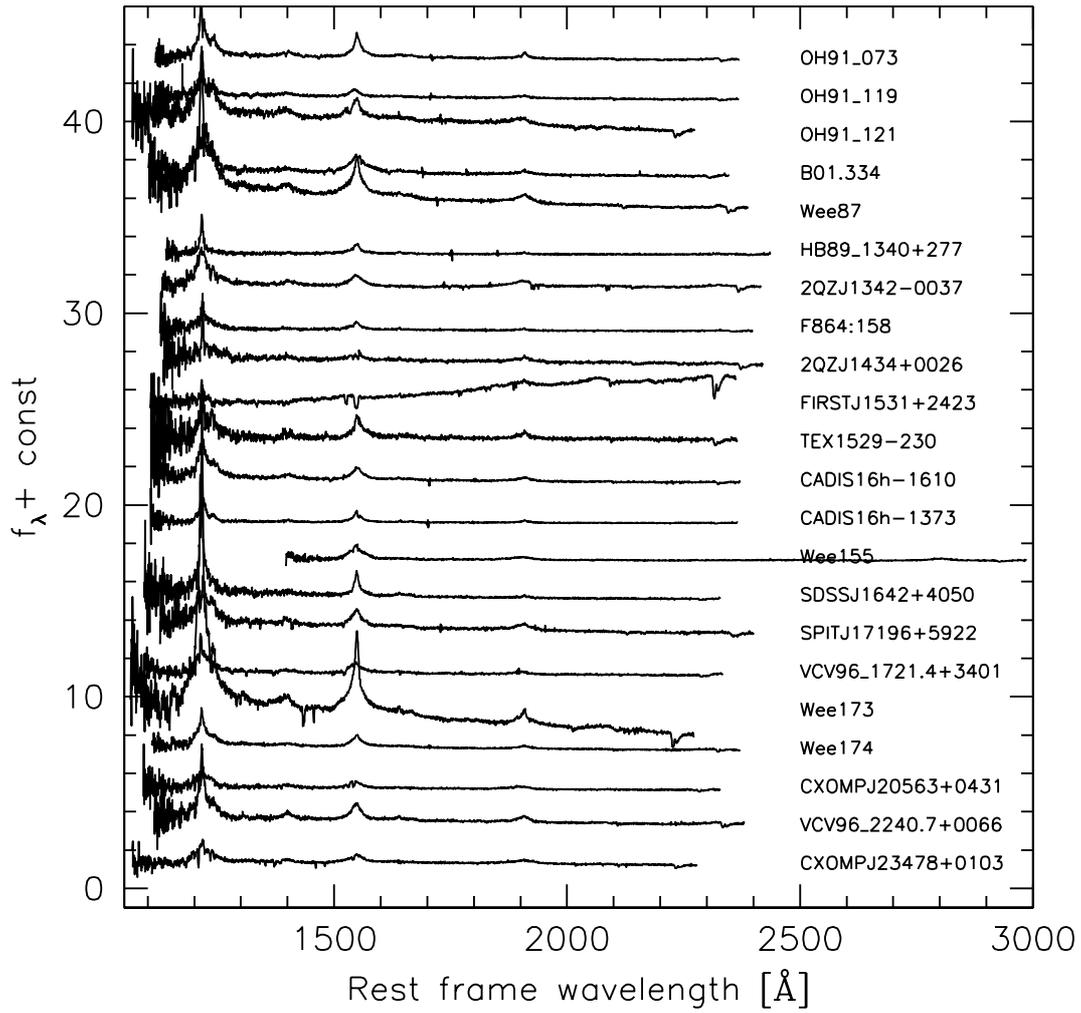}
\end{center}
\caption{The GTC quasar spectra after redshift correction. Abscissa is the rest frame wavelength in \AA, ordinate is specific flux in units of 10$^{-15}$ erg s$^{-1}$ cm$^{-2}$ \AA$^{-1}$. Spectra have been vertically  displaced by adding  steps of $\Delta f_{\lambda} = 2$\ (no step was added   for Wee 173 to avoid confusion with VCV96   from bottom to top.  \label{fig:overview} }
\end{figure*}

Figures \ref{fig:overview}, \ref{fig:fitsa} and 
\ref{fig:fitsb} show deredshifted spectra and individual fits for our GTC
sample where S/N varies from 5 -- 6 to 40. It  is easy to see how
the blend modelling becomes less certain for the lowest S/N examples.

\subsection{A low redshift comparison sample of luminosity analogues}

The significance of observations for high $z$  low luminosity quasars can 
be fully understood only if a suitable low-$z$\ control sample exists for 
the same luminosity range. Archived HST/FOS observations provide spectra 
with similar dispersion, S/N and luminosity. The archival spectra were
previously studied in the 4DE1 context \citep{bachevetal04, sulenticetal07}. 
The heterogenous nature of both the GTC/OSIRIS and HST/FOS samples 
means that neither is complete. 

The FOS sample in particular is strongly
biased towards radio loud sources. We randomly extracted  smaller
subsamples satisfying the  conditions: (1) absolute
magnitude distribution not statistically different from the GTC
sample; (2) consistent fraction of radio sources ($\approx$ 10\%  as
in the GTC sample). In practice sources were selected  from the 
sample of 130 sources given in  Tables 1 and 2 of \citet{sulenticetal07} available 
at {\sc vizier}\footnote{ J/ApJ/666/757/CIVlines at http://vizier.u-strasbg.fr/} 
in the absolute magnitude range $-25 \la M_\mathrm{V} \la -23$. This yielded 
42 sources, of which 23 were radio quiet. In order to produce a control 
sample  with a RL fraction closer to that found in optically-selected samples 
we defined  a pure RQ sample of 23 sources supplemented with 3 
randomly-selected RL sources that  yielded 26 sources in the 
control sample (the 26 low-$z$ sample will be hereafter referred to as 
the FOS-CS).

\section{Results}
\label{results}

Fig. \ref{fig:fitsa} and Fig. \ref{fig:fitsb} show the results of
{\sc specfit} \citep{kriss94} analysis  of the \lya, \civ\ and 1900 blends 
in individual Pop. A and B sources of the GTC sample. Table \ref{tab:meas}
reports quantities measured in the rest frame spectra of each source as
follows: rest-frame continuum flux level at 1450\AA\ and associated uncertainty;
total fluxes for the most prominent broad emission features (\/\lya, \nv,
\oiv + \siiv, \civ, \aliii, \siiii, \ciii); equivalent widths
for \lya\ and \civ \ (equivalent widths for other lines can be estimated
by scaling EW \civ\ by line \civ\ flux ratio). Data are presented in two groups: 
upper Pop. A and lower Pop. B.

Our {\sc specfit} analysis followed past work modelling H$\beta$\ and 
\civ\ where strongest results were obtained from studies of H$\beta$ (at $z <$ 0.7) for 
which the largest body of data exists \citep{marzianietal03a,marzianietal09,zamfiretal10}. 
As noted in \S \ref{anal},  past work identified three broad emission 
components (BLUE, BC and VBC)  which show different relative intensities in different 
sources and are not necessarily present in every individual line/source 
\citep{marzianietal10}. The ``classical'' relatively unshifted broad  \hb\ component 
is present in almost all sources and is assumed 
to be the most  reliable virial estimator after \mgii\  \citep{marzianietal13}. 
This is where the Population A-B distinction becomes important. Population A sources
involving the highest \lledd\ emitters show a Lorentzian BC component with FWHM
$\leq$ 4000 \kms\ plus, in extreme sources, a component on the blue side
of the BC. Pop. B sources, including most radio-loud quasars, show
a BC and a broader redshifted VBC component. In a few cases  we suspect
that the H$\beta$\ profile may be dominated by the VBC (e.g. PG 1416--129; 
\citet{sulenticetal00c}. Detailed studies of the \civ\ profile in low $z$ 
quasars are  more limited and rely on the $\approx$140
sources with HST archival spectra \citep{bachevetal04,sulenticetal07}.
Pop. A sources show a stronger \civ\ blue component which can be much
stronger than the BC (no VBC emission is seen). The well studied NLSy1
source IZw1 is a Pop. A (bin A3) prototype. In some sources a red asymmetry 
is seen in the CIV profile and is attributed to VBC emission--almost always 
involving Pop. B sources.

The estimated contribution of the blue component to the total flux of 
\civ\ is listed in  Table \ref{tab:meas} under the column labelled 
$F_\mathrm{blue}/F$.  Table \ref{tab:measprof} reports quantities measured 
from  the full broad line profiles (i.e. without separating line components).  
Following columns of Table \ref{tab:measprof}   list: FWHM \civ, and 
other profile shape parameters that provide a quantitative 
description of \civ\ (see \citealt{zamfiretal10} for 
definitions): asymmetry index, kurtosis, and centroid displacement 
(with respect to rest frame) at 1/4  and 1/2 fractional intensity. 

While we have applied our standard multicomponent {\sc specfit} analysis
to all GTC spectra, the S/N of the data make clear that  we could be
over-interpreting some of the spectra. We show all individual fits and
table derived quantities even though some are uncertain and not
directly relevant to the goals of the paper which involve high-low
redshift comparisons for Pop A and Pop. B quasars. Subtraction of a
blue component from \civ\ is well supported for Pop A sources which also
show a mean profile blueshift of $\sim$450 \kms. However Pop. B sources
show no systematic blueshift raising doubts about the
reality of a blue component in those sources. Modeling of broader Pop. B
line profiles is always more problematic. In the case of Pop. A sources
the presence of blue and BC components is well established and no evidence
of a VBC is found.

\subsection{\civ\ in Population A and B at high and low redshift}

\civ\ is the safest feature with which to effect a comparison between low
and high redshift. \civ\ is a strong collisional resonance doublet and {  is not
heavily blended } with other strong lines. The \civ\ profile shift
was chosen as a principal 4DE1 parameter because it showed a strong
population A-B difference. EW \civ\ and \civ\ profile asymmetry could
also serve as 4DE1 diagnostics \citep{sulenticetal07}.
Table \ref{tab:measprofb04} reports the virial broadening estimator
FWHM (see \S \ref{masses}) and measures of the \civ\ line profile 
(FWHM, asymmetry index, kurtosis and centroids at one quarter and 
half maximum; see \citealt{zamfiretal10} for definitions of parameters
given here) for the GTC and FOS-CS.

\civ\ shows striking similarity at high and low redshift in both
Pop A and B. The composite spectra show median equivalent width (EW),
asymmetry index (AI) and profile shift at half maximum (\chm) at low and
high $z$ that are almost identical (Table \ref{tab:measprofb04}). { We were 
able to distinguish Pop.  A and B on the basis of the same criteria 
employed at low-$z$ applied to the UV  (\S \ref{anal} and \citealt{negreteetal14}), } 
and  {\  quantitative} Pop.  A-B differences in EW and profile measures 
previously found at low redshift are confirmed in the high redshift GTC sample.  
\civ\ in Pop. A sources shows a systematically lower EW  than Pop. B as well 
as blue asymmetric and blueshifted profiles. Pop. B sources show a stronger high 
ionization spectrum without evidence for profile blueshifts/asymmetries. Pop. B 
shows an additional very broad and redshifted  component  with FWHM and shift 
values similar to those found for \hb\ at low redshift.

Figure \ref{fig:distributions} compares  the \civ\ equivalent width, 
centroid shift and FWHM distributions for the high and low redshift samples.
Population A and B are identified (Pop. B shaded). 
Low redshift samples \citep{bachevetal04,sulenticetal06}  show
lower Pop A equivalent widths and  FWHM values than Pop. B
(W(\civ)$_\mathrm{A} \approx $ 50 \AA\ vs. W(\civ)$_\mathrm{B} \approx$ 83
\AA\ and  FWHM (\civ)$_\mathrm{A} \approx$ 4060 \kms\ vs.
FWHM (\civ)$_\mathrm{B} \approx$  5810 \kms). Our high redshift sample
shows W(\civ)$_\mathrm{A} \approx$ 45 \AA\ vs W(\civ)$_\mathrm{B}
\approx$ 69 \AA\  similar to the differences found at low redshift where
Pop. A sources generally show a weaker high ionization line spectrum
\citep{marzianietal01,marzianietal10}. Note that there is again no
significant difference between the W(\civ) values between the GTC
and FOS-CS. The distribution of Figure \ref{fig:distributions} show
a few  (1 -- 3, depending on the random selection of radio sources) large equivalent width sources 
in the FOS-CS that have no
correspondence in the GTC sample . They are not numerous enough to make
the samples significantly different.  Median FWHM(\civ) measures
for the high redshift sample are 5010 \kms\ and 5830 \kms\ for
Pop. A and B, respectively roughly consistent with the low-$z$\
FOS-CS. However, there is a small discrepancy involving the FWHM \civ\
comparison where high $z$ Pop. A and B average composite widths are
about 27\%\ and 10\%\ broader than control sample (CS), respectively. These differences
are not statistically significant -- given the small sample size and the
errors involved. The difference disappears for Pop. B if medians are
considered  and is reduced to $\approx 15$\%\ for Pop. A. Analogous
considerations apply to the virial FWHM.

Considering the individual and composite fits  to the \civ\ profile, we do
not confirm a near ubiquity of blueshifts/asymmetries for type-1 quasars
\citep{richardsetal11}  but rather the same Pop. A/B dichotomy seen at low
redshift \citep{sulenticetal07} where Pop. B  sources show
red shifts/asymmetries or symmetric unshifted profiles.

\subsection{Comparing Pop. A and B UV spectra at high and low redshifts}
\label{compa}

Comparison of Pop. A and B sources at high and low redshift cannot extend
beyond the \civ\ line without use of composite spectra to enhance spectral
S/N. The GTC sample is not large enough ($N = 20$) to permit a fine
subdivision so we simply constructed the Pop. A and B median composites
that are shown in the upper panel of Fig. \ref{fig:popab} that involves
11 and 9 Pop. A and B sources respectively.

The best low $z$\ composites come from \citet{bachevetal04}. This sample was
large enough to permit construction of composite spectra for each of the
five most populated 4DE1 bins \citep{sulenticetal02}.  The bin composites
could then be combined to produce two higher S/N composites for Pop. A and
B. Low $z$\ Pop. A and B composites shown in the bottom panel of Fig.
\ref{fig:popab} were obtained by weighting sources from the 5 spectral 
subtypes by their relative frequency in the sample. These weighted-average
composites include all  \citet{bachevetal04} spectra. They are the best
control sample composites but include many sources that fall outside the
narrow luminosity range (almost all lower) of our GTC sample. In order to
check on these composites we also generated (lower S/N) composites from
our luminosity restricted FOS-CS control sample taken from
\citet{sulenticetal07}. The \citet{bachevetal04} and 
\citet{sulenticetal07} samples are largely overlapping with
\citet{bachevetal04} including 139 sources vs. 130 of 
\citet{sulenticetal07}. The 26 sources of CS are
included in both samples. The luminosity restricted FOS-CS  composites  
show no major differences from the ones obtained from the less restricted
\citet{bachevetal04} composites. Neither FOS-CS composite (except
for slightly weaker \niii\ emission in Pop. A). We therefore cautiously
adopt the higher S/N \citet{bachevetal04} composites for all comparisons.
The two bottom rows of Fig. \ref{fig:popabfits} (the panels labelled B04)
present results from {\sc specfit} analysis of the low-$z$
\citet{bachevetal04} composites for \lya, \civ, and the 1900 blend, and
the last two rows of  Table \ref{tab:measb04} report  measures derived
from this analysis (last two rows identified with the label B04).

If one assumes that source luminosity is the principal driver of quasar
structure and kinematics then we expect our low $z$ and GTC composites 
to be very similar.  In fact our work in the 4DE1 context suggests that  
\lledd\  is the principal driver \citep{marzianietal01}.
A comparison of this kind requires the Pop. A and B distinction
because quasar spectra are {\em not} the same and spectral 
differences at any fixed redshift are maximized using the Pop. A 
and B distinction which is driven by \lledd.  We are aware of no more 
effective distinction. Previous work on low redshift
samples  show that the two populations divide approximately 60\% /40\%,   
with RQ / RL quasars dividing  90 \% /10 \%, and  the wide majority of RLs 
belonging to Pop. B \citep{zamfiretal08}. If our samples are small or modest 
as in the present case, an overrepresentation of extreme sources 
would bias the comparison. There is no reason to expect (and no evidence for) such 
an overrepresentation.  If, for example, many of our sources were soft X-ray 
selected then we would  expect them to preferentially occupy 4DE1 spectral bins 
A2, A3, A4 which  also show the largest and most frequent \civ\ blueshifts (smallest \mbh\ 
and highest \lledd). Radio selected quasars would largely occupy Pop. B 
spectral bins. Table \ref{tab:id} suggests that few  of our quasars were 
selected in either way. If anything deep optical grism searches will favor 
quasars  with stronger (broader, higher EW) spectral lines.  This might 
disfavor extreme Pop. A quasars and indeed we  find only 2  candidate
extreme Pop. A (i.e., narrow line Seyfert 1) sources in the GTC sample 
{(Q1232-1113 and CXOMPJ20563+0431)}. Hypothesized to be
as younger quasars, we might expect to find more of them at  $z \approx$ 2.3. 
High $z$  analogues of I Zw 1 (albeit much higher luminosity) were 
identified in one of our companion surveys  
\citep{dultzinetal11,negreteetal12,marzianisulentic13,marzianisulentic14}.

Table \ref{tab:measb04} reports normalized fluxes and equivalent widths
for Pop. A and B composite (average and median) spectra. The
spectra are shown in Fig. \ref{fig:popab}.  Note that the measures come
from the composite spectra using the same {\sc specfit} analysis
(Fig. \ref{fig:popabfits}, top rows) applied to individual sources. 
Given the sizes of both samples some differences due to outlying/peculiar 
sources could affect average values/composites;  however the mean and median
composites yield values that are  mutually consistent. Values for Pop. A 
and B composite spectra are consistent with values derived for the 11 and 9 
individual spectra.

In conclusion, following the 4DE1 formalism we attempt a 
first-order separation of our high-$z$\ and low-$z$\ samples into Population  
A and B quasars. In the UV rest wavelength range studied here we find  
evidence for spectral differences between Pop A and B especially in measures 
of CIV which is strong and not blended with other strong lines. 
Table \ref{tab:measb04} compares high and low redshift median composite 
spectra for Pop. A and B. Both low- and high-$z$\ Pop. B samples show stronger  
high ionization spectra  and stronger narrow lines than Pop. A, while no major difference 
is found between GTC and the B04 composites.

\subsection{Metallicity difference between high and low $z$?}
\label{metals}

One advantage of comparisons involving UV spectra is that they provide 
several line ratios that can serve as metallicity indicators. Most results
of the past decade point toward solar and supersolar metallicity
even at very high redshifts \citep[e.g.][]{hamannferland93,kurketal07,willottetal10}. Of
course these studies involved extremely  luminous sources.  The
GTC sample spectra have high enough S/N to permit reasonable
measures of metallicity indicators (\nv/\civ, \siiv+\oiv/\civ).  
The ratio involving nitrogen and carbon scale with   metallicity
by virtue of the secondary  enrichment of nitrogen in massive stars 
\citep{hamannferland93,hamannetal02}. The ratio \siiv+\oiv/\civ\ involves 
ionic species of two alpha $\alpha$ elements relative to carbon. The dependence 
of \siiv+\oiv/\civ\ on $Z$\ has been calibrated
by photoionization computations \citep{nagaoetal06} and widely used in recent 
quasar studies involving metal abundances \citep[e.g.,][]{wangetal12,shinetal13}. The
values of the ratios obtained from the fluxes in Table \ref{tab:meas}
indicate a large spread (at least 0.6dex) in both the  \nv/\civ\ and
(\siiv+\oiv/\civ)\ ratios (middle and bottom panel of Fig.\ref{fig:metaldistr}). 
The largest sample presently available for comparison is  provided 
by \citet{shinetal13} who measured \nv/\civ\ and \siiv+\oiv/\civ\ from
IUE and HST spectra of PG quasars available in the MAST archive. 
Fig. \ref{fig:metaldistr} compares the GTC distribution of \nv/\civ\ (middle panel) 
and (\siiv+\oiv)/\civ\ (lower panel) with \citet{shinetal13}.   
Fig. \ref{fig:metals} compares GTC measures for \nv/\civ\ (upper panel) and
(\siiv+\oiv)/\civ\ with those of \citet{shinetal13}  as a function of
source bolometric luminosity. The GTC sample overlaps the high-$L$ \ part
of the \citet{shinetal13} sample, where metal richer sources are found
\citep{juarezetal09}. The correlation between $Z$-sensitive ratio
\nv/\civ\ and $L$\ in Fig. \ref{fig:metals} was seen previously
\citep{nagaoetal06} although its origin  is unclear (\S \ref{metalsdisc}). 
In order to make a meaningful comparison between the GTC and low-$z$ samples 
we applied a restriction to luminosities larger than $10^{45}$ \ergss. In this
case, there is a  consistent difference in the distribution of the ratios,
supported by K--S tests, in the sense  that the GTC sample sources
show lower values. 

Metallicity is estimated to be solar or slightly sub-solar in the lowest
$Z$ cases.  We tentatively identify at least four quasars (F864-158, Q
1340+27, Q 1640+40 and CADIS 16h-1610) that  appear to show  sub-solar or solar
metallicities following the normalization of \citet{nagaoetal06}.
A low $Z$\ for these sources is also suggested  by their large
\civ/\ciii\ ratios (upper panel of Fig.\ref{fig:metaldistr}). This ratio is 
usually not considered as a metallicity
indicator because it is sensitive to ionization level, density (the
\ciii\ line is emitted in an inter-combination transition with a well
defined critical density), and because it is the ratio between lines of
two different ionic species of the same element. However, for a fixed
density (below the critical density of \ciii) and ionization parameter, 
{\sc cloudy} \citep{ferlandetal13} photoionization simulations show a remarkable 
dependence on metallicity for the ratios \ciii/\civ\ and
(\aliii+\siiii+\ciii)/\ciii\ (Fig. \ref{fig:ratios}). These trends are
explained by an increase of electron temperature with decreasing
metallicity: at $Z  \sim 0.01 Z_\odot$, $T_\mathrm{e} \approx 27000$K, at
$Z \sim 5 Z_\odot$, $T_\mathrm{e} \approx 16000$\ (assuming $\log U$ =
-1.75 $\log$\ \nh = 10). The higher $T_\mathrm{e}$, due to lower cooling
rate of the BLR at low $Z$,  increases the collisional excitation rate of
\civ, and therefore favors this line over lower excitation metal
lines. It is interesting to note that this interpretation is analogous
to ascribing the change in \oiiiopt\ strength in Galactic
and extragalactic \hii\ regions to differences in metal content
\citep[e.g.,][]{searle71}. 

An additional line of evidence supports the idea of  lower chemical
abundance in GTC sources in Pop. A. Fig. \ref{fig:compara}
shows a comparison of the difference in line width of the
Pop. A composite spectra for GTC and FOS sources on an expanded
scale to emphasize weak lines. The low-$z$ FOS composite is much richer
in faint narrow and semi-broad features. Two emission lines coming
from two different ionic species of nitrogen are detected in quasar
spectra: \niii\ and \niv.  We estimate an upper limit of $\approx$ 0.5\AA\ 
for \niv\ (the line is not visible) and  W(\niii)$\approx$0.6\AA\ for 
the GTC median composite. Both lines are clearly 
seen in the low-$z$  composites with W(\niv)$\approx$0.8\AA\ 
and W(\niii)$\approx$1.6\AA.  The weakness or even absence of 
both of them  in the GTC composites implies lower nitrogen 
abundance in the high z sources. The \ion{Si}{ii}\l 1264, 
\ion{O}{i}\l1304 and \oiiiuv\ lines  (from $\alpha$ elements), as 
well as the \feii\ UV 191 blend  are certainly detected in the FOS 
composite.  The weaker UV 191 \feii\ multiplet in the GTC spectrum 
points toward a clear role for excitation conditions and chemical 
abundance. Both standard diagnostic line ratios and the comparison 
of composites are evidence for a metallicity redshift gradient.

\subsection{\mbh\ and \lledd\ at high and low $z$.}
\label{masses}

Our first attempt to estimate black hole masses and Eddington ratios 
revealed no significant difference between population A and B sources. We 
considered virial FWHM values for five UV lines (\lya, \civ, \aliii, \siiii, and \ciii) as 
well as an average FWHM of  all five lines. In a few cases not all the lines 
could be measured. The first line is dangerous for many reasons.
The second is our strongest and least contaminated  line but different studies 
suggest it is not reliable as a virial estimator \citep{sulenticetal07,netzeretal07}. 
\ciii\ is seriously blended and similarly FWHM \ciii\ shows no correlation with 
FWHM \hb.  Recent studies using high S/N spectra \citep{negreteetal13a,negreteetal14} suggest 
that \aliii\ and \siiii\ are likely the safest virial estimators in UV spectra of quasars. 
\aliii\ is preferred because it is on the blue edge of the $\lambda$1900 emission 
line  blend. 

We do not report \mbh\ and \lledd\ estimates for individual sources  but follow the 
procedure developed  in \citet{sulenticetal02} and \citet{marzianietal09} which
involves producing high S/N composite spectra in the 4DE1 context.
This was already done for our low-$z$~control sample \citep{bachevetal04}
where composites for the five most populated bins were generated. The size
of the GTC sample allows only a simple binning into Pop. A and B sources.  
Composite spectra are shown in Fig. \ref{fig:popab} with {\sc specfit} results in Fig.
\ref{fig:popabfits}. The GTC binning was partially validated
in an earlier section where \civ\  measures for Pop. A and B sources at high- and 
low-$z$  were compared and found to be very similar. Pop. A -- B differences 
previously established for the low-$z$ sample were also confirmed in the GTC 
sample. Our best chance for  making a reliable comparison of \mbh\ and
\lledd\ lies with comparison of the high and low-$z$ composites where the
range of source luminosities of both  samples is small and similar.
Use of composites yields spectra with S/N high enough to resolve the 1900 
blend and measure FWHM \aliii\ and \siiii\ with reasonable accuracy. We avoid 
using the other lines as virial estimators.

Armed with a virial broadenin?g estimator we computed \mbh\  from the 
standard relation of \citet[][see \citealt{marzianisulentic12} for a recent review on \mbh\ 
derivation in quasars]{vestergaardpeterson06}. This scaling law is preferred over a more recent 
formulation \citep{shenliu12} because the latter applies most directly to sources with 
higher luminosity than the ones in our samples. Table \ref{tab:mlm} presents estimates 
of \mbh\  and \lledd\  derived using virial FWHM measures listed in Table \ref{tab:measprofb04}. 
Values are given for the four median composite spectra displayed in  Fig.\ref{fig:popab}. 
Median source luminosities are also presented in Table \ref{tab:mlm}.
Table \ref{tab:measprofb04} also lists FWHM \civ\ values for comparison; making it clear that 
use of \civ\ as a virial estimator would yield \mbh\ estimates much larger than those derived 
from FWHM \aliii\ and FWHM \siiii. It is difficult to fully estimate effects of sample 
bias in the composite spectra. If the GTC sample is close to random we reasonably expect
most sources to occupy 4DE1 bins A1, A2 and B1. The RL fraction (n=2
usually pop. B) $\approx$10\%\ is the random expectation. We also identify two candidate
extreme pop A sources (possible bin A3). All of these observations coupled with the similarity
of Pop. A and B median \civ\ measures, as well as, pop A and B median differences 
suggest that our samples are reasonably well matched.
Bolometric corrections were applied following \citet{elvisetal94} to
compute \lledd. We confirm the trend found in low $z$ samples where \mbh\ is smaller in 
Pop. A (than Pop. B) sources by 0.67 dex and 0.27 dex for low $z$ and GTC samples
respectively. Median \lledd\  values for Pop. A  are 0.33 and 0.46 dex higher than
Pop. B at low and high $z$.

\section{Discussion}
\label{discuss}

There are likely significant numbers of quasars at $z = 2.3$ with
luminosities (L$_\mathrm{bol} \sim10 ^{46}$\ \ergss) similar to the most 
luminous local quasars and SDSS-III/BOSS has recently greatly 
increased their numbers. One assumes that such quasars are increasingly 
numerous at all redshifts higher than z$\approx$0.6 \citep{glikmanetal11,ikedaetal12}. 
The GTC sample shows that their spectra are similar to low $z$\ analogues. 
Our previously  defined 4DE1 formalism \citep{sulenticetal00b,sulenticetal07} identified
two quasar  populations based on optical, UV and X-ray measures. We made
a first attempt at identifying Pop. A and B sources in the GTC sample
using previously established criteria including: presence/absence
of a C{\sc iv}1549 blueshift/asymmetry, EW C{\sc iv} (low for Pop. A and
high for Pop. B), low/high FWHM \civ\ and line ratios in the 1900 \AA\
blend \citep{bachevetal04,sulenticetal07}. On this basis, high and low
accretors ($\propto$ Eddington ratio) could be tentatively identified.
Eddington ratio appears to be the principal driver of 4DE1 differences
\citep[e.g.,][]{marzianietal01,boroson02,yipetal04,kuraszkiewiczetal09}.
We find 11 Pop. A and 9 Pop. B sources (omitting the lower $z$ and the
red quasar) essentially the same fractional division as observed locally. 
So far the low $L$\ quasar population at $z \approx$ 2.5 appears similar to
the low $z$\ one. The Pop. A  -- B distinction and spectral differences are 
preserved. Observational and physical parameters of GTC and the control 
sample (FOS-CS)  are found to be consistent.

\subsection{A population of moderately accreting quasars at high redshift: are
there real evolutionary effects?}
\label{evol}

The optical luminosity function for quasars (QLF) based on 2dF or 2QZ is
well fit by a pure luminosity evolution model at redshifts lower than
the redshift of the peak in the quasar population
\citep[e.g.,][]{boyleetal00}. At $z \approx 0$ and 2.5 we find turnover absolute
magnitude $M_\mathrm{B}^{\ast}$= --22 and --26 respectively. Thus our samples in the range
--23 to --25 are very luminous at low $z$ and underluminous at $z \approx $ 2.5.
Sources in this luminosity range are still undersampled at $z  >  2 $\ but
the first iteration of the BOSS survey (SDSS III: e.g., \citealt{palanquedelabrouilleetal13}) 
designed to overcome
this deficit requires a double power-law fit to the QLF  with pure
luminosity evolution up to $z \approx 1 - 2$ and a breakdown at higher $z$
\citep[e.g.,][]{richardsetal06}.  At $z \approx  2 - 2.5$ the onset of the monster
quasar population, absent locally, is well developed. The similarity of
the $z \approx $ 2.1 -- 2.3 sample to the local analogues in the control sample
supports the interpretation that we are not sampling the monsters but
rather the underlying  quasar population that grows much more slowly.

Simple expectations for an evolving universe involve an  \mbh\ decrease
and  an  \lledd\ increase at higher $z$ \citep[e.g., ][]{kollmeieretal06}.
A systematic decrease in Eddington ratio after $z \approx$ 2 is what is
believed to cause the fading of the quasar population
\citep{cavalierevittorini00}. However, this effect is dominated by the
evolution of the most luminous ``monster'' quasars.  If we restrict
attention to the less luminous but more numerous population below
the OLF turnover luminosity  it is at least conceivable that \mbh\ could
be systematically smaller (and the Eddington ratio higher) at an age of
$\approx$ 3Gyr after the Big Bang than in the local Universe. 

The luminosity function can be written as
\begin{equation}
\Phi(L, z) dL = \left[ \int \Psi(M_\mathrm{BH}, z) P\left(\frac{L}{M_\mathrm{BH}} | M_\mathrm{BH}, z \right) dM_\mathrm{BH} \right] dL
\end{equation}

where $\Psi(M_\mathrm{BH}, z)$\ is the mass function and $P$\ the probability of having
a given Eddington ratio at mass \mbh\ at $z$. GTC and FOS-CS  have statistically
indistinguishable luminosity distribution; however, this does not rule
out the possibility of systematic differences: for example, systematically
lower masses and higher \lledd \ at high $z$. 

Our best chance for making a reliable comparison of \mbh\ and
\lledd\ lies with comparison of the high and low-$z$\ composites where the
range of source luminosities is small, and similar, for the two samples.
Use of composites make it possible to resolve the 1900 blend and measure
FWHM \aliii\ and \siiii\ with higher accuracy. We assigned a luminosity for the FOS-CS 
composite from the weighted average of the luminosity of each spectral type 
used to build the composite itself. We found that \mbh\ derived from median composites 
at low and high $z$ are consistent.  

We also build a distribution using the FWHM \hb\  (at low-$z$, for the FOS-CS) and the ``virial FWHM"  
the GTC sample. Black hole mass values computed following the photoionization method of 
\citet{negreteetal13a} are  consistent with scaling-laws values, with average difference 
$\delta \log $ \mbh $\approx 0.02 \pm$ 0.21. We infer consistent ranges in \mbh\  
(8 $\la \log M_\mathrm{BH} \la 9$) and \lledd\ at high and low $z$, if   the absolute statistical 
uncertainty of the single-epoch mass estimates ($\pm 0.66$ dex) is taken into account. The \lledd\ 
systematic difference between Pop. A and B is confirmed at low- and high-$z$  by both the distribution 
of \lledd\ and by the \lledd\ values estimated from the composite spectra (Table \ref{tab:mlm}).  
Differences between composite and sample medians are believed to be mainly due to the different 
assumptions employed to compute \mbh\ and \lledd\ at low and high $z$.  

Our analysis therefore does not provide evidence for a large difference in 
$\Psi(M_\mathrm{BH}, z)$\ and $P(\frac{L}{M_\mathrm{BH}}, z)$, implying no major effect 
within the constraints of our small sample. The GTC data allowed us to verify for the 
first time the presence of a population of moderately accreting quasars
with masses  and Eddington ratios similar to those seen in the local
Universe. This is not ruling out a systematic evolution in the
average Eddington ratio with redshift since the massive \mbh\
that were shining at high $z$\ have now disappeared and the
GTC sample luminosity places them in  the high luminosity tail of the
local quasar luminosity function \citep{grazianetal00,croometal04}.
The situation depicted for the quasars at $z \approx$ 2.3 indeed  appears
in close analogy with the ones in the local Universe, where most active
black holes have masses in the range  $10^{8}- 10^{9}$\ solar masses,
while less massive black holes $10^{6}- 10^{7}$\ \msol\ can be accreting at a
higher pace (i.e., they are the local extreme Pop. A sources). In other
words, the observation of a similar spread in \lledd\ in the GTC and
low-$z$ samples is, within the limit of the present data, consistent with
an anti-hierarchical black hole growth scenario
\citep{hasingeretal05,brandtalexander10}: black holes in the mass range
$10^{8}- 10^{9}$\ \msol\ are expected to have already acquired a large
faction of their mass at $z \approx$ 2.4 \ and be entering a phase of
slower growth \citep[e.g.,][]{marconietal06}. Smaller mass and higher
Eddington ratio sources may become more frequent earlier
\citep{kollmeieretal06,netzeretal07,trakhtenbrotetal11}, or may be simply
have been undetected in optical surveys (or even in the soft X-ray domain)
at   $z \approx 2.4$. This latter case should be  seriously taken into
account  since the one object that is heavily obscured, FIRST/F2M
J153150.4+242317, shows properties that are most likely of extreme Pop. A (see discussion 
in Appendix \ref{red}).
Clearly a larger and less biased sample is needed to gain more constraints
on the quasar demographics at $z \approx 2.4$.

\subsection{Do Pop. B sources also show \civ\ blueshifts?}
We used estimates of \civ\ centroid shift and asymmetry in our
separation of high $z $ population A and B sources. Previous work on 
low redshifts sample found the preponderance of blueshifts in 
population A sources. The individual shift values listed in Table \ref{tab:measprof} 
and composite profile values in Table \ref{tab:measprofb04} fully confirm this population A-B 
difference.  We find a few less consistent values in Table \ref{tab:measprof}  which 
could be explained in several ways: 1) sources misclassified as population 
A or B, 2) sources with lower S/N spectra where the shift measure is 
less certain or 3) sources with a somewhat unusual  ratio of the BLUE and 
BC components. The \civ\ line profile in population A sources is 
composed of at least two components: 1) BLUE- usually attributed to a wind 
or outflow and 2) BC-a relatively symmetric and unshifted component 
assumed to be analogous to the primary component seen in H$\beta$\ (i.e. 
the classical BLR used as virial \mbh\ estimator). This is the reason why 
Pop B sources lacking the BLUE component show FWHM \civ\ values more in 
agreement with FWHM H$\beta$. Pop A sources without a significant blueshift 
may involve sources where \civ\ is dominated by emission from the unshifted 
classical BLR component. The overall consistency of the GTC measures 
with low $z$\ results is impressive. 

Analysis of \civ\ profiles in large SDSS samples of quasars
\citep{richardsetal11} suggest that \civ\ blueshifts are quasi ubiquitous 
among the RQ majority of quasars. We do not find this at low redshift where
only Pop. A sources show a blueshift. In this paper we also find
an absence of pop B blueshifts in the higher redshift GTC sample.
The Pop. A-B  difference motivated the inclusion of  \chm\ as the principal 
4DE1 diagnostic involving high ionization broad lines
\citep{sulenticetal00a}. Clearly the Pop. B result for the GTC sample
is consistent with zero blueshift. Taken at face value the low redshift 
results, and now the GTC sample, suggest that about half the quasars
show no \civ\ blueshift.  Are our results really in conflict with 
\citet{richardsetal11}?  Figure 5 of that paper shows the  distribution 
of \civ\ shift vs. $\log L_{\nu}$(1550\AA) for a large SDSS sample.
Of course the S/N of the vast majority of these spectra are inferior 
to the worst of our data. The centroid of the \citet{richardsetal11} 
distribution lies near $\log L_{\nu}$= 30.8 \ergss\ Hz$^{-1}$\ and 
\chm(\civ)= -800 \kms\ for RQ quasars. \civ\ blueshift decreases 
rapidly for lower luminosity quasars and our samples concentrate
near $\log L_{\nu} \approx$30  \ergss\ Hz$^{-1}$ where sources are (as
already noted) rare in the SDSS database. The rare equivalent luminosity
SDSS quasars in Figure 5 of  \citet{richardsetal11} show \civ\ shifts 
between --200 (red) and +1000 (blue) \kms. There is general consistency  
allowing for the large uncertainties for most of these shift measures 
and also considering that \citet{richardsetal11} do not distinguish
between Pop. A and B sources. They do however distinguish between
radio-quiet (RQ) and radio-loud (RL)  sources. RL sources are
largely Pop. B sources in the 4DE1 context  and RL sources show a 
\civ\ centroid blueshift in their Figure 5 (in the L range of our study)
of only about 150 \kms.  If RL can be used as a population B surrogate 
then our results are again roughly consistent with SDSS as explored 
in \citet{richardsetal11}. The blueshift results for luminous quasars do not extend 
  to the lower luminosity quasars studied here,  which might be
an indication that higher luminosity objects are more likely to have a
strong wind component.

\subsection{Is there a \civ\ {\em evolutionary} Baldwin effect?}

Since the 1970s there has been interest in the possibility that \civ\ 
measures might provide a way to use quasars  as standard candles for 
cosmology\citep[c.f.][]{bianetal12}. Interest was sparked by discovery  
of an  apparently strong anticorrelation between EW \civ\ and source 
luminosity in a sample of RL quasars (Baldwin 1977). EW CIV showed a 
change from  $\log$ W(\civ) $\approx$ 2.1  at $\log L_{\nu} 
$(1450)$\approx$ 30.0 \ergss\ Hz$^{-1}$\  to $\log$ W(\civ) $\approx$ 0.9 
at $\log L_{\nu} $(1450)$\approx$ 31.9 \ergss\ Hz$^{-1}$\ (presented 
with appropriate caution). The history of follow-up studies for this ``Baldwin Effect'' 
\citep{sulenticetal00a} revealed an anti-correlation  
between measures of correlation strength and sample size (decreasing 
correlation strength with increasing sample size). More recently 
Eigenvector studies have found that the anticorrelation is most likely 
intrinsic since quasars at a fixed redshift also show it 
\citep{bachevetal04,baskinlaor04,marzianietal08a}. Figure 
\ref{fig:distributions} (top panels) show the distributions of
W(\civ) for our high and low redshift samples. There is no
statistically significant  difference  according to a K-S test. 
We see the Pop. A-B differences mentioned earlier but no difference 
in the EW CIV range between the high and low $z$ samples   We 
find no evidence for a $z$ dependent W(\civ) decrease in the GTC sample. 

If the GTC measures are compared with the data of \citet{kinneyetal90}, the
GTC values are found to be similar to typical values for low-$z$ quasars with the 
same luminosity, arguing against an evolutionary Baldwin effect. Overlaying the GTC 
equivalent width values onto the more recent data of \citet{bianetal12} shows that 
our sample is consistent with the weak anticorrelation detected in large 
samples. Small samples like the GTC are however prone to statistical 
fluctuation.  Considering that the Baldwin effect is a rather weak correlation, 
there is no point in claiming a detection (or a non detection) of a Baldwin 
effect unless the sample exceeds $\sim 100$\ quasars \citep{sulenticetal00a}.  
We can just comment that we find  the full range of observed W(\civ) values 
even in samples with a  restricted luminosity and redshift range: as mentioned, 
the equivalent width dispersion is here not due to a luminosity correlation since 
we are studying a sample with essentially fixed luminosity.  If W(\civ)
correlates with luminosity-related parameters  it is most likely the
source Eddington ratio, i.e., the real effect might be what has been
called  the ``intrinsic'' Baldwin effect \citep[e.g.,][]{marzianietal08a,marzianietal06}.

\subsection{Possible interpretation of lower chemical abundance at high $z$ }
\label{metalsdisc}

Several lines of evidence suggest lower chemical abundances in the line
emitting gas of the GTC sample with respect to the control sample. This
effect -- first seen in our sample -- was probably never detected before
because of the strong correlation between $z$ and $L$ in flux limited
samples.  It seems  unlikely that the strength of inter combination lines
with low critical density could be considered a manifestation of the
``disappearance'' of the  NLR observed in some  high-luminosity sources
\citep{netzeretal04}, since the GTC sample is of ordinary luminosity.

Previous work pointed toward correlation between $Z$ and redshift,
luminosity, \mbh, and Eddington ratio. {A trend involving increasing $Z$
with redshift pointed out in the early mid-1990s has been disproved as
associated with  Malmquist bias }.  At present,
the better defined correlation of metal content appears to be with
luminosity \citep{shinetal13}. This correlation  is especially strong, and
is confirmed by  diagnostics based  on narrow lines
\citep{nagaoetal06b,nagaoetal10}. \citet{dietrichetal09} found that  gas
metallicity of the broad-line region is super-solar with  3 $Z/Z_{\odot}$
in luminous, intermediate-redshift quasars, from measures of the
\niii/\oiiiuv\ and \nv/\civ\ emission line ratios. Also \citet{shinetal13}
show a well defined correlation between $Z$ and luminosity for PG quasars
at low $z$.  A $Z$--\mbh\ correlation has been also claimed
\citep{matsuokaetal11}, in analogy with what found in galaxies
\citep{matteucci12}.  Last, the possibility of a connection between $Z$
and Eddington ratio has been also explored. Although a connection between $Z$ and 
\lledd\ is found in the 4DE1 context, as discussed below, 
the physical origin  of this correlation remains unclear, also because the data shown by
\citet{shinetal13} are still insufficient to prove a statistically unbiased correlation.

We must preliminarily point out that great care should be exerted in
analyzing data based on the \nv/\civ\ and \nv/\heiiuv\ ratios because: (1)
the intensity of the \nv\ line is very difficult to estimate unambiguously
unless a reliable model of the \lya\ wings are build as done by
\citet{shinetal13} and in the present paper; (2) the \heiiuv\ profile is
often shelf-like, the \heiiuv\ lines is blended with much stronger \civ\
and is therefore difficult to measure accurately especially in low S/N
conditions.

The present data suggest a metallicity decrease with $z$ in the same
luminosity range.\footnote{Note that this effect is not detectable in the $Z$-sensitive 
ratios of Tab. \ref{tab:measb04} since the median combination of spectra tend to cancel a trend  that is seen 
in a minority of sources.} Within the limits of our sample (a weak anti correlation
can be spurious in a small sample: also, the lower $Z$ also rely an even
smaller number of sources),   lowest metallicity sources have $Z$ values
around solar or slightly sub-solar, while other sources show super solar
enrichment consistent with the quasar population at the same luminosity.
There is a simple   interpretation for this finding. The GTC quasars are
accreting gas that reflects the chemical composition of the host galaxy;
it is known that there is a clear metallicity decrease with $z$: for fixed
host mass 10$^{11}$ \msol, the O/H ratio would decrease from 2.5 solar,
to 1.5 solar, with a steep decrease with lower host masses
\citep{savaglioetal05}.  For the typical \mbh\ of our sample, the \mbh\ --
bulge mass expected at $z\approx 2.35$
\citep{merlonietal10,schulzewisotzki14} involves stellar masses $\log
M_{\star} \sim 11$ [\msol]. In this case we may well expect metallicity
around solar. 

The important point here is that we do not have evidence of
accretion of ``pristine'' gas: the metallicity is not highly sub-solar. At
the same time, the enrichment is, at least in several cases, not as strong
as the one of many luminous quasars: the accreting gas diagnostic ratios
are not demanding any enrichment associated with circumnuclear star
formation \citep{sanietal10,negreteetal12}, and may be well ascribed to
gas whose metal enrichment has followed the typical processes of the host
galaxy of the interstellar medium. In other words, the black hole is being
fed by gas whose composition appear normal or close to normal for host galaxies at their
cosmic age.

Do these finding allow us to make some more general consideration on the
correlations between $Z$ and physical parameters? In the 4DE1 context,
there is probably no relation between \mbh\ and $Z$: the most massive
sources in the local Universe, those of Pop. B, do not show evidence of
particular enrichment: the metal-content  sensitive lines (\feii, \aliii,
\nv) are all the weakest along the 4DE1 sequence. At the other end of the
sequence are the highest accretors: they show strong \feii\ emission 
and \aliii/\siiii$\ga$ 0.5 and \siiii $\ga$ \ciii\ in the UV 
\citep{marzianisulentic13}, and evidence of a large increase in $Z$, 
possibly demanding circumnuclear star formation with a top-loaded
IMF and/or a special timing in the active nucleus evolution
\citep{negreteetal12}.  Therefore, a metallicity trend with \lledd\ might be expected.
Significant metal enrichment  is consistent with a scenario in which the
rapidly-accreting phase of the quasar black hole follows  gas
accumulation, collapse and star formation in the circumnuclear regions
\citep[see
e.g.,][]{hopkinsetal06,sandersetal09,tengvielleux10,trakhtenbrotetal11}.
The quasars of the GTC sample do not belong to extreme sources. Their most
frequent spectral type is likely A1 or B1, with at most 2 sources whose
properties are more consistent with A2/A3. Eddington ratio values are
modest, and mass inflow rates are estimated to be in the range 0.3 -- 1  \msol\ yr$^{-1}$.
Therefore, the GTC quasars show properties that are consistent with the
$Z$ -- \lledd\ relation suggested by the 4DE1 contextualization.

\section{Conclusion}

We have analyzed GTC spectra for 22 sources near redshift
$z \approx$ 2.35 using techniques similar to those employed
for low-$z$ quasars observed in the UV by HST-FOS. The latter
provide an effective comparison sample of luminosity analogues.
The properties of the GTC sample are quite similar to the low
redshift quasars in the sense that we observe roughly equal
numbers of Population A and B high/low accretors. The same
Pop. A-B median properties (and Pop. A-B differences) are seen
at high and low redshift -- this comparison relies on the strong
\civ\ line previously selected as a 4DE1 diagnostic. We find
no evidence for an ``evolutionary'' Baldwin effect involving
\civ. We find evidence that the \civ\ blueshift, apparently
quasi-ubiquitous among high luminosity quasars, is present in
only half of both high and low redshift quasars in the moderate
luminosity range studied here. It is a high accreting
population A property. If Pop. A sources are on average younger by
virtue of their smaller inferred \mbh\ (and larger \lledd\  values), our high
redshift sample does not show any significant increase in the fraction
of ``younger''\  Pop.  A sources. However, there is a luminosity selection
operating here that lets us miss low \mbh\ sources with   \mbh\ significantly below $10^8$\ \msol\ even if
they are radiating at Eddington limit (with our magnitude limits 
no source below $\log$ \mbh\ $\approx 7.5$ should be detectable even with \lledd $\approx$1).

Evidence for an excess of lower metallicity quasars in the high redshift
sample is provided by two diagnostic ratios based on strong emission lines
as well as by the overall strength of fainter metal lines in Pop. A
sources. These finding are consistent  with the evolution of metal content
in the stellar populations and presumably in the interstellar medium of
the host galaxy.

\begin{acknowledgements}

Part of this work was supported by Junta de Andaluc\'{\i}a through Grant TIC-114 and Proyecto
de Excelencia P08-FQM-4205 as well as by the Spanish Ministry for Science and Innovation through 
Grant AYA2010-15169. PM wishes to thank the IAA for supporting her visit in March 2014. 
DD acknowledges support from grant PAPIIT107313, UNAM. Based on observations made with the Gran 
Telescopio Canarias (GTC), instaled in the Spanish Observatorio del 
Roque de los Muchachos of the Instituto de Astrof\'{\i}sica de Canarias, in 
the island of La Palma. We thank all the GTC Staff, and especially René Rutten and Antonio Cabrera, 
for their support with the observations. We would like to thank Josefa Masegosa 
for all the fruitful discussions on the subject. We also thank the referee for 
many useful comments which helped to significantly improve the 
presentation of the GTC survey. This research has made use of the VizieR catalogue 
access tool, CDS, Strasbourg, France. The original description of t
he VizieR service was published in A\&AS 143, 23. This research has also made use 
of the NASA/IPAC Extragalactic Database (NED), which is operated by 
the Jet Propulsion Laboratory, California Institute of Technology, under contract with the 
National Aeronautics and Space Administration. 

\end{acknowledgements}

\bibliographystyle{aa}

\clearpage



\begin{sidewaystable*}
\setlength{\tabcolsep}{5pt}
\begin{center}
\caption{Source Identification and Basic Properties\label{tab:id}}
\begin{tabular}{lcccccclllll}\hline\hline
\multicolumn{1}{l}{NED Identification} & \multicolumn{1}{c}{$m_V^\mathrm{a}$}  &\multicolumn{1}{c}{$z$}  & \multicolumn{1}{c}{$\delta z$} &\multicolumn{1}{c}{$M_\mathrm{B}^\mathrm{b}$}   &  Discovery   & Ref. & Alternate Name &
\multicolumn{1}{l}{Notes} \\ \hline
\object{OH91 073}	&	20.9	&	2.2630	&	0.0016	&	-24.4	&	grism/grens	&	1	& NVSS J121848-110332, Q 1216-1046 &	NVSS 38.9$\pm$1.6 mJy, \\
&&&&&&&&  RL serendipitous	\\ 
\object{OH91 119}	&	20.9	&	2.2660	&	0.0005	&	-24.4	&	grism/grens	&	1	&	Q 1232-1059	\\
\object{OH91 121}	&	21.0	&	2.4065	&	0.0027	&	-23.8	&	grism/grens	&	1	&	Q 1232-1113	\\
B1.0334	&	21.6	&	2.2999	&	0.0012	&	-23.7	&	X-ray	&	2	&	\object{SDSS J123742.52+621811.6}	\\
\object{Wee 87}	&	20.8	&	2.2418	&	0.0022	&	-24.3	&	grism/grens	&	3	&	HB89 1257+357	\\
\object{HB89 1340+277}	&	21.4	&	2.1842	&	0.0023	&	-23.7	&	grism/grens	&	3	&	Q 1340+2744	\\
\object{2QZ J134206.3-003702}	&	20.3	&	2.2115	&	0.0028	&	-24.4	&	color (2dF)	&	4	&	SDSS J134206.34-003701.2	\\
\object{F864:158}	&	21.7	&	2.2297	&	0.0023	&	-23.4	&	color 	&	5	&	SDSS J134423.95-002846.4	\\
\object{2QZ J143400.0+002649}	&	20.8	&	2.2054	&	0.0005	&	-24.3	&	color (2dF)	&	6	&	 	\\
\object{FIRST J153150.4+242317}	&	20.5	&	2.2841	&	0.0011	&	-24.7	&	IR	&	7	&SDSS/F2M J153150.4+242317 &Heavily obscured; \\
&&&&&&&&  RL serendipitous	\\
\object{TXS 1529-230}	&	21.4	&	2.2800	&	0.0020	&	-23.8	&	Radio	&	8	& PMN J1532-2310 & NVSS 138.9$\pm$4.2 mJy	\\
\object{CADIS 16h-1610}	&	20.8	&	2.2717	&	0.0016	&	-24.2	&	color + FP?	&	9	&	SDSS J162359.21+554108.7	\\
\object{CADIS 16h-1373}	&	21.9	&	2.2780	& 	0.0018	&	-23.2	&	color + FP?	&	9	&	SDSS J162421.29+554243.0	\\
\object{Wee 155}	&	21.0	&	1.5981	&	0.0049	&	-24.2	&	grism/grens	&	3	& 	HB89 1634+332&$z\approx$ 2.37 in catalogues	\\
\object{SDSS J164226.90+405034.3}	&	20.4	&	2.3258	&	0.0005	&	-25.0	&	grism/grens	&	10	&	Q 1640+4056 &	\\
SPIT 17196+5922	&	20.7	&	2.2255	&	0.0015	&	-24.1	&	IR/Spitzer	&	11	&\object{SDSS J171941.24+592242.1}	&	\\
\object{VCV96 1721.4+3401}	&	20.9	&	2.3244	&	0.0055	&	-24.1	&	grism/grens	&	12	&	Q 1721+3400	\\
\object{Wee 173}	&	20.2	&	2.4144	&	0.0016	&	-25.3	&	grism/grens	&	3	&	HB89 1834+509	\\
\object{Wee 174}	&	20.6	&	2.2720	&	0.0008	&	-24.8	&	grism/grens	&	3	&	HB89 1835+509	\\
\object{CXOMP J205620.5-043100}	&	21.2	&	2.3299	&	0.0032	&	-23.6	&	X-ray	&	13	&	 	CXO J205620.5-043100\\
\object{VCV96 2240.7+0066}	&	20.4	&	2.2579	&	0.0019	&	-24.5	&	grism/grens	&	12	&	Q 2240+0066 	\\
\object{CXOMP J234752.5+010306}	&	20.5	&	2.4043	& 	0.0014	&	-24.3	&	X-ray	&	13	&	SDSS J234752.55+010305.0	\\
\hline
\end{tabular}
\end{center}
\begin{list}{}{}
\item[$^\mathrm{a}$]{Indicative  $V$ magnitudes as reported by \citet{veroncettyveron10}.  }
\item[$^\mathrm{b}$]{Absolute B magnitude at $z=0$ as tabulated by \citet{veroncettyveron10}.}
\item[]{1: \citet{osmerhewett91}; 2: \citet{brandtetal01}; 3: \citet{weedman85}; 4: \citet{croometal01}; 5: \citet{boyleetal91}; 6: \citet{croometal04}; 7: \citet{glikmanetal07}; 8: \citet{griffithetal94}; 9: \citet{wolfetal99}; 10: \citet{cramptonetal88}; 11: \citet{papovichetal06}; 12: \citet{cramptonetal85}; 13: \citet{silvermanetal05}}
\end{list}
\end{sidewaystable*}

\newpage
\clearpage

\begin{sidewaystable*}
\setlength{\tabcolsep}{5pt}
\begin{center}
\caption{Log of Observations\label{tab:obs}}
\begin{tabular}{llccccc}\hline\hline
 \multicolumn{1}{l}{QSO Identification} & \multicolumn{1}{l}{Date} & \multicolumn{1}{c}{ET} & 
 \multicolumn{1}{c}{Exp. N.} &\multicolumn{1}{c}{S/N at} &\multicolumn{1}{c}{Slit width} &
 \multicolumn{1}{c}{seeing} \\
 \multicolumn{1}{l}{}  & \multicolumn{1}{l}{Observation} & \multicolumn{1}{c}{Time (s)} & 
 \multicolumn{1}{c}{} & \multicolumn{1}{c}{1450\AA} &\multicolumn{1}{c}{(arcsecs)} &
 \multicolumn{1}{c}{(arcsecs)} \\ 
 \hline
$[\rm OH91]$ 073       & 26-03-2011 & 2400 & 4 & 14 & 1.0  & 1.55 \\
$[\rm OH91]$ 119       & 26-03-2011 & 2400 & 4 & 10 & 1.0  & 1.3   \\
$[\rm OH91]$ 121       & 07-04-2011 & 2400 & 4 & 20 & 1.0  & 1.0   \\
 B1.0334               & 07-04-2011 & 2400 & 4 & 10 & 1.0  & 1.15  \\
Wee 87                 & 07-04-2011 & 2400 & 4 & 30 & 1.0  & 0.99  \\
$[\rm HB89]$ 1340+277  & 02-05-2013 & 2400 & 4 &  5 & 1.0  & 1.0   \\
 2QZ J134206.3-003702  & 02-05-2011 & 2400 & 4 & 12 & 1.0  & 0.98  \\
F864:158               & 06-05-2011 & 2400 & 4 &  5 & 1.0  & 1.3   \\
2QZJ143400.0+002649    & 05-05-2013 & 2600 & 4 & 11 & 1.23 & 1.27    \\
FIRST J153150.46+242317.7 & 02-05-2011 & 2400 & 4 &  9 & 1.0  & 0.80 \\ 
TEX 1529-230           & 07-06-2011 & 2400 & 4 &  6 & 1.0  & 1.56  \\
CADIS 16h-1610         & 08-07-2013 & 3250 & 5 & 12 & 1.23 & 0.97  \\
CADIS 16h-1373         & 14-06-2013 & 3250 & 5 & 10 & 1.23 & 0.82  \\
Wee 155                & 02-05-2011 & 2400 & 4 &  5 & 1.0  & 1.10  \\
 SSDS J164226.90+405034.3 & 08-04-2011 & 2400 & 4 &  6 & 1.0  & 1.25  \\
 SPIT 17196+5922       & 06-04-2011 & 2400 & 4 & 11 & 1.0  & 1.01  \\
$[\rm VCV96]$ 1721.4+3401  & 06-05-2013 & 2600 & 4 &  7 & 1.23 & 1.15  \\
Wee 173               & 05-05-2013 & 2600 & 4 & 40 & 1.23 & 0.99  \\
Wee 174               & 11-06-2013 & 2600 & 4 & 21 & 0.8  & 1.2   \\
 CXOMP J205620.5-043100   & 04-07-2011 & 2400 & 4 &  8 & 1.0  & 1.15  \\
$[\rm VCV96]$ 2240.7+0066  & 03-07-2011 & 2400 & 4 & 14 & 1.0  & 1.20  \\
CXOMP J234752.5+010306  & 11-06-2013 & 2600 & 4 & 12 & 0.8  & 0.8   \\
 \hline
\end{tabular}
\end{center}
\end{sidewaystable*}

\newpage
\clearpage

\begin{sidewaystable*}
\setlength{\tabcolsep}{6pt}
\begin{center}
\caption{Measured Quantities \label{tab:meas}}
\begin{tabular}{lcccccccccccrrrr}\hline\hline
\multicolumn{1}{l}{NED Identification} & \multicolumn{1}{c}{$f_\lambda^\mathrm{a}$}  &\multicolumn{1}{c}{$\delta f_\lambda$}& \multicolumn{2}{c}{Ly$\alpha$} &\multicolumn{1}{c}{N{\sc v}}     &\multicolumn{1}{c}{O{\sc iv}]+Si{\sc iv}} & \multicolumn{3}{c}{\civ} & \multicolumn{1}{c}{Al{\sc iii}}  &\multicolumn{1}{c}{Si{\sc iii}]} &\multicolumn{1}{c}{C{\sc iii}]} &\multicolumn{1}{c}{$\lambda$1900} \\
\cline{4-5} \cline{8-10}
&& & W [\AA]$^\mathrm{b}$ & $F^\mathrm{c}$ & $F^\mathrm{c}$ & $F^\mathrm{c}$ &  W [\AA]$^\mathrm{b}$ & $F^\mathrm{c}$ & $F_\mathrm{blue}/F$ & $F^\mathrm{c}$ & $F^\mathrm{c}$ & $F^\mathrm{c}$ & $F^\mathrm{c}$\\ \hline
\\ \hline
\multicolumn{14}{c}{Population A}\\ \hline
\\
OH91 119	&	0.35	&	0.04	&	58	&	23.3	&	9.6	&	4.0	&	36	&	10.8	&	0.38	&	0.7	&	1.8	&	3.2	&	5.7	\\
OH91 121	&	1.24	&	0.06	&	55	&	80.4	&	23.9	&	14.0	&	37	&	42.5	&	0.20	&	3.5	&	9.6	&	5.6	&	18.8	\\
Wee 87	&	1.29	&	0.06	&	90	&	153.3	&	27.0	&	10.5	&	51	&	59.8	&	0.09	&	2.7	&	5.4	&	13.8	&	21.9	\\
F864:158	&	0.16	&	0.03	&	137	&	27.4	&	5.9	&	2.0	&	72	&	11.2	&	0.20	&	0.5	&	1.1	&	1.8	&	3.4	\\
2QZ J143400.0+002649	&	0.61	&	0.06	&	68	&	52.1	&	7.9	&	4.4	&	26	&	15.6	&	0.11	&	1.1	&	4.4	&	4.5	&	10.0	\\
SPIT 17196+5922	&	0.76	&	0.07	&	66	&	65.5	&	16.2	&	6.0	&	48	&	32.3	&	0.11	&	2.0	&	5.1	&	7.8	&	14.9	\\
Wee 173	&	2.36	&	0.21	&	89	&	261.7	&	34.0	&	17.0	&	44	&	100.5	&	0.09	&	1.0	&	6.4	&	15.9	&	23.3	\\
Wee 174	&	0.44	&	0.02	&	70	&	37.4	&	7.8	&	3.6	&	47	&	19.7	&	0.06	&	0.2	&	1.3	&	4.0	&	5.5	\\
 CXOMP J205620.5-043100 	&	0.26	&	0.04	&	89	&	24.8	&	8.8	&	4.3	&	51	&	12.9	&	0.31	&	1.7	&	4.1	&	2.8	&	8.6	\\
VCV96 2240.7+0066	&	0.66	&	0.05	&	101	&	73.7	&	11.1	&	8.0	&	50	&	29.6	&	0.19	&	1.5	&	3.2	&	8.2	&	13.0	\\
CXOMP J234752.5+010306	&	0.36	&	0.03	&	144	&	49.9	&	4.8	&	2.9	&	46	&	17.4	&	0.10	&	1.8	&	2.5	&	5.1	&	9.4	\\
\\ \hline
\multicolumn{14}{c}{Population B}\\ \hline
\\												
OH91 073	&	0.41	&	0.03	&	87	&	44.1	&	10.2	&	6.8	&	75	&	27.4	&	0.08	&	1.0	&	1.5	&	4.7	&	10.6	\\
B1.0334         	&	0.40	&	0.05	&	73	&	41.9	&	9.6	&	7.6	&	118	&	39.0	&	0.14	&	1.1	&	1.5	&	5.2	&	11.8	\\
HB89 1340+277	&	0.12	&	0.03	&	252	&	20.5	&	1.4	&	1.8	&	141	&	14.4	&	0.17	&	0.5	&	0.5	&	1.3	&	2.7	\\
2QZ J134206.3-003702	 	&	0.48	&	0.04	&	72	&	47.7	&	6.5	&	6.1	&	47	&	19.9	&	0.20	&	2.8	&	3.4	&	8.5	&	18.9	\\
TEX 1529-230 	&	0.51	&	0.08	&	72	&	36.7	&	21.2	&	6.5	&	64	&	31.6	&	0.02	&	0.6	&	2.6	&	6.8	&	13.4	\\
CADIS 16h-1610	&	0.38	&	0.03	&	77	&	42.2	&	9.3	&	6.8	&	59	&	19.5	&	0.10	&	1.4	&	2.6	&	5.4	&	11.8	\\
CADIS 16h-1373	&	0.11	&	0.02	&	277	&	27.3	&	5.8	&	1.5	&	129	&	13.2	&	0.09	&	0.6	&	0.8	&	2.6	&	5.2	\\
SDSS J164226.90+405034.3	&	0.31	&	0.06	&	132	&	60.1	&	4.2	&	2.9	&	79	&	19.4	&	0.07	&	0.3	&	0.7	&	1.3	&	2.8	\\
VCV96 1721.4+3401	&	0.27	&	0.04	&	77	&	27.8	&	7.4	&	4.3	&	90	&	19.5	&	0.32	&	0.8	&	1.9	&	2.9	&	6.5	\\
\hline
\end{tabular}
\end{center}
\begin{list}{}{}
\item[$^\mathrm{a}$]{Rest frame specific flux at 1450 \AA\  in units of 10$^{-15}$ erg  s$^{-1}$ cm$^{-2}$\AA$^{-1}$.}
\item[$^\mathrm{b}$]{Rest frame equivalent width in \AA.  }
\item[$^\mathrm{c}$]{Rest frame   flux   in units of 10$^{-15}$ erg  s$^{-1}$ cm$^{-2}$.}
\end{list}
\end{sidewaystable*}

\newpage
\clearpage

\begin{sidewaystable*}
\setlength{\tabcolsep}{6pt}
\begin{center}
\caption{Measured \civ\ Line Profile Quantities  \label{tab:measprof}}
\begin{tabular}{lcccccccccccrrrr}\hline\hline
\multicolumn{1}{l}{NED Identification}  &\multicolumn{1}{c}{FWHM(\civ)}& \multicolumn{1}{c}{A.I.} &\multicolumn{1}{c}{Kurt.}     &\multicolumn{1}{c}{$c(\frac{1}{4})$} & \multicolumn{1}{c}{$c(\frac{1}{2})$} \\
&   [\kms] & & & [\kms] & [\kms]\\ 
\hline
\\ \hline
&\multicolumn{3}{c}{Population A}&\\ \hline
\\
OH91 119	&	 4510	$\pm$	260	&	-0.17	$\pm$	0.06	&	0.41	$\pm$	0.05	&	-1270	$\pm$	210	&	-1170	$\pm$	130	\\
OH91 121	&	 		6990	$\pm$	370	&	-0.18	$\pm$	0.06	&	0.37	$\pm$	0.07	&	-1060	$\pm$	320	&	-1160	$\pm$	190	\\
Wee 87	&	 	4610	$\pm$	390	&	-0.14	$\pm$	0.09	&	0.31	$\pm$	0.04	&	-390	$\pm$	360	&	-50	$\pm$	190	\\
F864:158	&	 	4880	$\pm$	540	&	-0.26	$\pm$	0.07	&	0.28	$\pm$	0.04	&	-830	$\pm$	290	&	-380	$\pm$	270	\\
2QZ J143400.0+002649	& 	4800	$\pm$	930	&	-0.23	$\pm$	0.07	&	0.26	$\pm$	0.03	&	-1130	$\pm$	320	&	-500	$\pm$	470	\\
SPIT 17196+5922	 &	5160	$\pm$	400	&	-0.11	$\pm$	0.08	&	0.33	$\pm$	0.04	&	-480	$\pm$	340	&	-330	$\pm$	200	\\
Wee 173	&	                 4230	$\pm$	350	&	-0.21	$\pm$	0.11	&	0.27	$\pm$	0.04	&	-1130	$\pm$	470	&	-310	$\pm$	170	\\
Wee 174	&	 	         5070	$\pm$	360	&	-0.04	$\pm$	0.08	&	0.34	$\pm$	0.04	&	-210	$\pm$	350	&	-200	$\pm$	180	\\
CXOMP J205620.5-043100 	 &	6240	$\pm$	280	&	-0.07	$\pm$	0.06	&	0.49	$\pm$	0.04	&	-820	$\pm$	270	&	-950	$\pm$	140	\\
VCV96 2240.7+0066 	&	5300	$\pm$	340	&	-0.08	$\pm$	0.07	&	0.37	$\pm$	0.05	&	-490	$\pm$	310	&	-490	$\pm$	170	\\
SDSS J234752.55+010305.0	 	&	5600	$\pm$	620	&	-0.19	$\pm$	0.08	&	0.28	$\pm$	0.03	&	-900	$\pm$	390	&	-260	$\pm$	310	\\
\\
 \hline
&\multicolumn{3}{c}{Population B}&\\ \hline
\\
OH91 073	 &	4170	$\pm$	250	&	-0.02	$\pm$	0.11	&	0.38	$\pm$	0.05	&	130	$\pm$	360	&	190	$\pm$	130	\\
B1.0334	 	&	7580	$\pm$	580	&	0.08	$\pm$	0.09	&	0.31	$\pm$	0.04	&	530	$\pm$	650	&	-10	$\pm$	290	\\
HB89 1340+277	 	 	&	4620	$\pm$	280	&	-0.13	$\pm$	0.18	&	0.35	$\pm$	0.07	&	-860	$\pm$	740	&	-360	$\pm$	140	\\
SDSS J134206.34-003701.2	 &	6880	$\pm$	440	&	0.00	$\pm$	0.07	&	0.38	$\pm$	0.04	&	-220	$\pm$	410	&	-250	$\pm$	220	\\
TEX 1529-230		&	5840	$\pm$	340	&	0.11	$\pm$	0.10	&	0.39	$\pm$	0.05	&	1040	$\pm$	490	&	640	$\pm$	170	\\
CADIS 16h-1610	& 	5930	$\pm$	340	&	-0.03	$\pm$	0.08	&	0.40	$\pm$	0.05	&	10	$\pm$	370	&	110	$\pm$	170	\\
CADIS 16h-1373	 	&	5460	$\pm$	310	&	-0.01	$\pm$	0.08	&	0.41	$\pm$	0.05	&	-10	$\pm$	310	&	40	$\pm$	150	\\
SDSS J164226.90+405034.3	 	&	3180	$\pm$	180	&	0.00	$\pm$	0.07	&	0.42	$\pm$	0.04	&	-100	$\pm$	180	&	-110	$\pm$	90	\\
VCV96 1721.4+3401	&	 	8740	$\pm$	490	&	0.12	$\pm$	0.06	&	0.43	$\pm$	0.04	&	-120	$\pm$	410	&	-650	$\pm$	240	\\
\hline
\end{tabular}
\end{center}
\end{sidewaystable*}
\newpage
\clearpage

\begin{sidewaystable*}
\setlength{\tabcolsep}{6pt}
\begin{center}
\caption{Measured Quantities on Composite Spectra \label{tab:measb04}}
\begin{tabular}{lcccccccccccrrrr}\hline\hline
\multicolumn{1}{l}{NED Identification} & \multicolumn{2}{c}{Ly$\alpha$} &\multicolumn{1}{c}{N{\sc v}}     &\multicolumn{1}{c}{O{\sc iv}]+Si{\sc iv}} & \multicolumn{3}{c}{\civ} & \multicolumn{1}{c}{Al{\sc iii}}  &\multicolumn{1}{c}{Si{\sc iii}]} &\multicolumn{1}{c}{C{\sc iii}]} &\multicolumn{1}{c}{$\lambda$1900} \\
\cline{2-3} \cline{6-8}
  & W [\AA]$^\mathrm{b}$ & $F^\mathrm{c}$ & $F^\mathrm{c}$ & $F^\mathrm{c}$ &  W [\AA]$^\mathrm{b}$ & $F^\mathrm{c}$ & $F_\mathrm{blue}/F$ & $F^\mathrm{c}$ & $F^\mathrm{c}$ & $F^\mathrm{c}$ & $F^\mathrm{c}$\\ \hline
Pop. A Average	&	 	78	&	87 	&	19.1	&	8.4	&	46	&	42.9	&	0.10	&	1.9	&	7.3	&	8.5	&	17.7	\\
Pop. B Average 	&	 	84	&	103.0	&	38.3	&	13.9	&	77	&	69.3	&	0.09	&	3.6	&	7.0	&	13.3	&	30.5	\\
\\
Pop. A Median	&	 	74	&	80.0	&	16.7	&	9.0	&	45	&	42.7	&	0.18	&	1.8	&	5.8	&	6.6	&	14.1	\\
Pop. B Median	&	 	121	&	131.5	&	26	&	13.0	&	69	&	62.2	&	0.07	&	3.1	&	3.9	&	14.5	&	26.0	\\
\\
Pop. A B04	&	 	87	&	107.7	&	15	&	13.9	&	51	&	49.7	&	0.13	&	2.5	&	5.4	&	11.4	&	19.3	\\
Pop. B B04	&		95	&	115.9	&	27	&	11.5	&	83	 &	77.5	&	0.11	&	2.6	&	3.5	&	13.0	&	23.5	\\
\hline
\end{tabular}
\end{center}
\begin{list}{}{}
\item[$^\mathrm{a}$]{Rest frame specific flux at 1450 \AA\  in units of ergs s$^{-1}$ cm$^{-2}$\AA$^{-1}$.}
\item[$^\mathrm{b}$]{Rest frame equivalent width in \AA.  }
\item[$^\mathrm{c}$]{Rest frame   flux   in units of ergs s$^{-1}$ cm$^{-2}$.}
\end{list}
\end{sidewaystable*}


\begin{sidewaystable*}
\setlength{\tabcolsep}{6pt}
\begin{center}
\caption{Measured Line Profile Quantities on Composite Spectra \label{tab:measprofb04}}
\begin{tabular}{lcccccccccccrrrr}\hline\hline
\multicolumn{1}{l}{NED Identification} & \multicolumn{1}{c}{Virial FWHM$^\mathrm{a}$} &\multicolumn{1}{c}{FWHM(\civ)$^\mathrm{b}$}& \multicolumn{1}{c}{A.I.$^\mathrm{c}$} &\multicolumn{1}{c}{Kurt.$^\mathrm{c}$}     &\multicolumn{1}{c}{$c(\frac{1}{4})^\mathrm{c}$} & \multicolumn{1}{c}{$c(\frac{1}{2})^\mathrm{c}$} \\
&  [\kms] & [\kms] & & & [\kms] & [\kms]\\ 
\hline
Pop. A Average	&	3470:	&	5220	$\pm$	430	&	-0.12	$\pm$	0.08	&	0.32	$\pm$	0.04	&	-730	$\pm$	370	&	-440	$\pm$	220	\\
Pop. B Average	&	5080	 	&	6290	$\pm$	380	&	0.01	$\pm$	0.09	&	0.39	$\pm$	0.05	&	10	$\pm$	460	&	-20	$\pm$	190	\\
\\
Pop. A Median	&	2880	$\pm$	320$^\mathrm{a}$	&	5010	$\pm$	390	&	-0.10	$\pm$	0.10	&	0.31	$\pm$	0.04	&	-570	$\pm$	430	&	-270	$\pm$	200	\\
Pop. B Median 	&	4430	 	&	5830	$\pm$	350	&	0.02	$\pm$	0.10	&	0.38	$\pm$	0.05	&	350	$\pm$	450	&	270	$\pm$	170	\\
\\
Pop. A B04	&	2430	$\pm$	100	&	4060	$\pm$	400	&	-0.15	$\pm$	0.09	&	0.31	$\pm$	0.04	&	-770	$\pm$	310	&	-500	$\pm$	200	\\
Pop. B B04	&	4300	&	5810	$\pm$	450	&	0.00	$\pm$	0.12	&	0.36	$\pm$	0.06	&	10	$\pm$	590	&	-10	$\pm$	230	\\

\hline
\end{tabular}
\end{center}
\begin{list}{}{}
\item[$^\mathrm{a}$]{Virial FWHM measured on \aliii\ and \siiii\ in accordance with the 
finding of \citet{negreteetal13a}. The uncertainty is the standard deviation computed on the 
two lines FWHM. If no uncertainty is given, the fit was obtained with the same FWHM value for 
both  \aliii\ and \siiii.  }
\item[$^\mathrm{b}$]{FWHM of the \civ\ line measured on the full profile, i.e., including 
BC and VBC and BLUE when appropriate.}
\item[$^\mathrm{c}$]{Measures of the \civ\ full line profile 
asymmetry index, kurtosis and centroids at one quarter and 
half maximum on; see \citealt{zamfiretal10} for definitions of parameters.}
\end{list}
\end{sidewaystable*}

\newpage
\clearpage

\begin{table*}
\setlength{\tabcolsep}{5pt}
\begin{center}
\caption{\mbh\ and \lledd\ estimates from composite spectra \label{tab:mlm}}
\begin{tabular}{llcccccl}\hline\hline
 \multicolumn{1}{l} {Spectrum} & \multicolumn{1}{c}{$\log \lambda L_\lambda^\mathrm{a}$} &   \multicolumn{1}{c}{$\log L$} &\multicolumn{1}{c}{$\log$\mbh} & 
 \multicolumn{1}{c}{$\log$\lledd}  \\
& \multicolumn{1}{c}{[\ergss]} &\multicolumn{1}{c}{[\ergss]} & \multicolumn{1}{c}{[\msol]} \\ \hline
Pop. A median		&	45.49	&45.99	&8.37	&-0.56\\
Pop. B median		&	45.30	&45.80&	8.64&	-1.02	\\			
\\
Pop. A B04 composite&	45.03&	46.03&	8.19&	-0.35\\
Pop. B B04 composite&	45.46&	46.47&	8.86	&-0.62\\
 \hline
\end{tabular}
\end{center}
\begin{list}{}{}
\item[$^\mathrm{a}$]{V luminosity for the B04 composites and 1450 \AA\ $\lambda L_\lambda$ for the GTC medians.}
\end{list}
\end{table*}

\clearpage

\clearpage

\begin{figure*}
\includegraphics[scale=0.2]{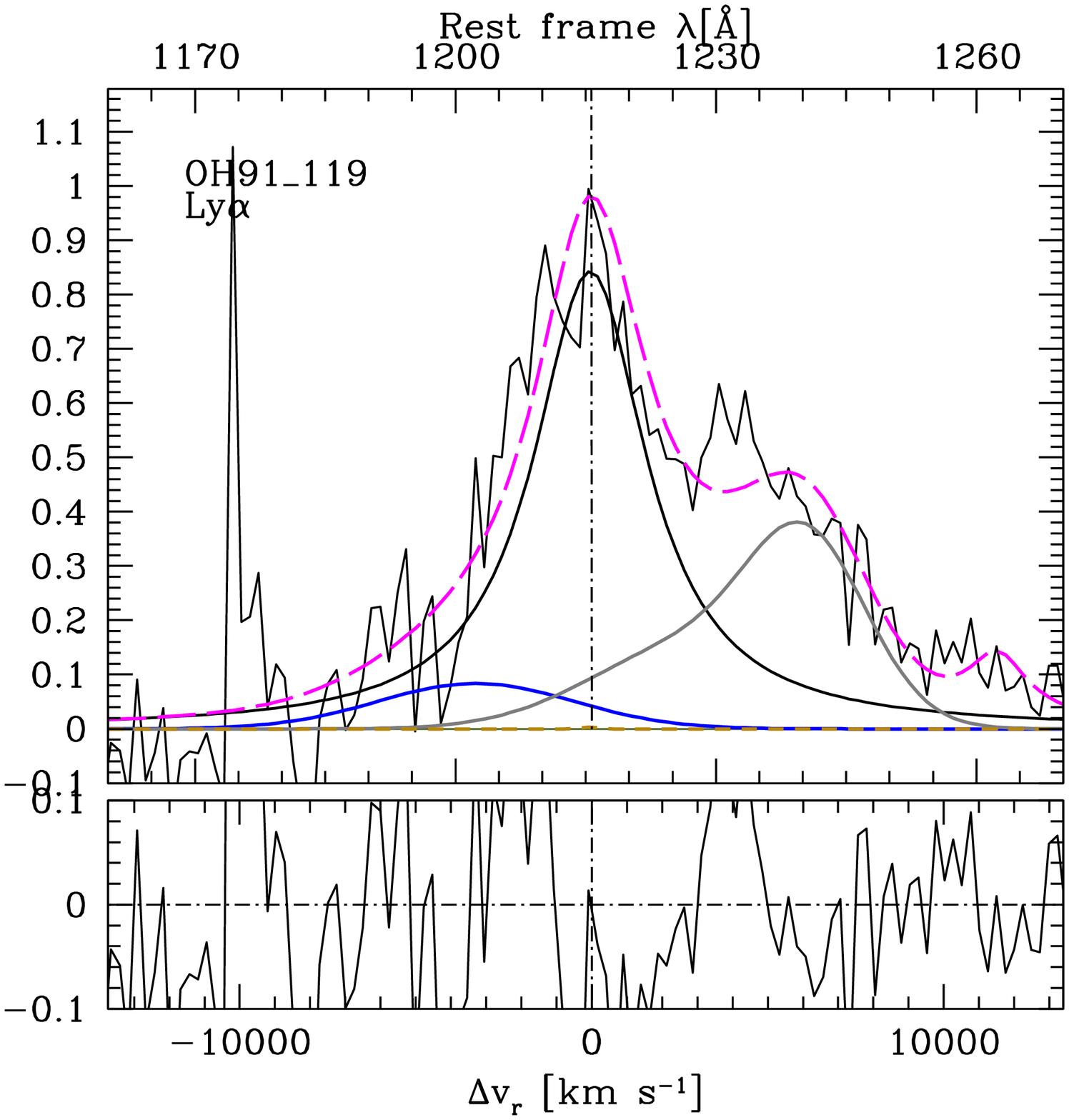}
\includegraphics[scale=0.2]{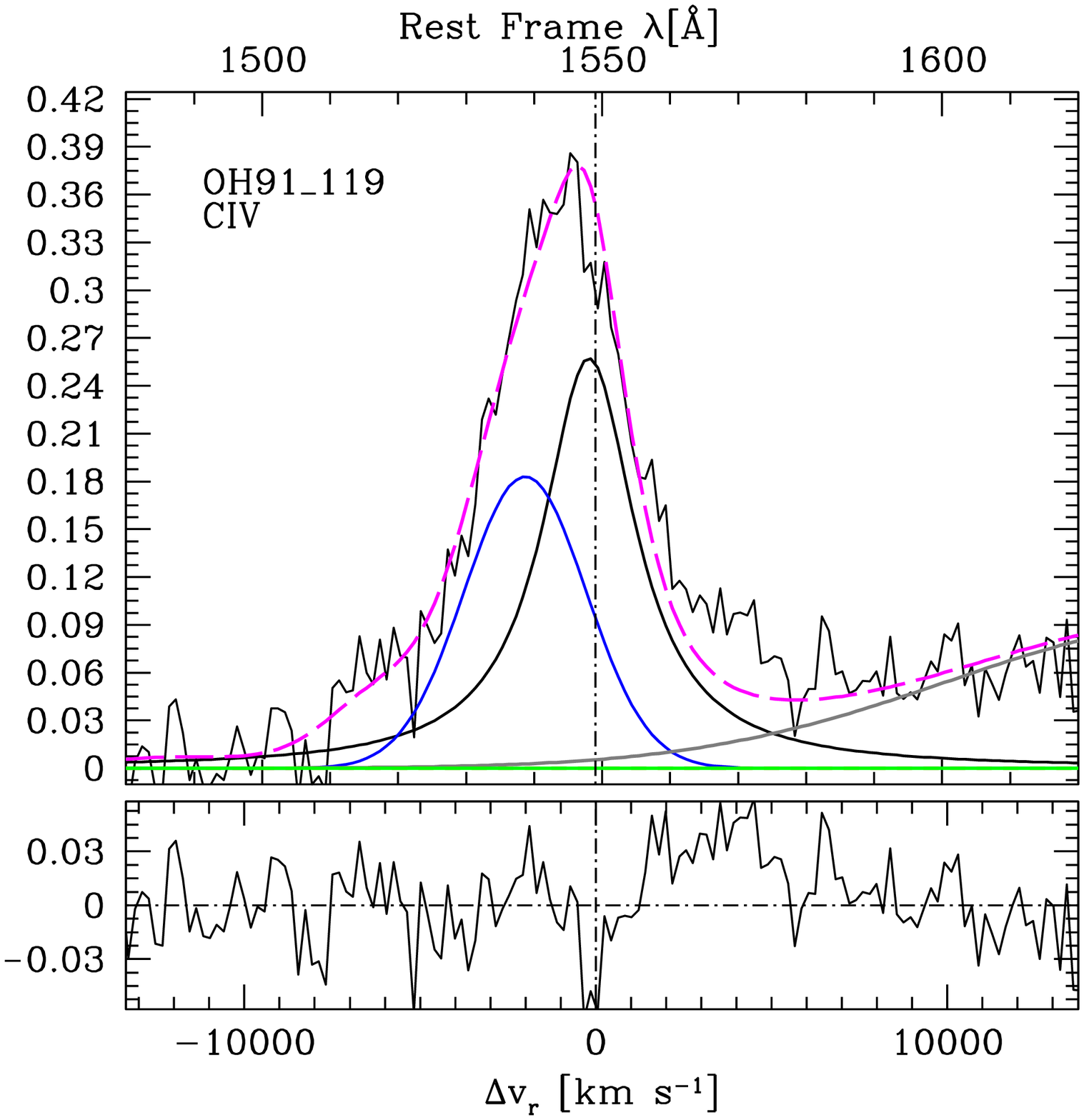}
\includegraphics[scale=0.2]{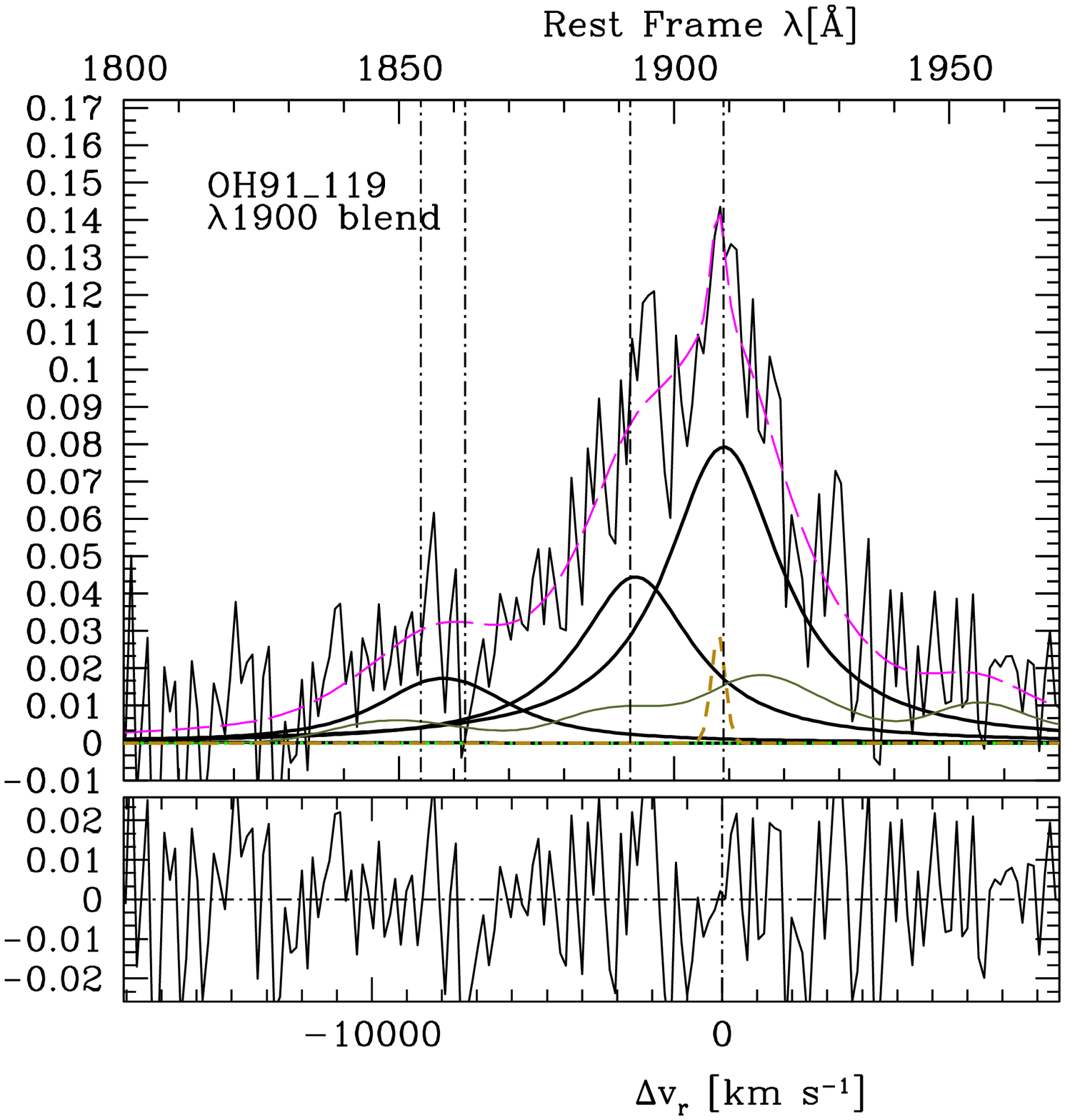}\\
\includegraphics[scale=0.2]{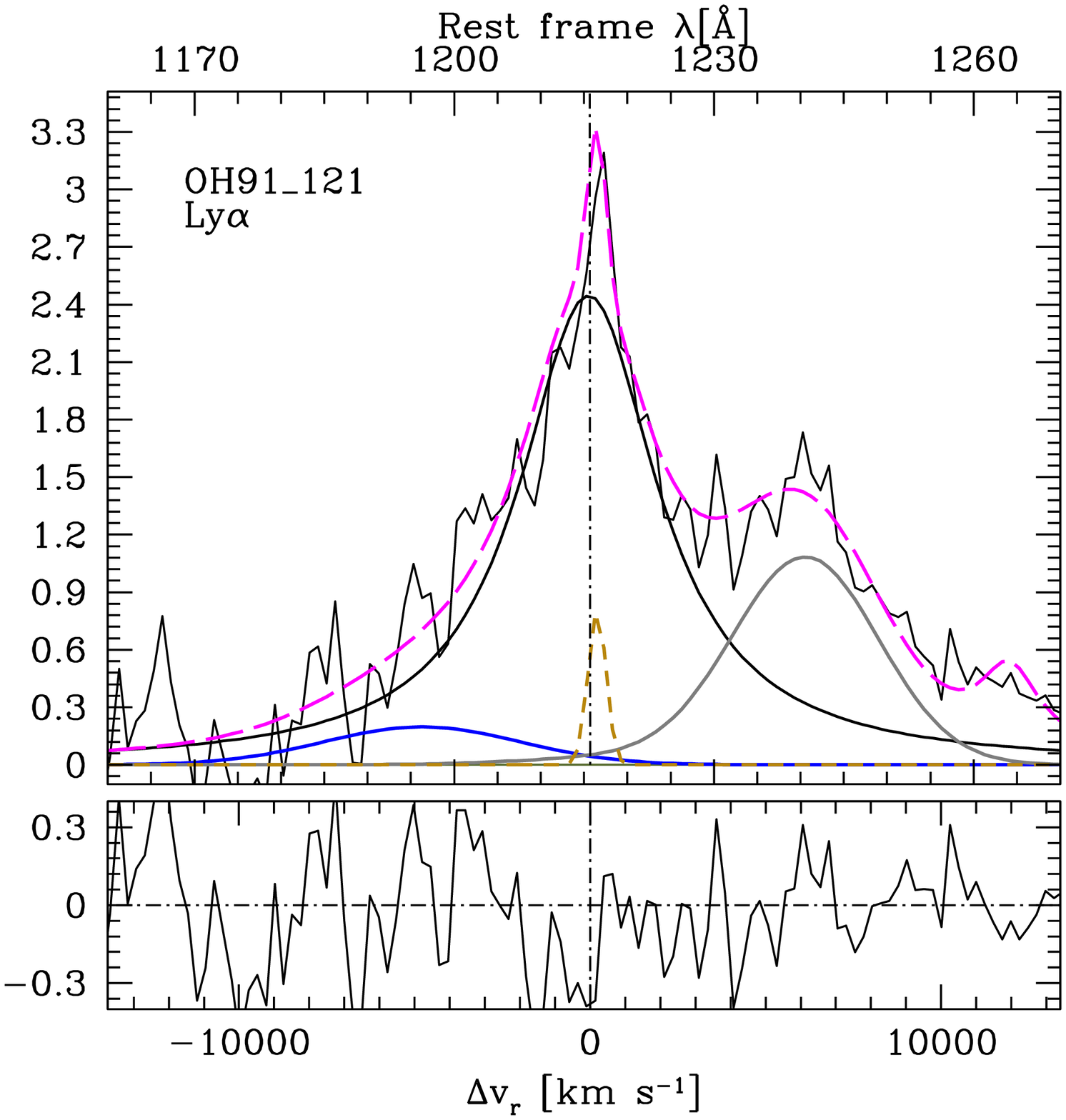}
\includegraphics[scale=0.2]{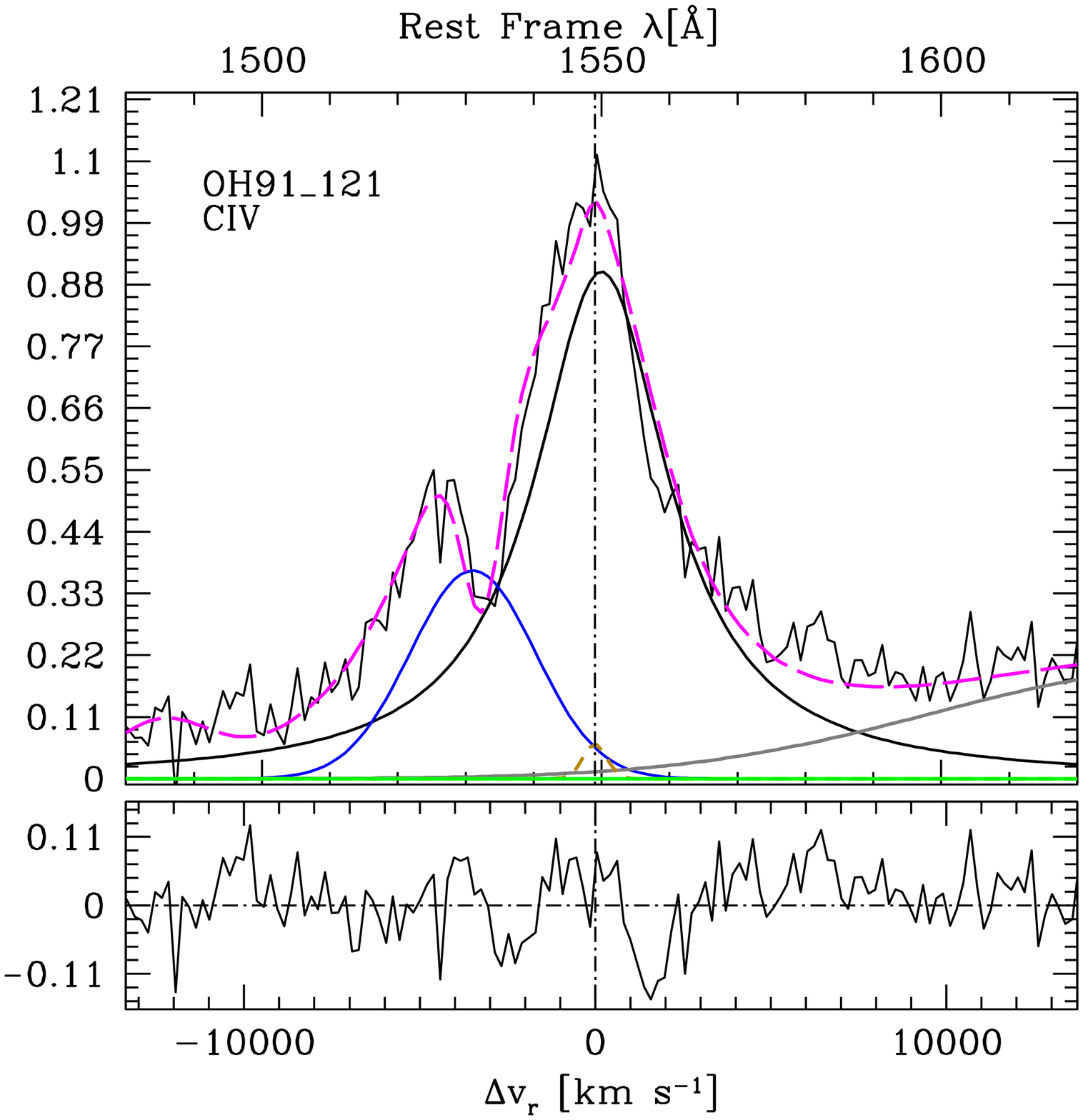}
\includegraphics[scale=0.2]{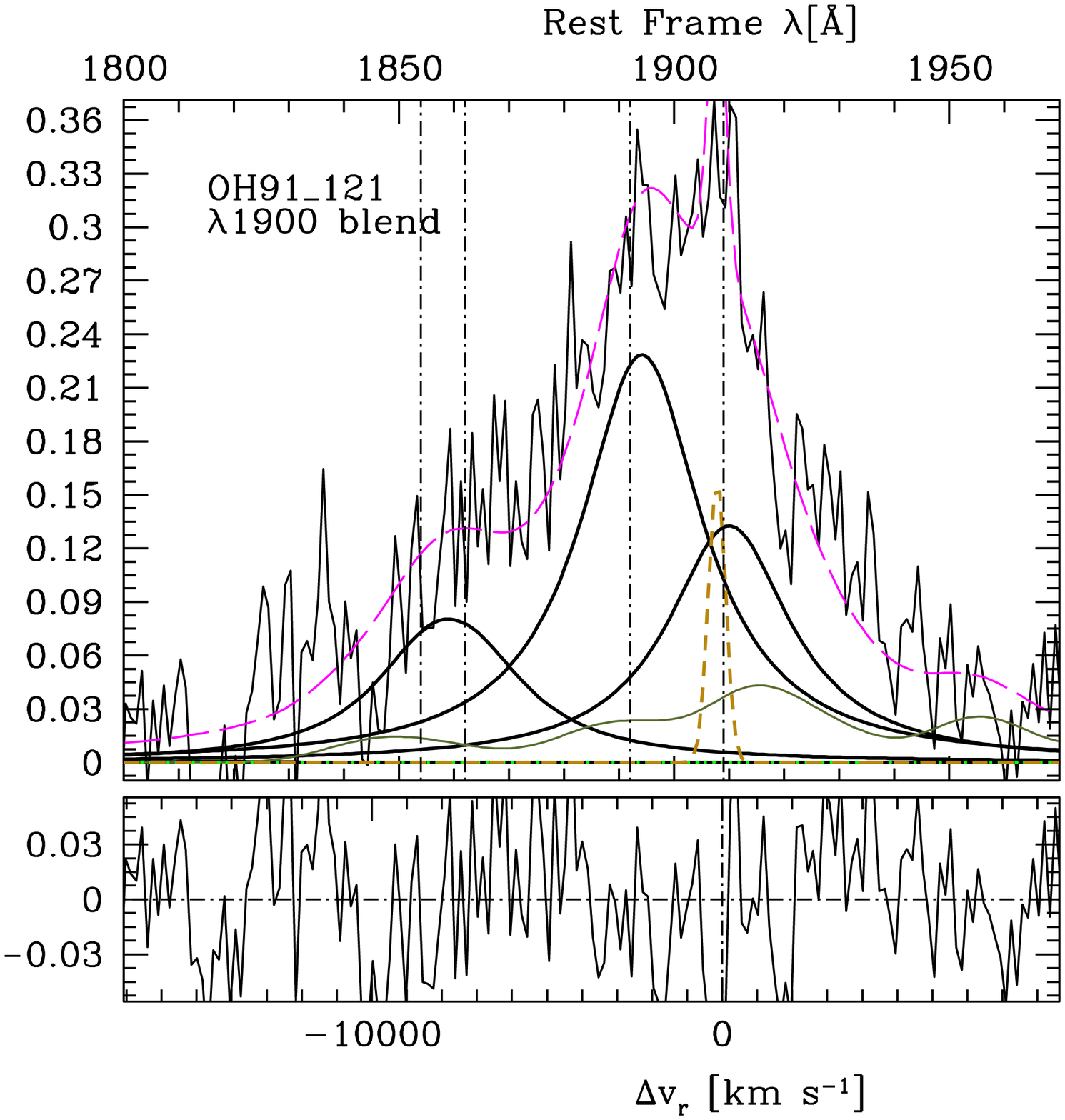}\\
\includegraphics[scale=0.2]{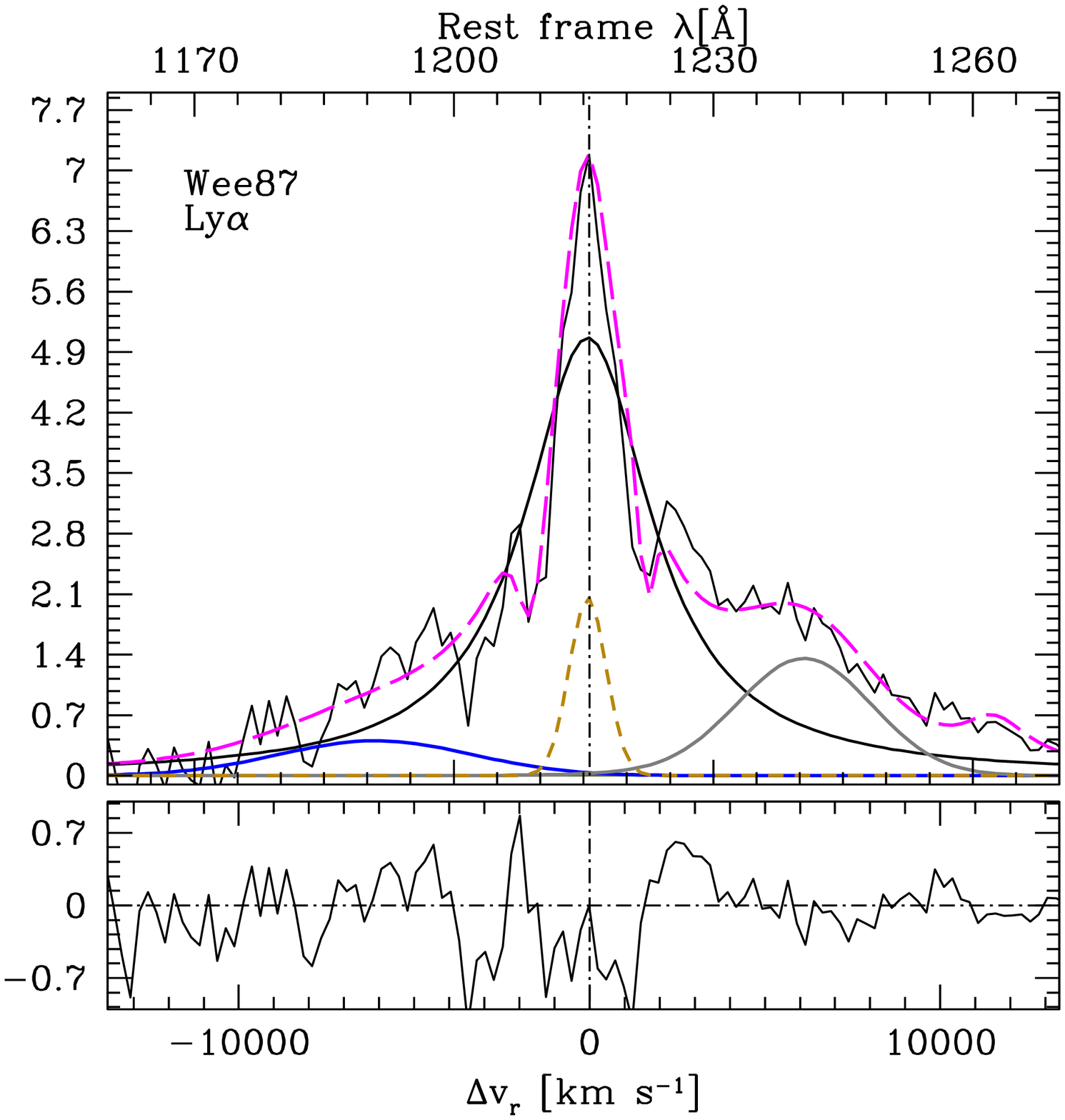}
\includegraphics[scale=0.2]{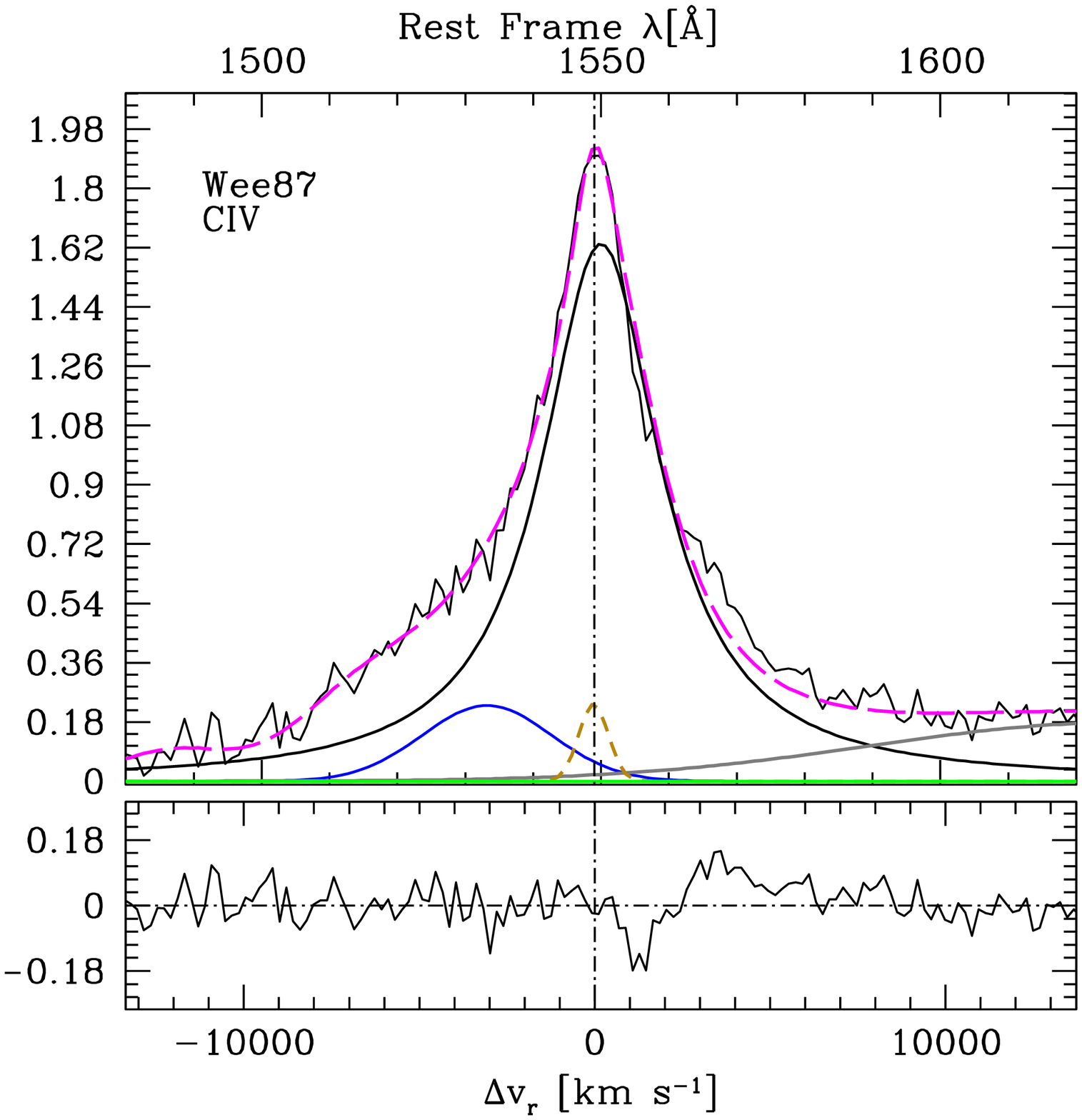}
\includegraphics[scale=0.2]{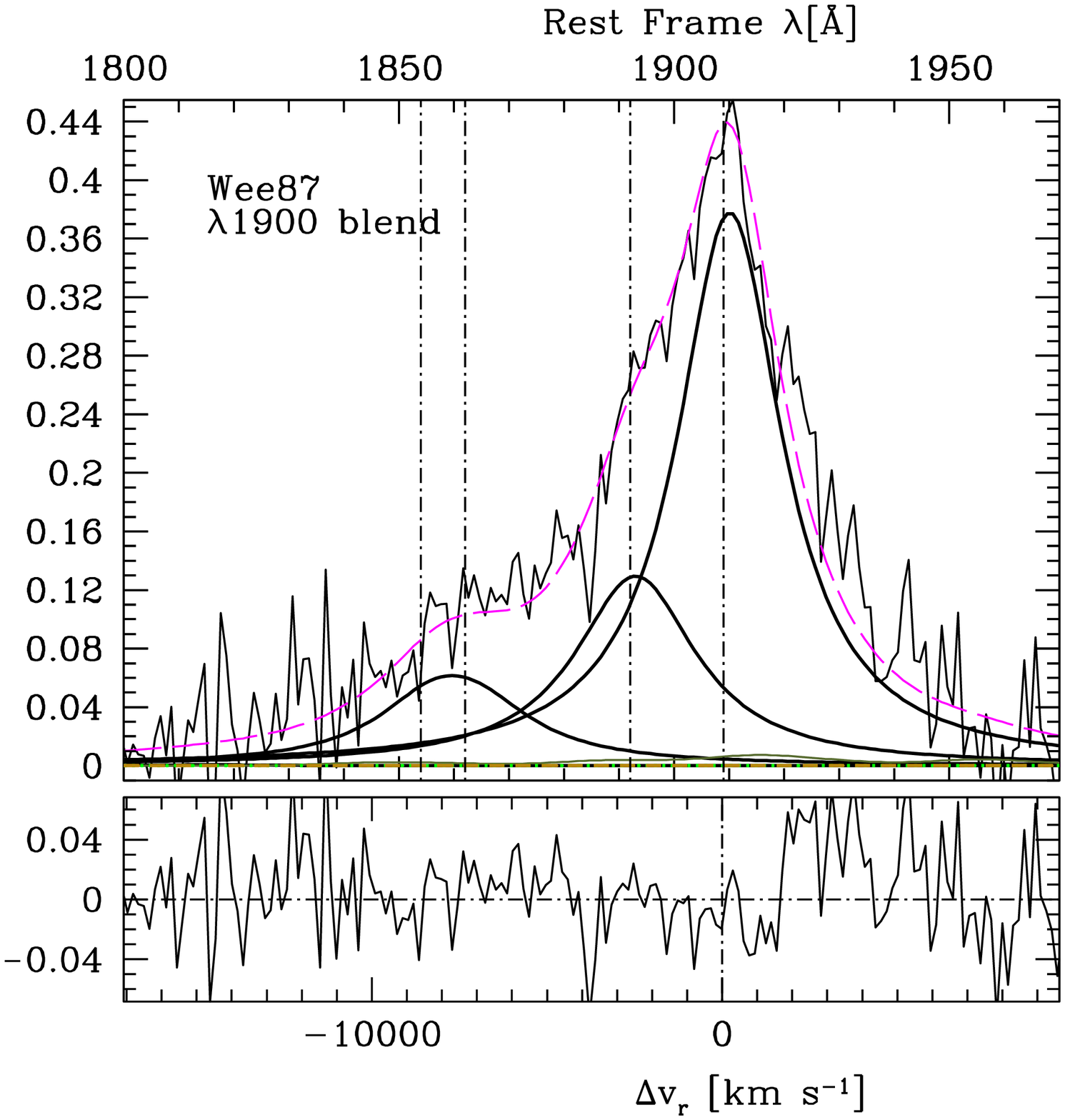}\\
\includegraphics[scale=0.2]{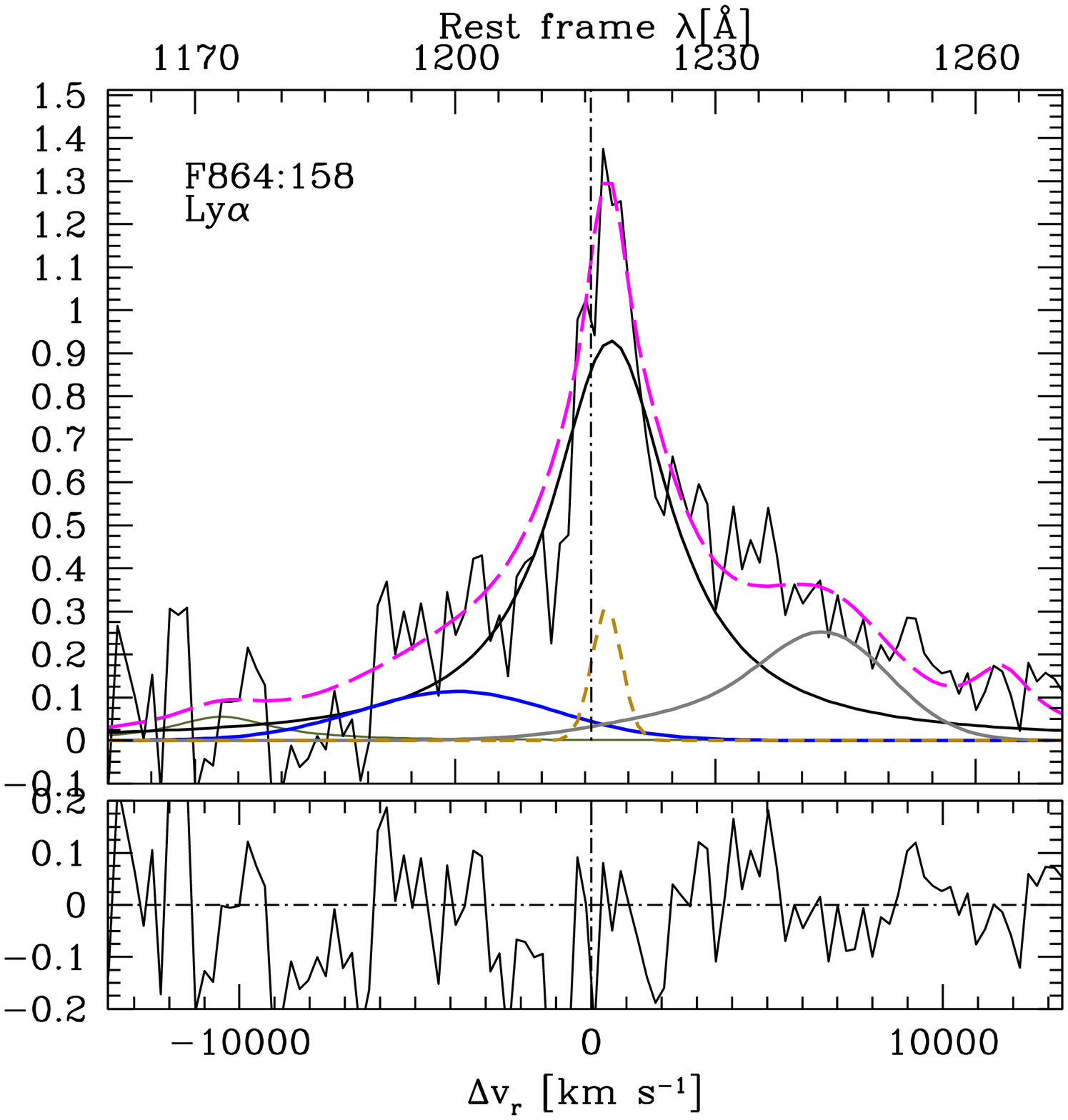}
\includegraphics[scale=0.2]{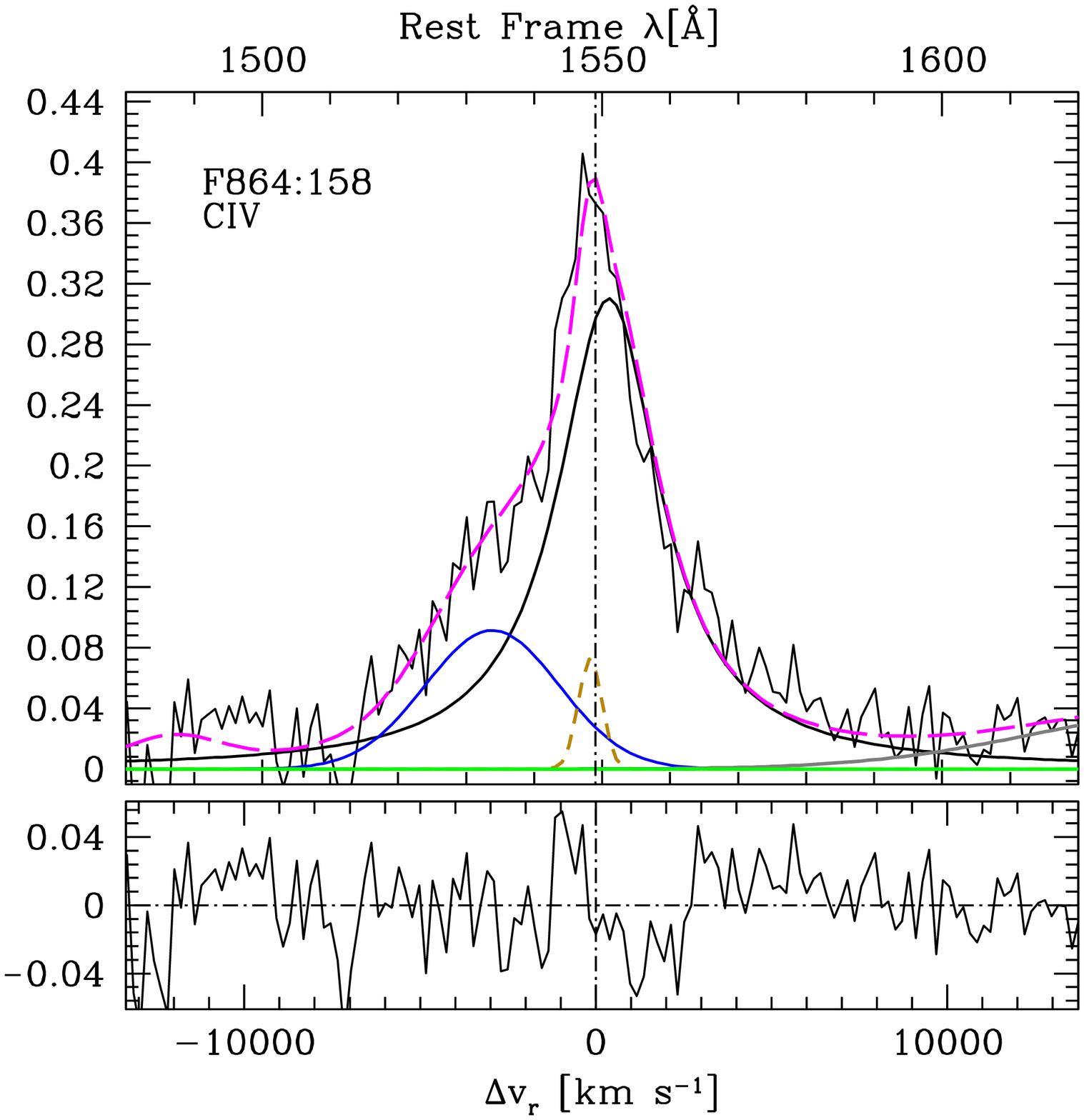}
\includegraphics[scale=0.2]{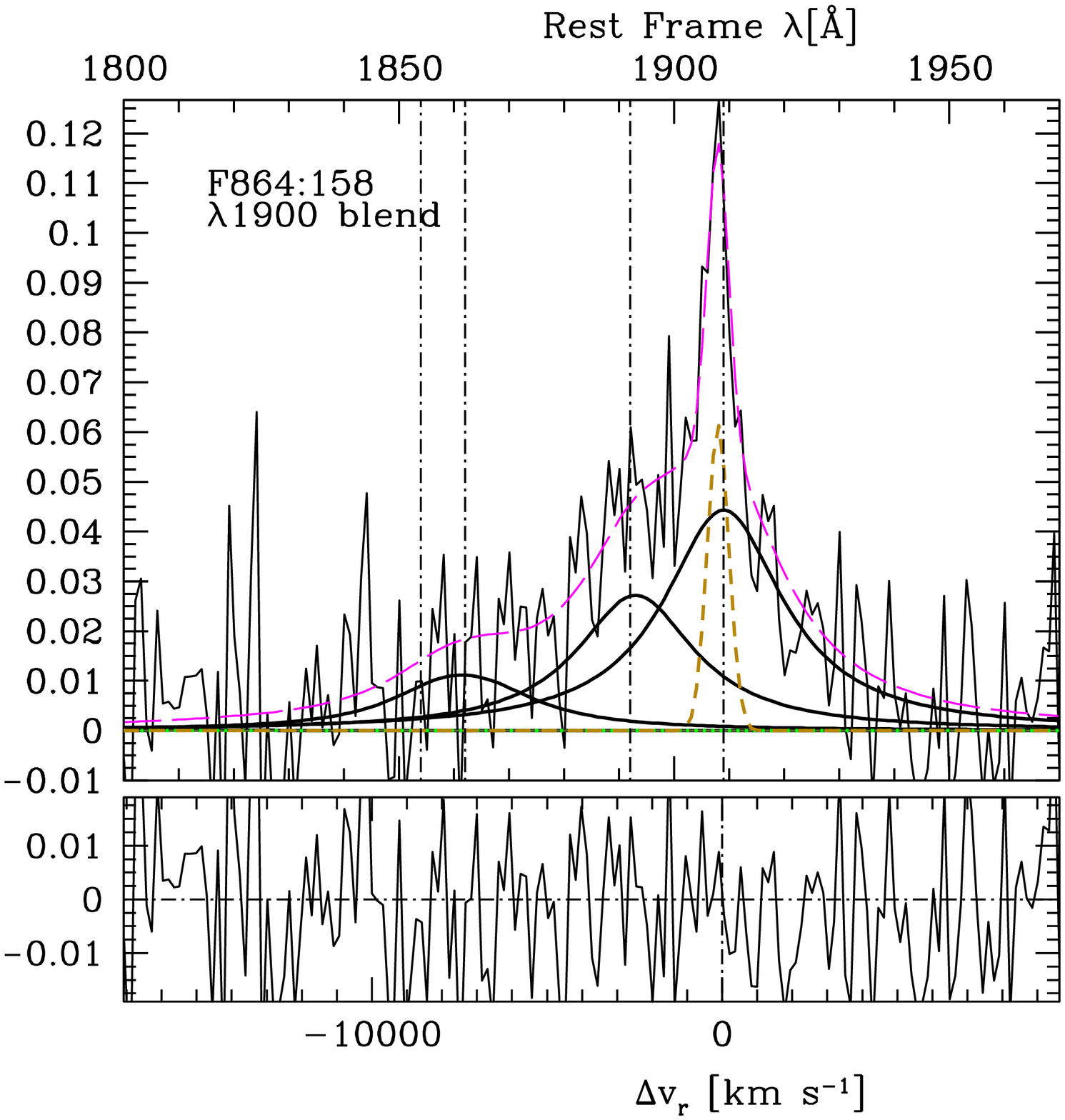}\\
\caption{Results of line fit analysis on the \lya\ (left panel) \civ\ (middle) and 1900 \AA\ emission 
features for Pop. A sources. The lines are shown after continuum subtraction. The magenta dashed line 
show the full model of the emission  features and   intervening narrow absorptions if present. The thick black lines show the broad components of the 
prominent emission lines. The thick blue line traces the BLUE component. The \nv\ line on the 
red side of the \lya\ profile is traced by a grey line that includes its BC and BLUE. Orange lines 
are narrow line components.  The lower panel of each frame shows the residuals between the line 
model and the observations as a function of radial velocity from rest frame (for \ciii\ in the case 
of the 1900 blend).    \label{fig:fitsa}}
\end{figure*}

\addtocounter{figure}{-1}

\begin{figure*}
\includegraphics[scale=0.2]{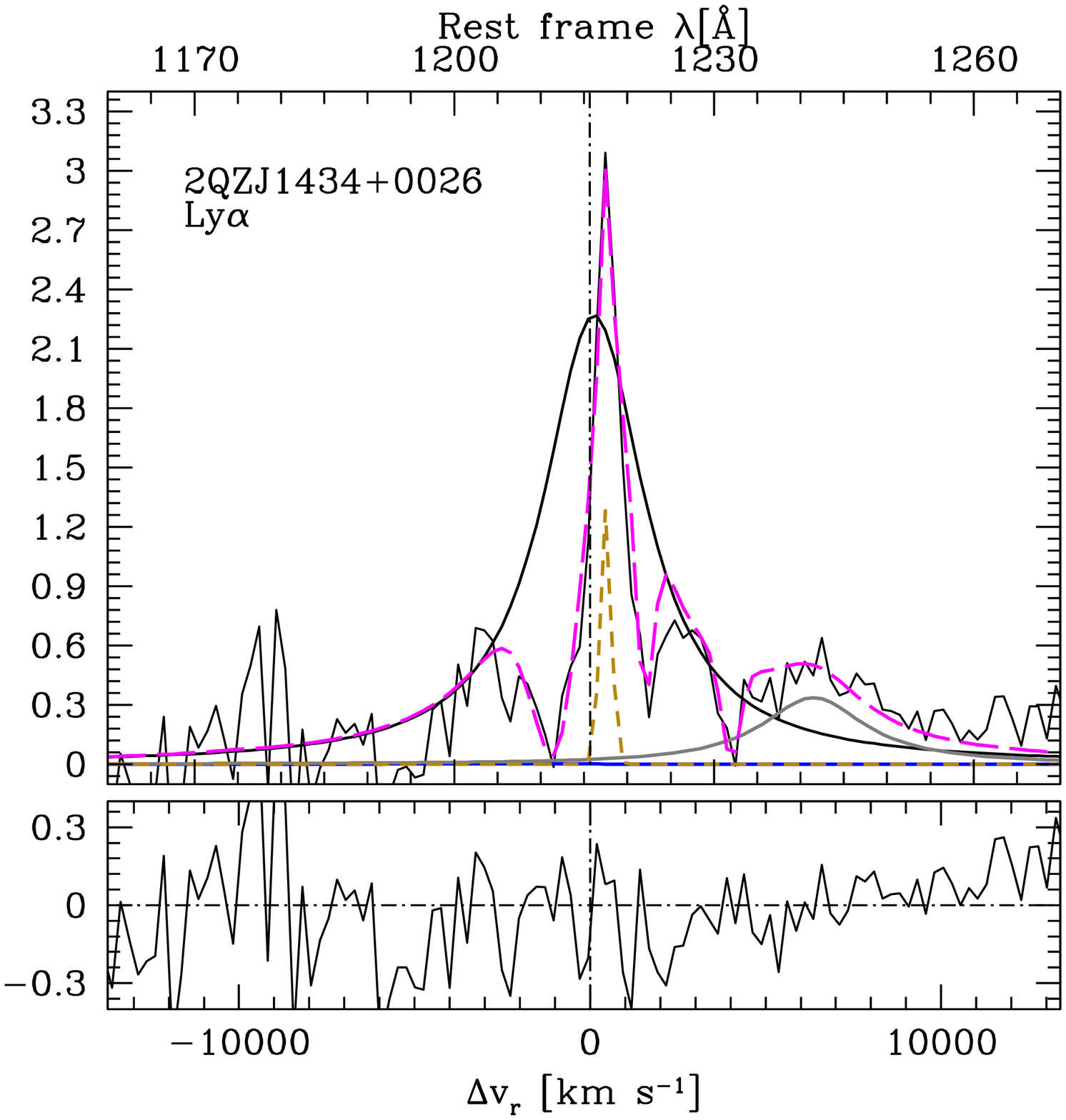}
\includegraphics[scale=0.2]{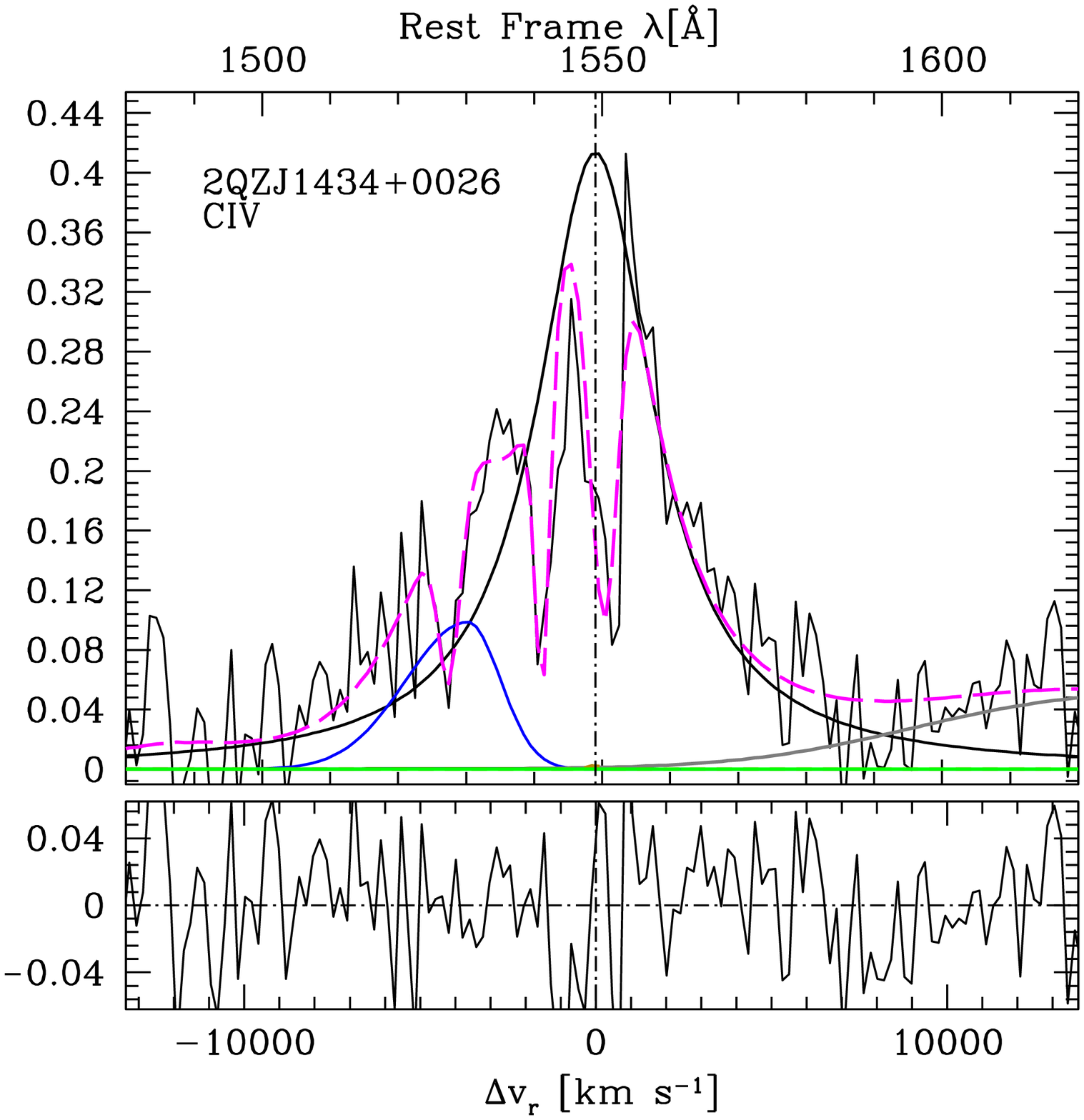}
\includegraphics[scale=0.2]{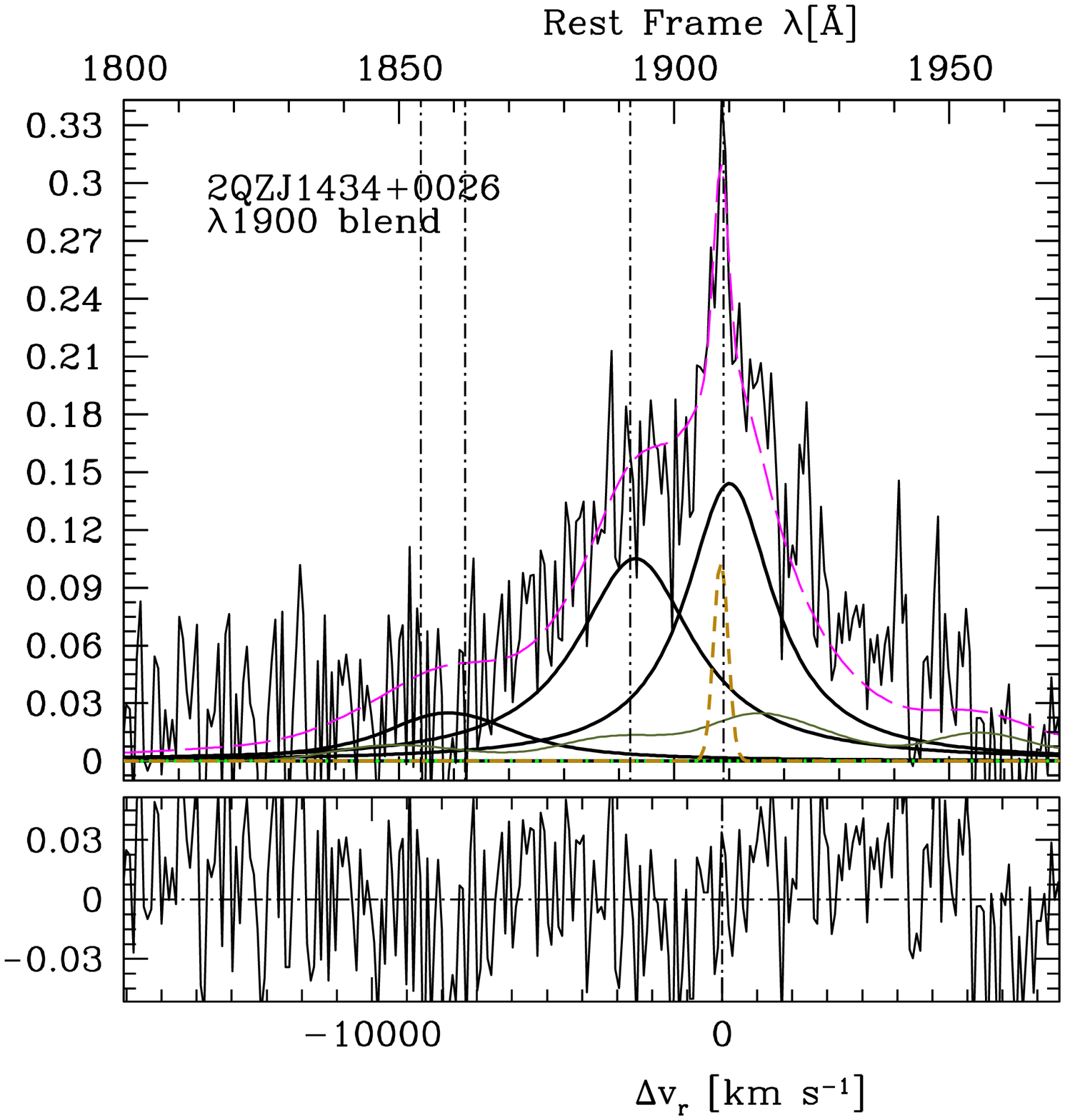}\\
\includegraphics[scale=0.2]{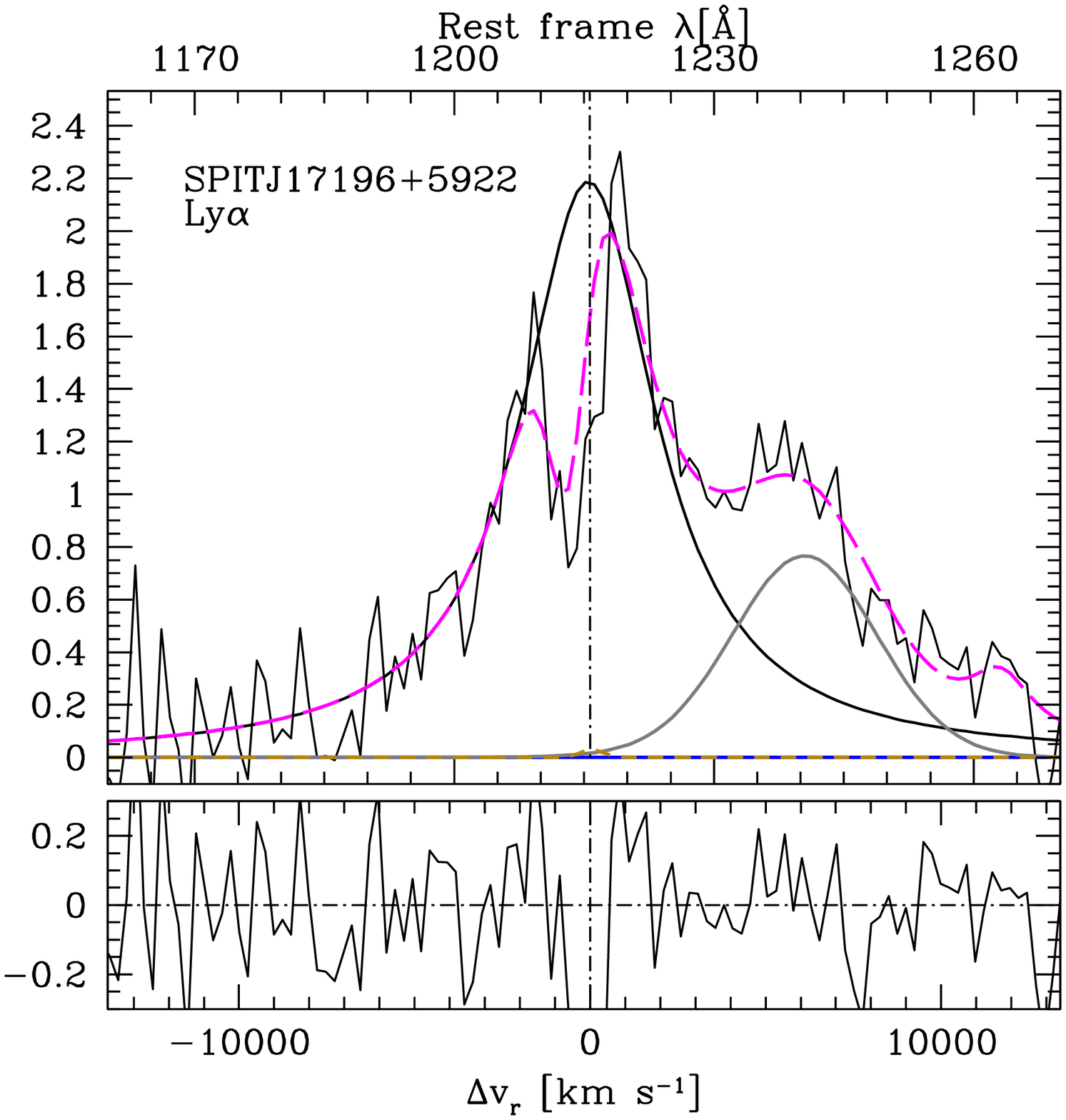}
\includegraphics[scale=0.2]{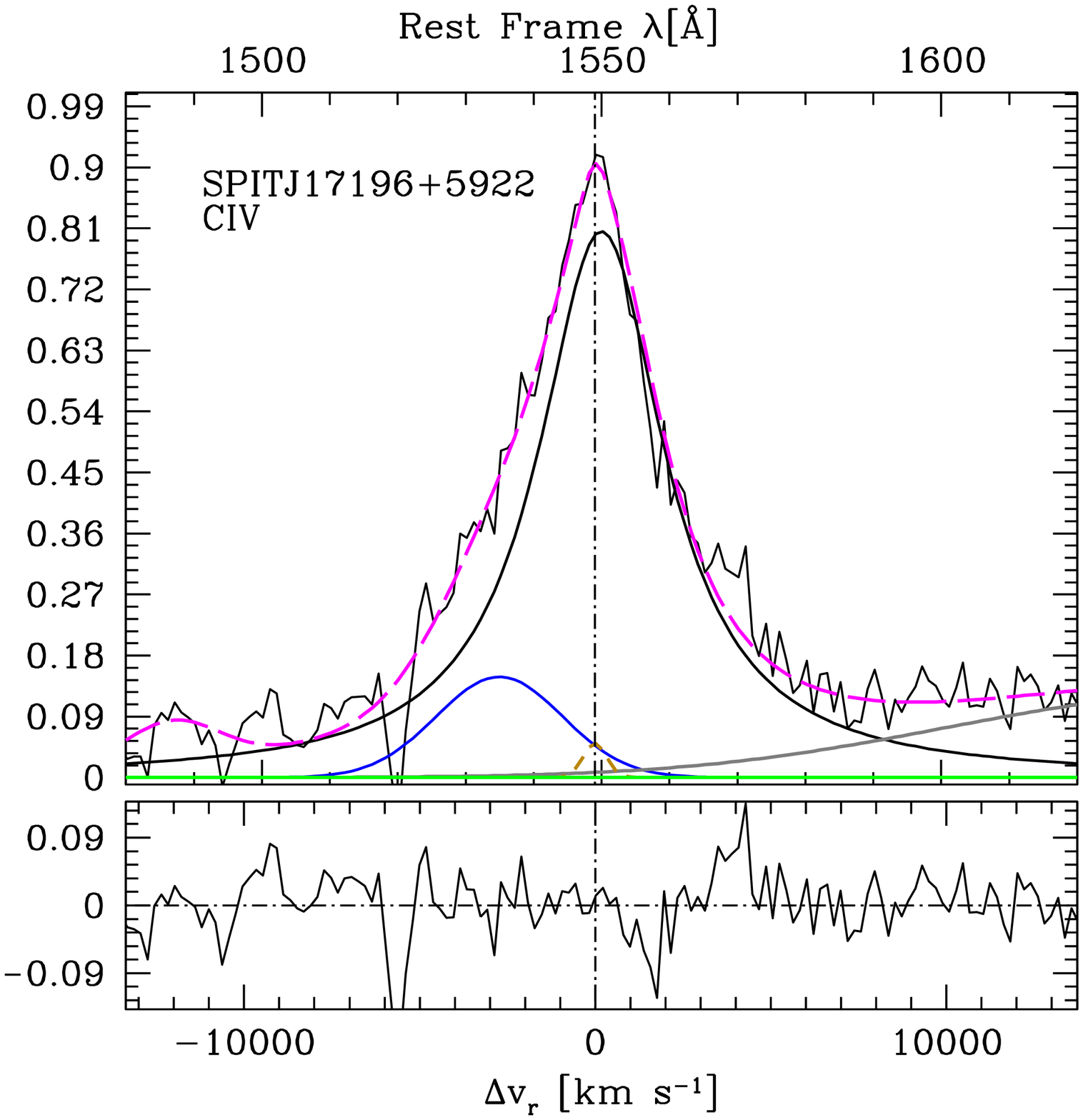}
\includegraphics[scale=0.2]{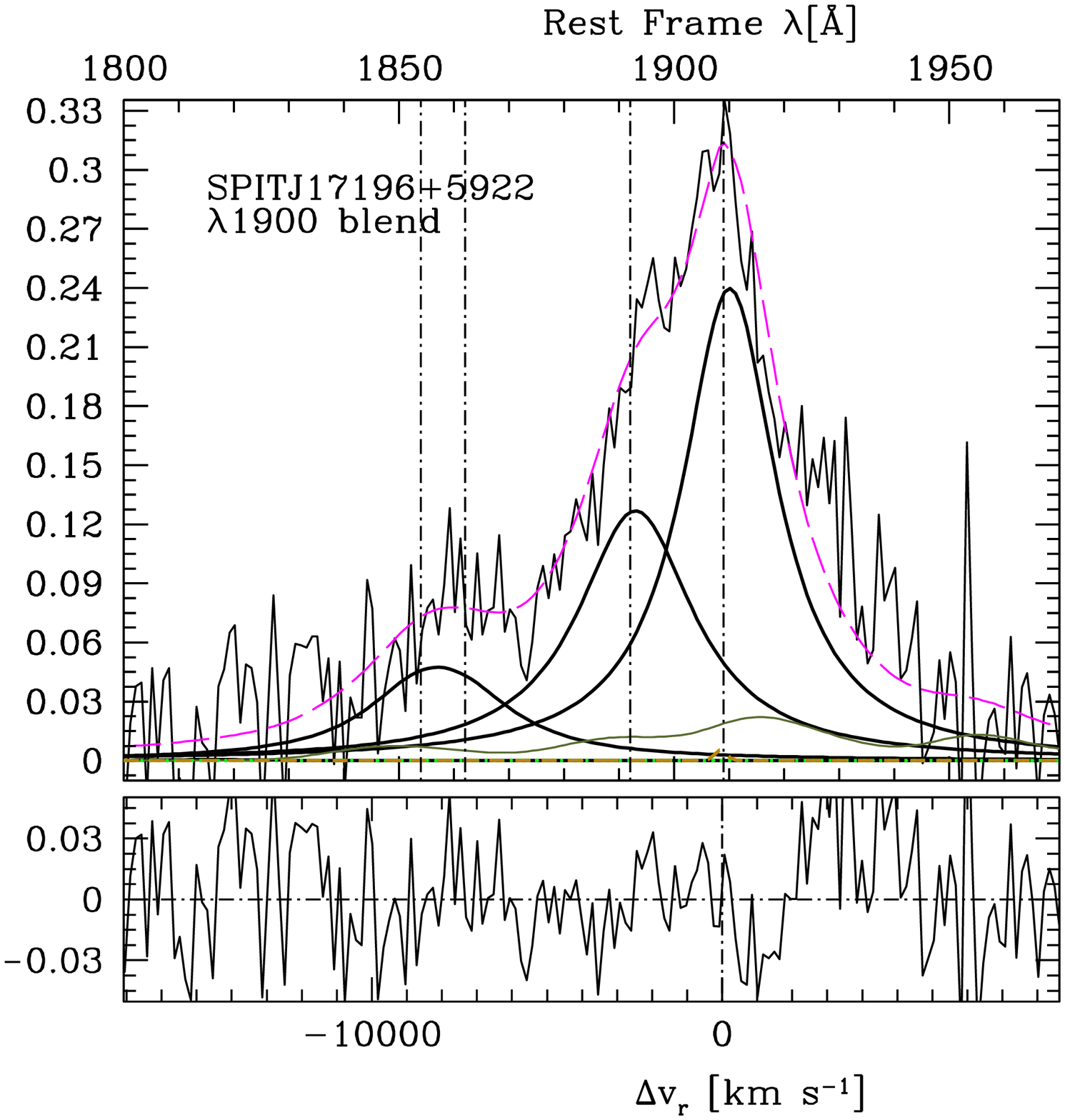}\\
\includegraphics[scale=0.2]{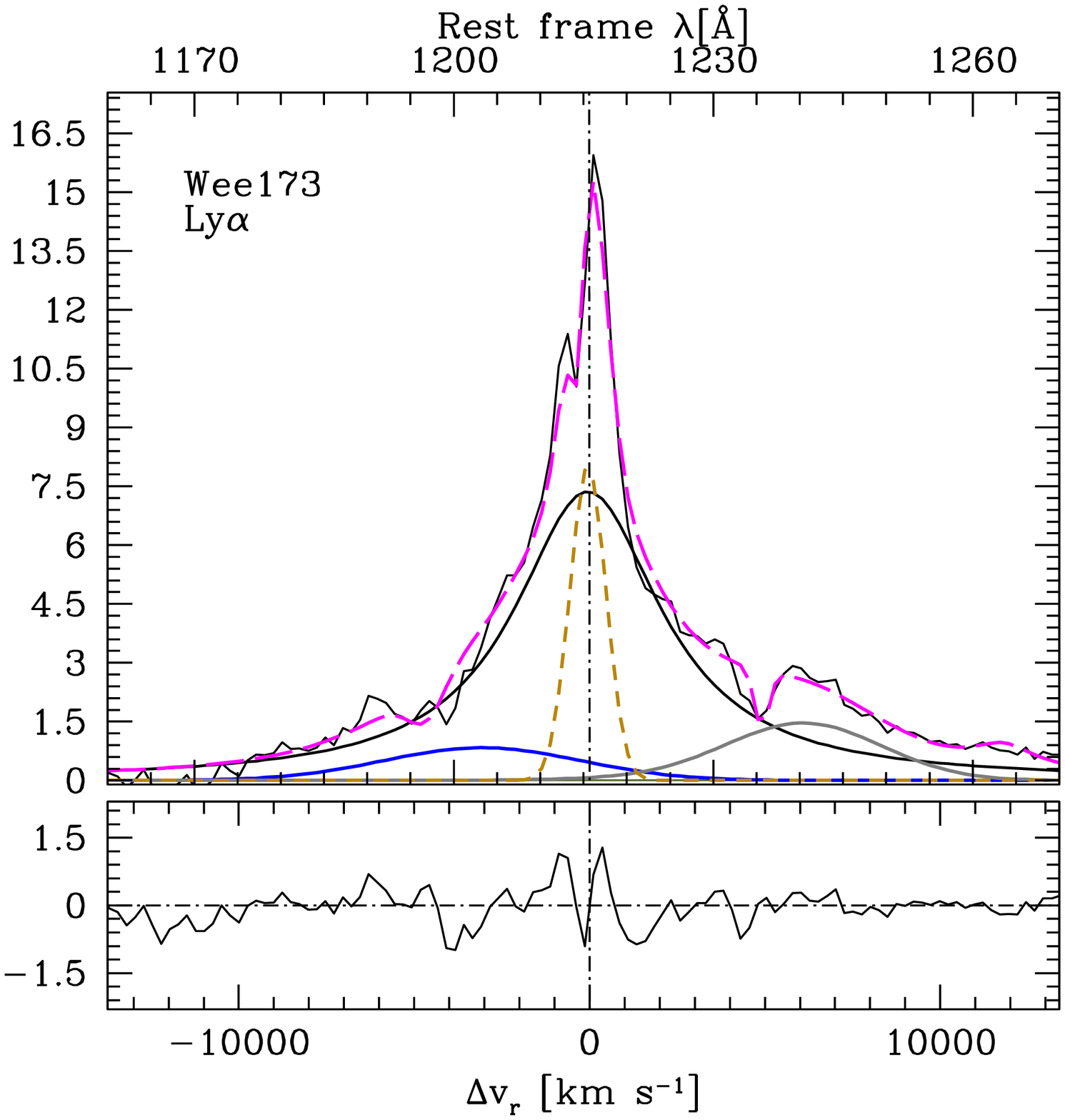}
\includegraphics[scale=0.2]{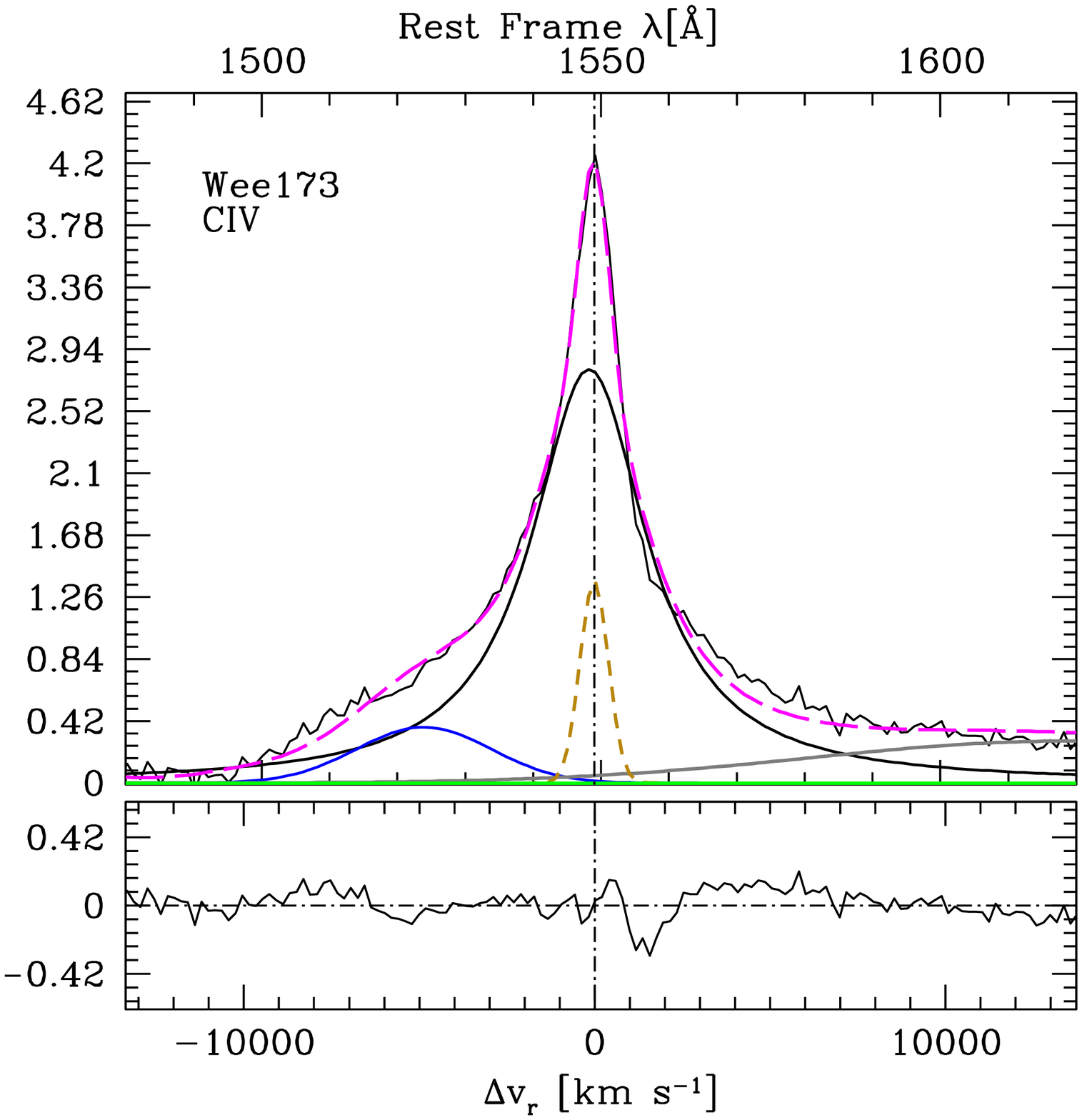}
\includegraphics[scale=0.2]{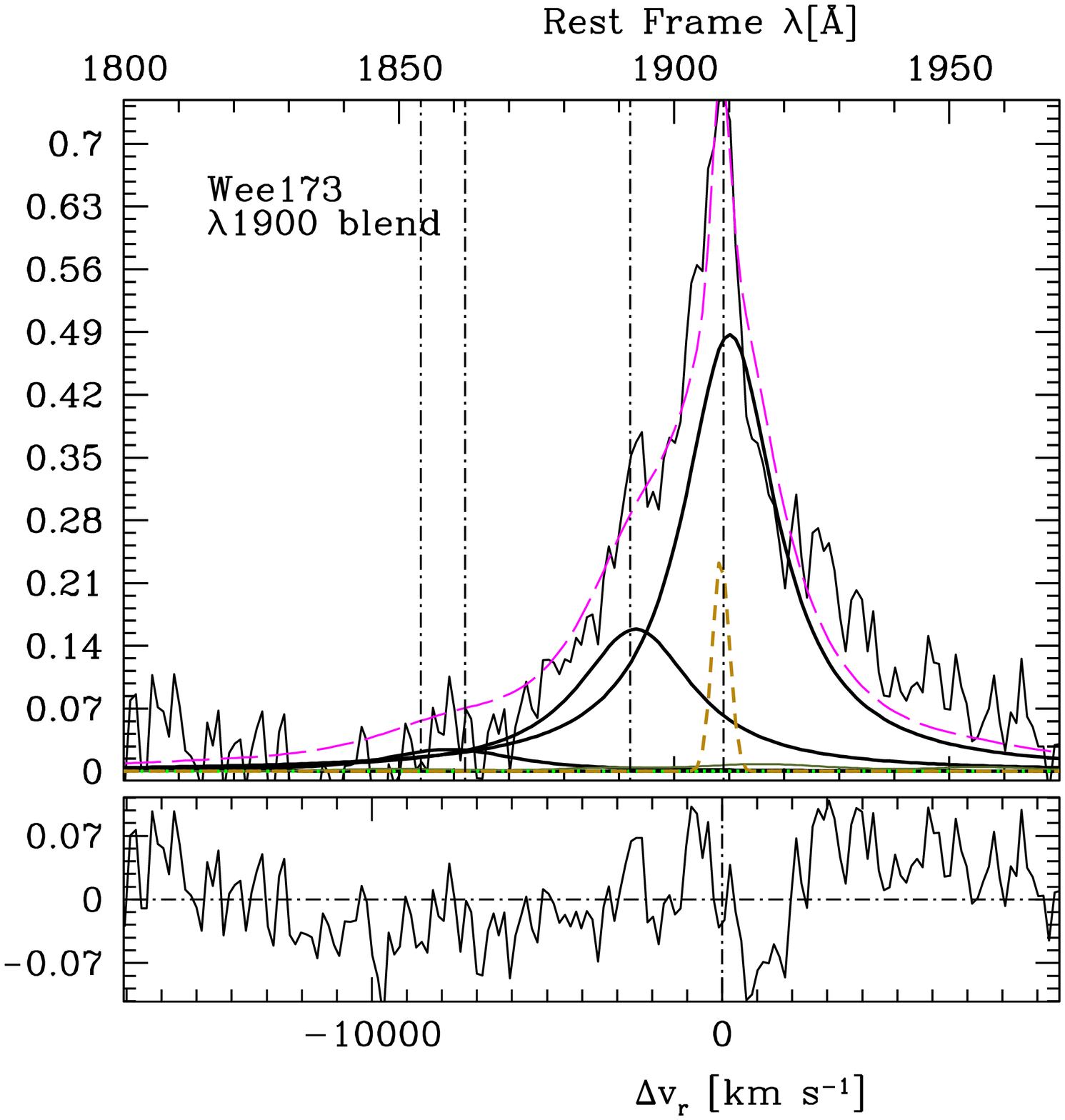}\\
\includegraphics[scale=0.2]{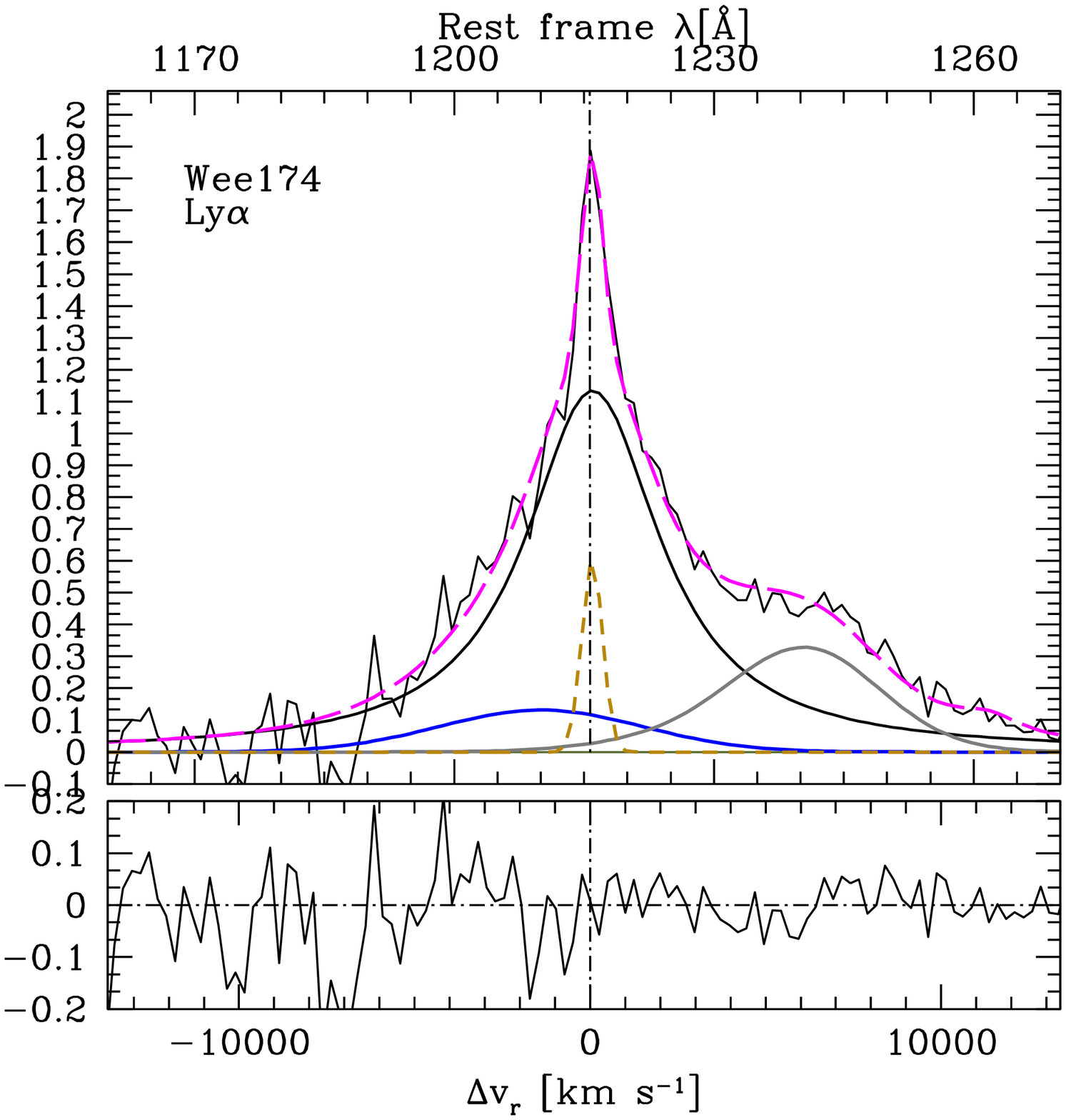}
\includegraphics[scale=0.2]{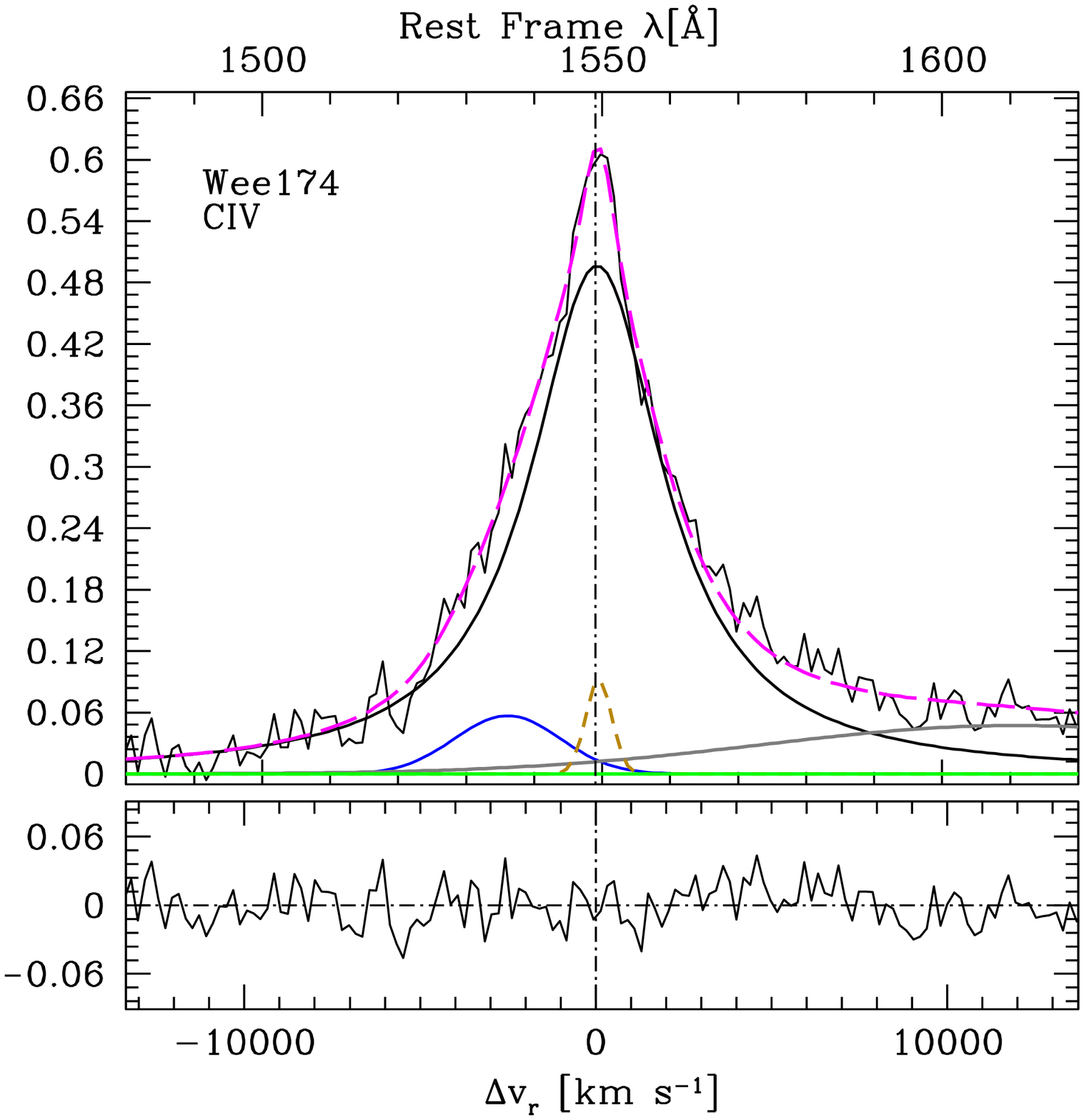}
\includegraphics[scale=0.2]{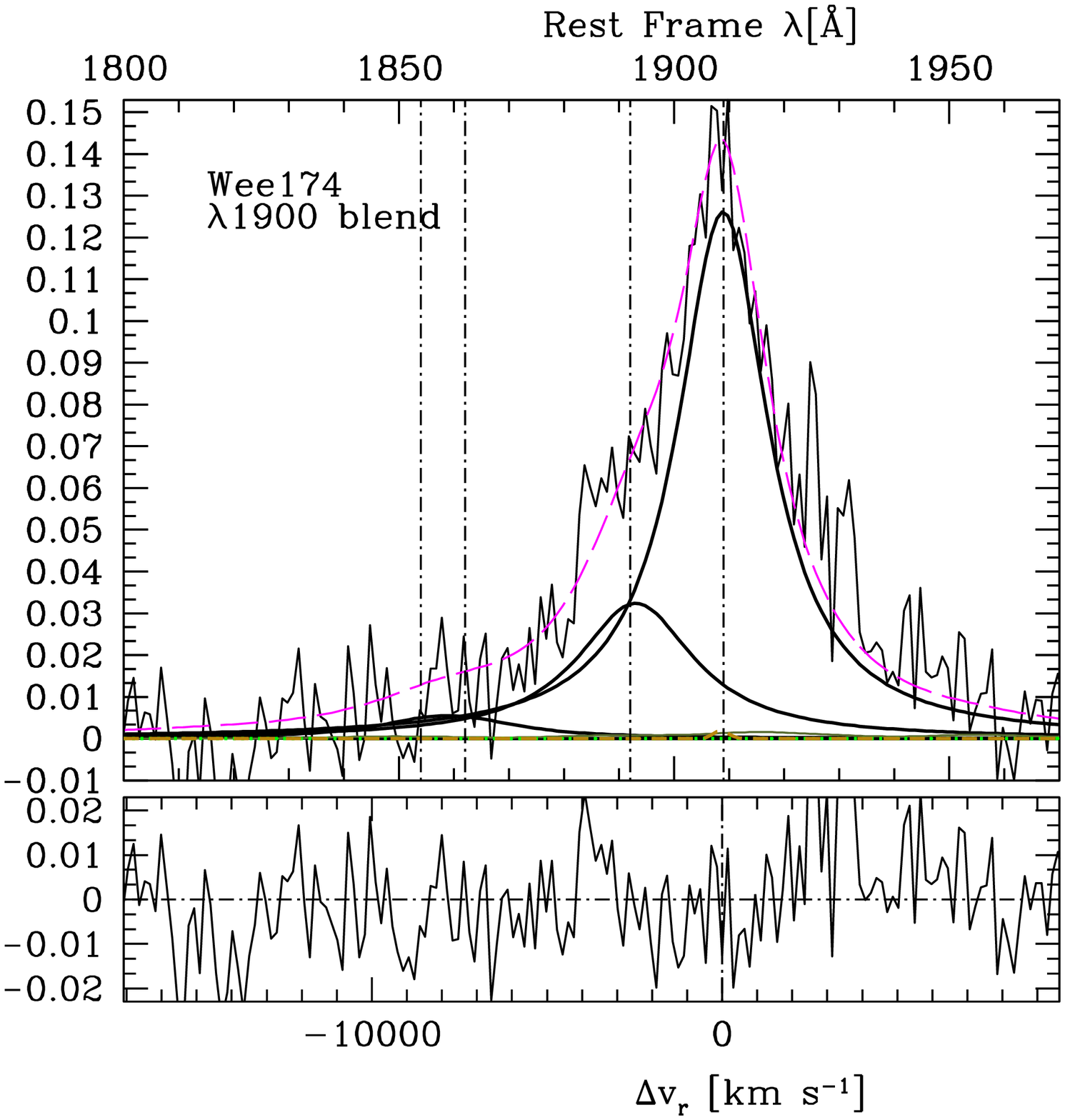}\\
\caption{(cont.)}
\end{figure*}

\addtocounter{figure}{-1}

\begin{figure*}
\includegraphics[scale=0.2]{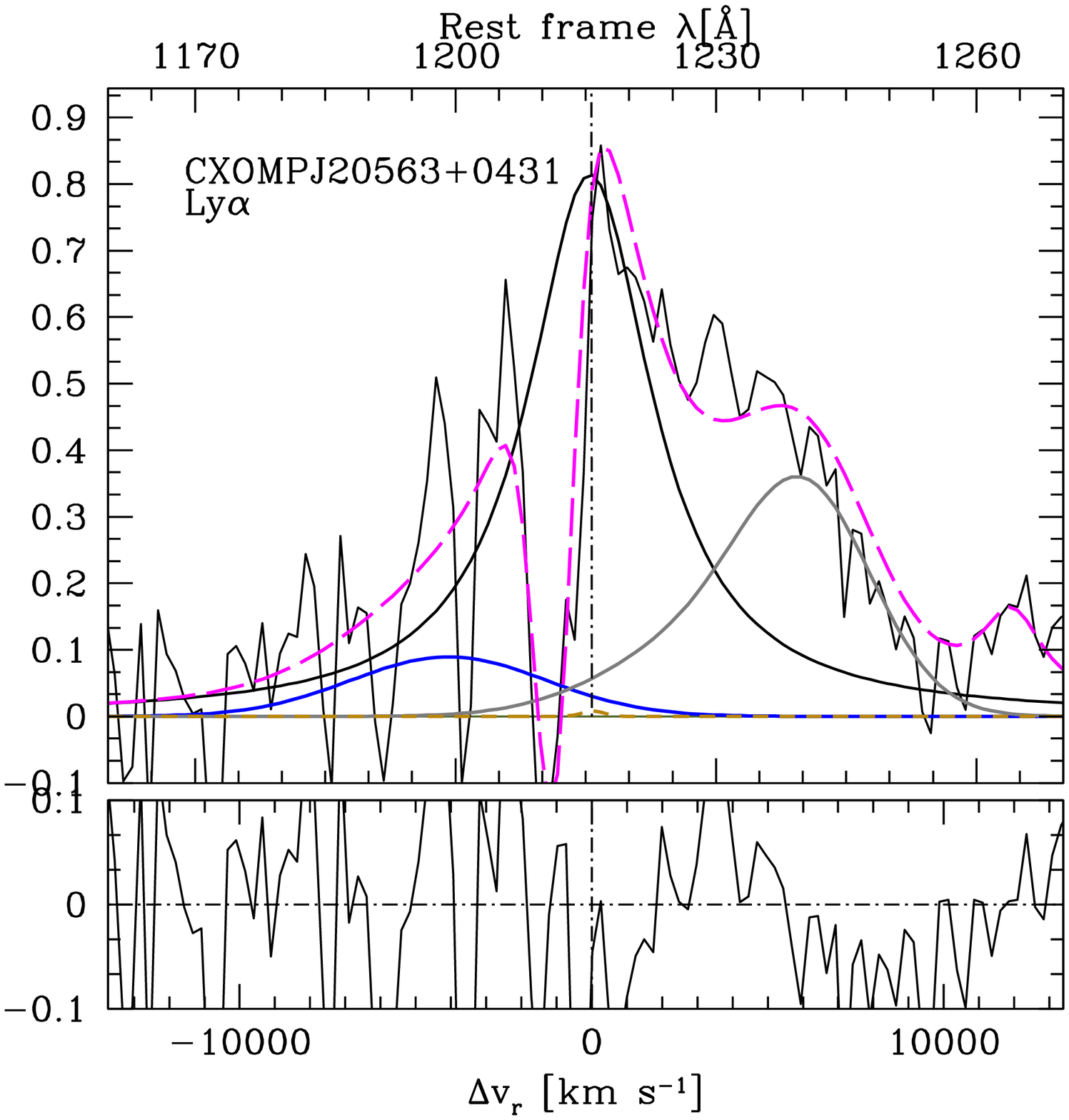}
\includegraphics[scale=0.2]{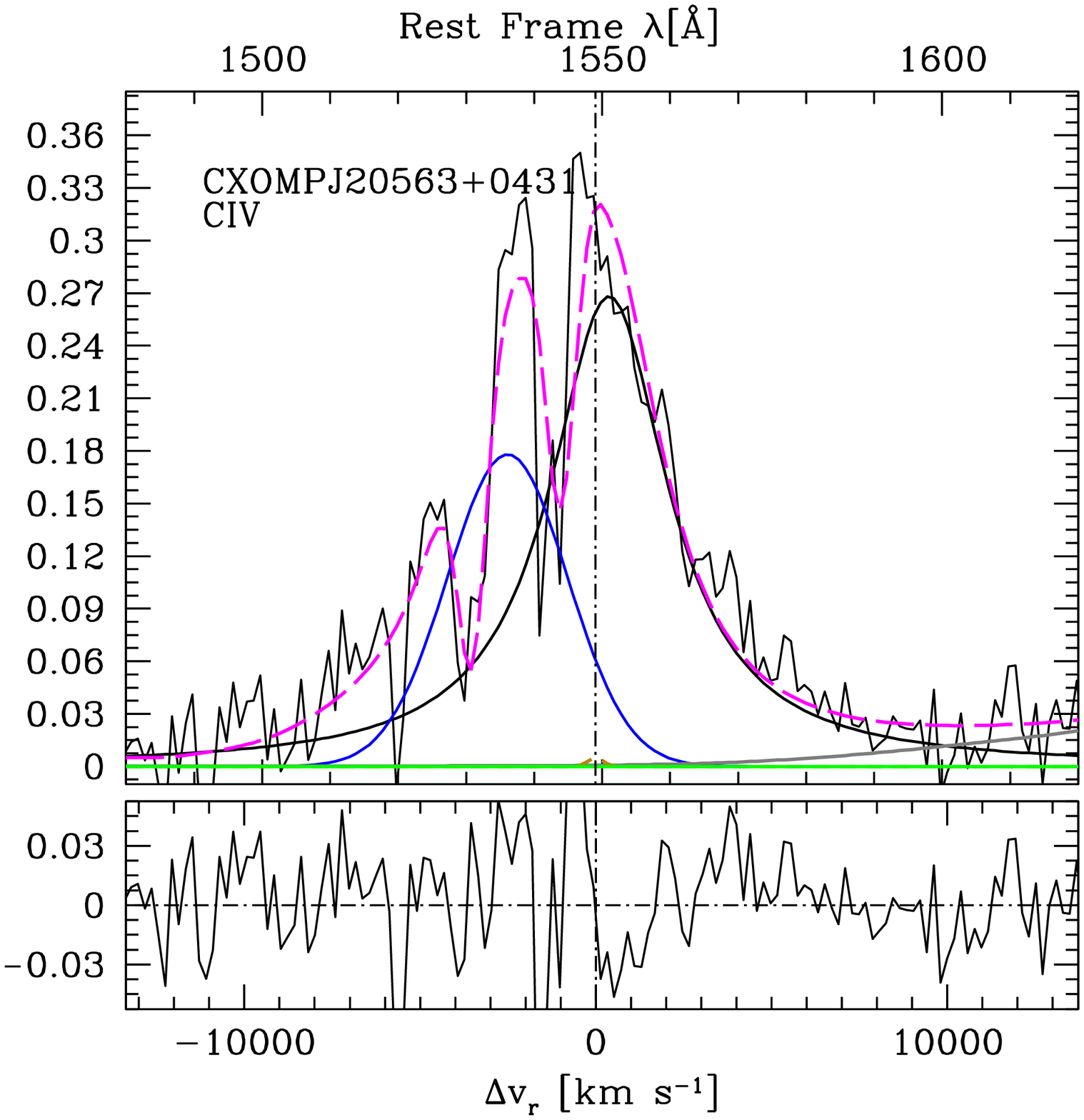}
\includegraphics[scale=0.2]{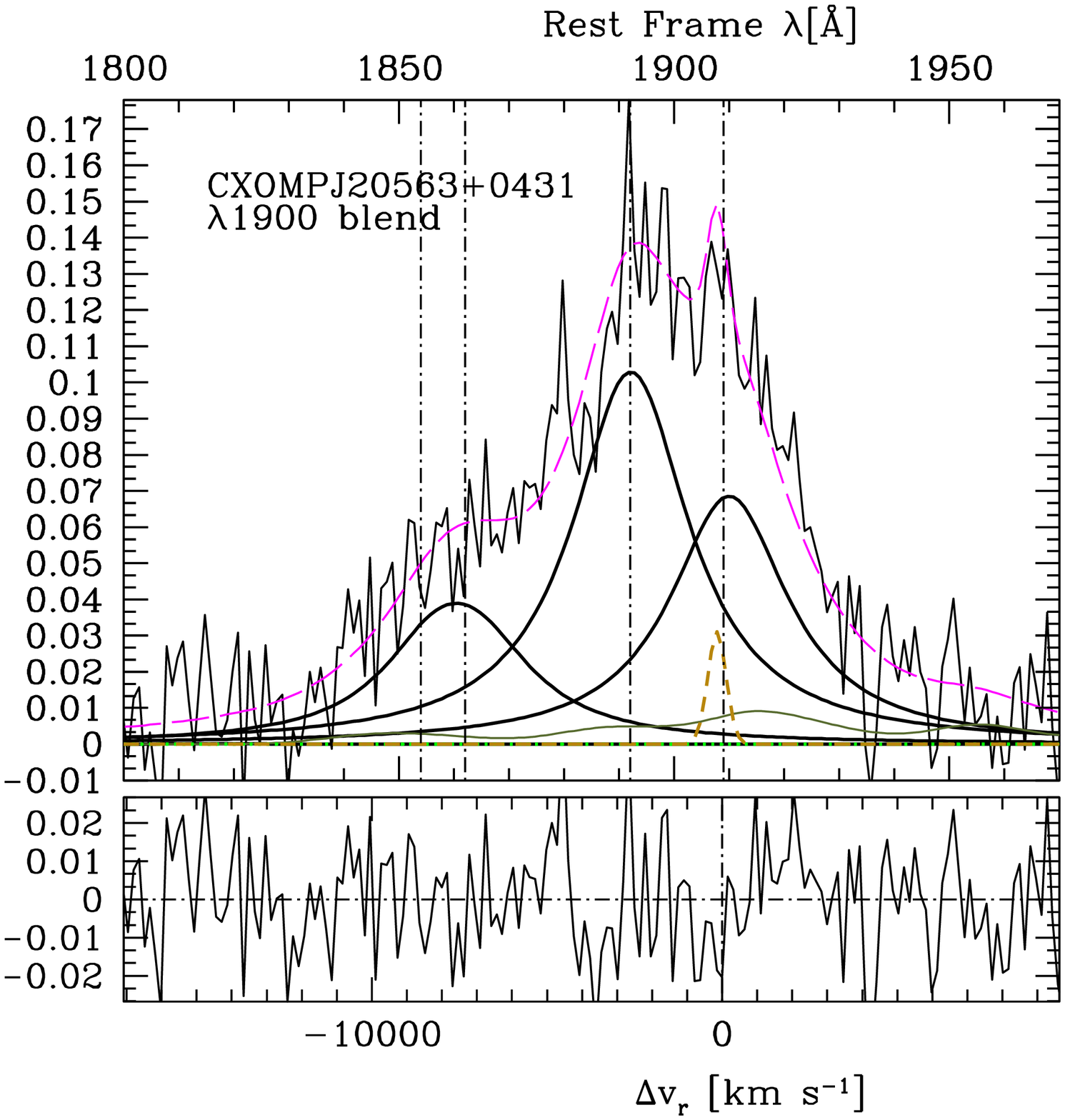}\\
\includegraphics[scale=0.2]{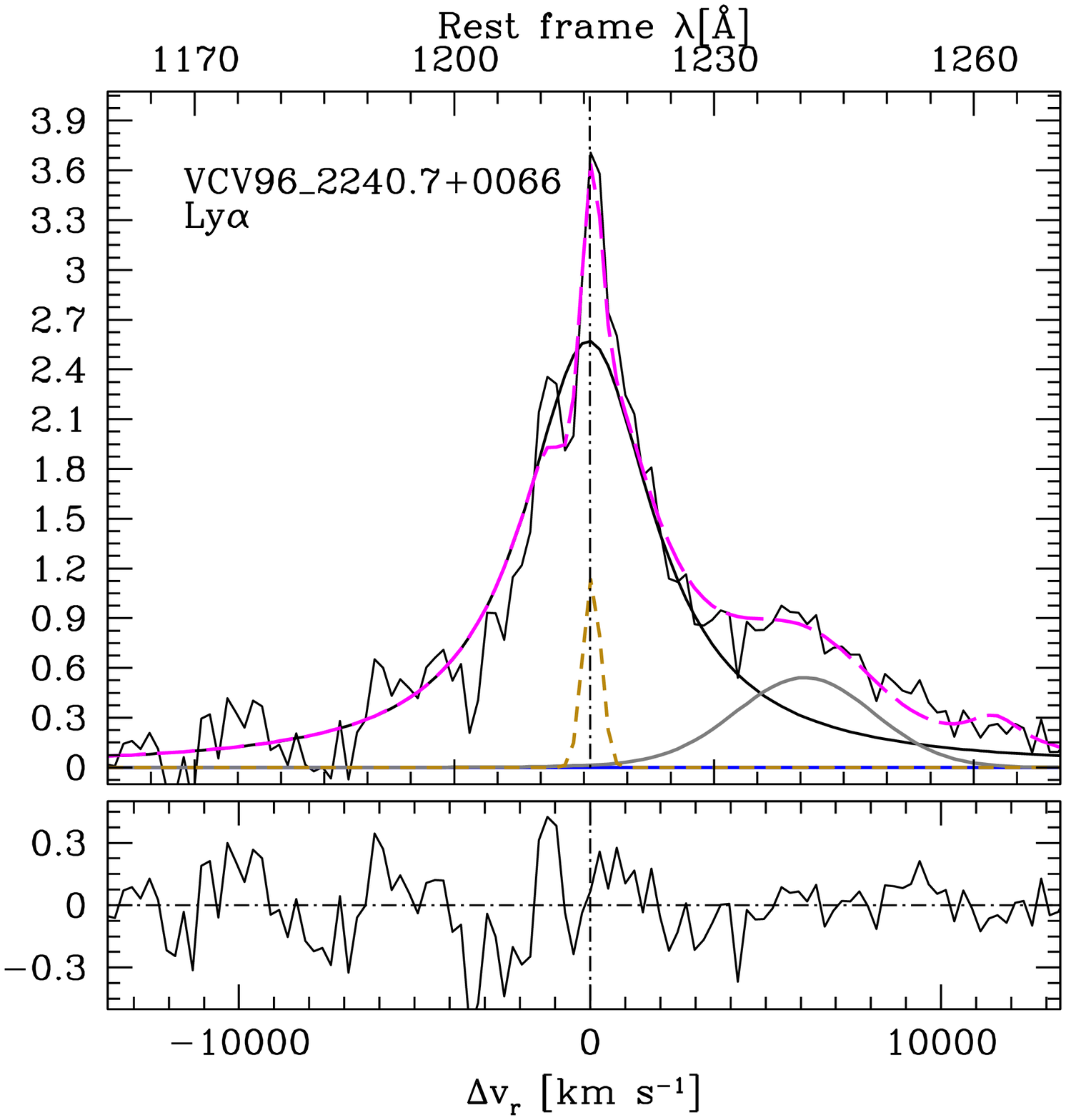}
\includegraphics[scale=0.2]{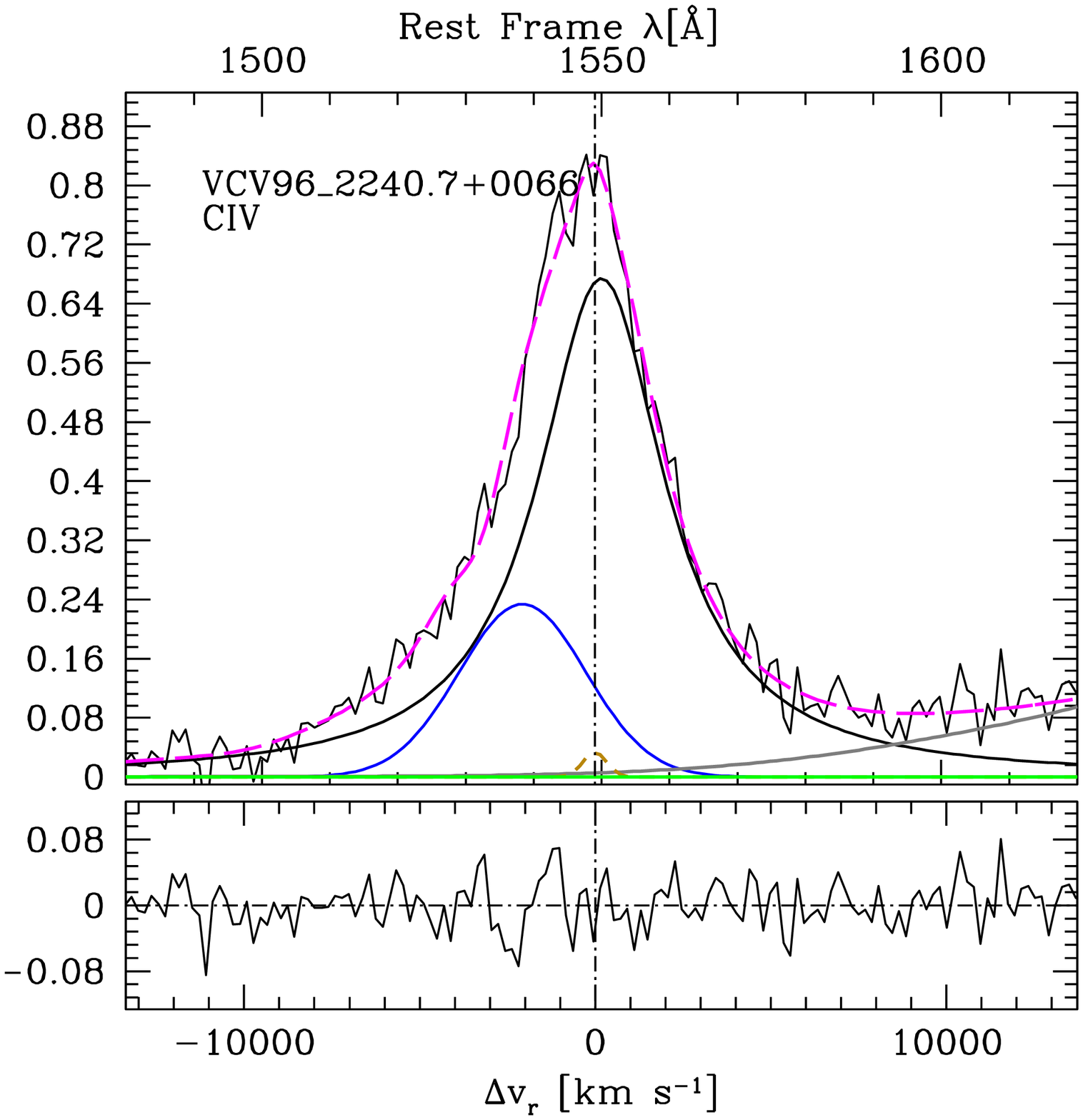}
\includegraphics[scale=0.2]{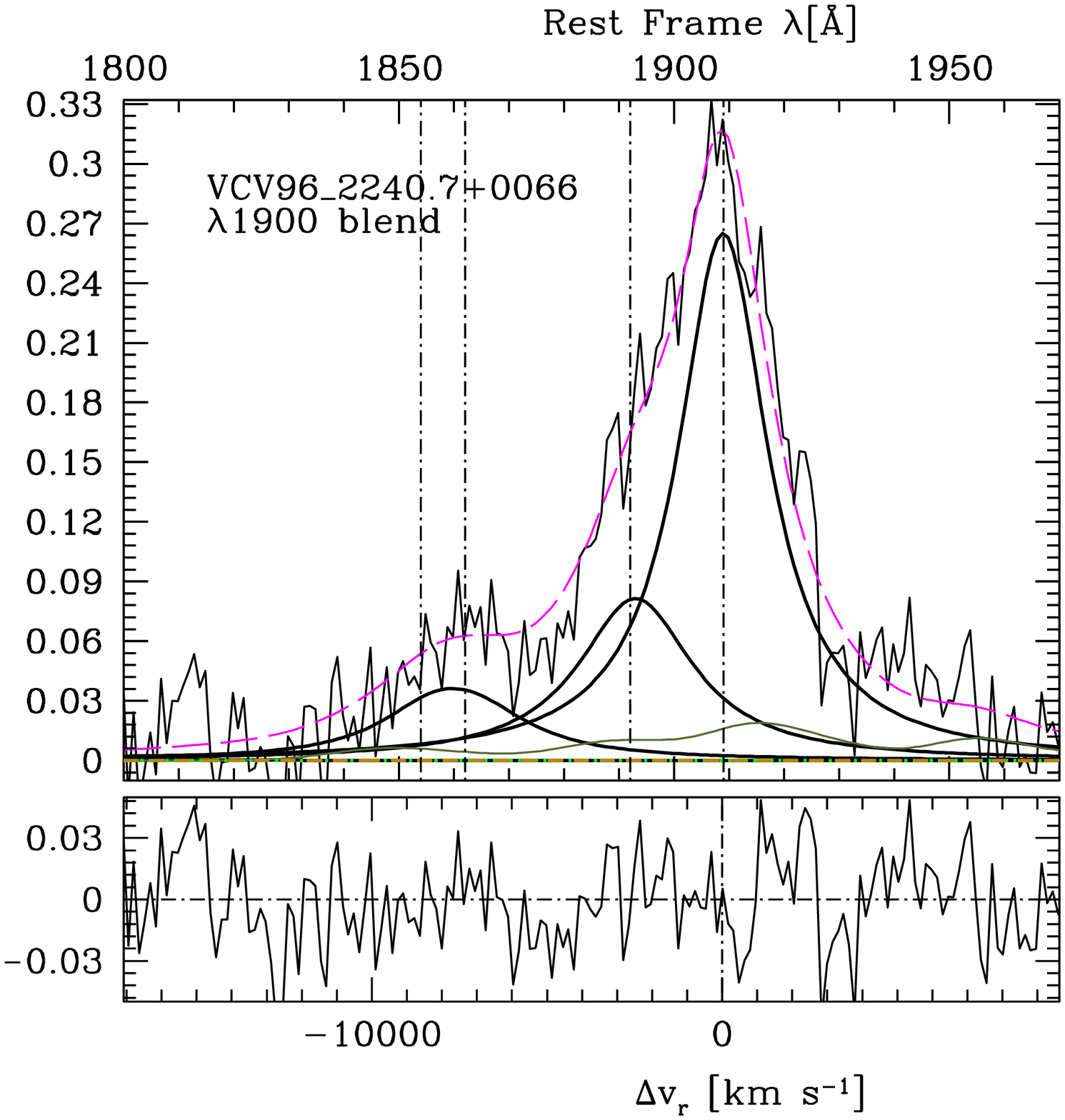}\\
\includegraphics[scale=0.2]{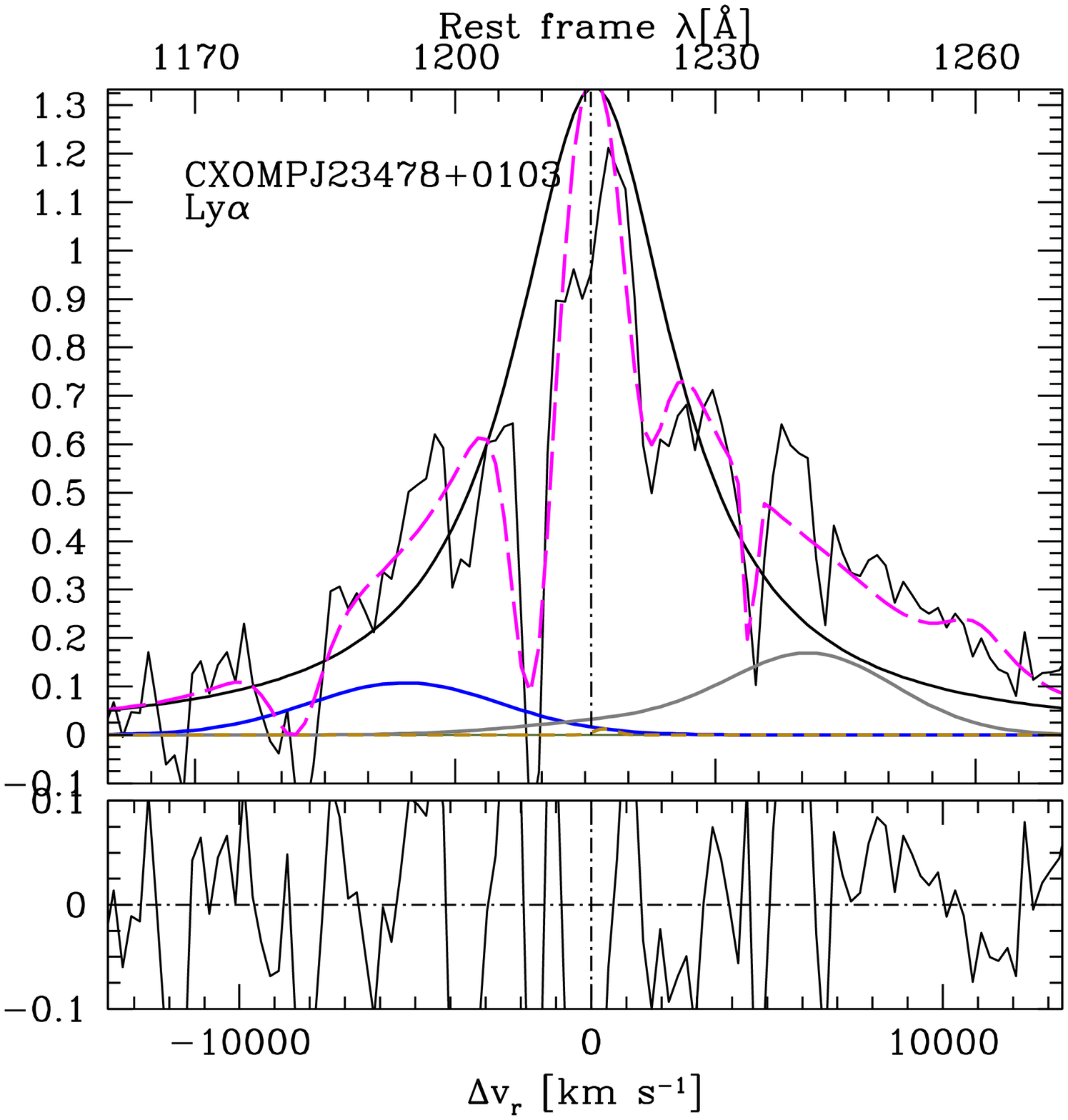}
\includegraphics[scale=0.2]{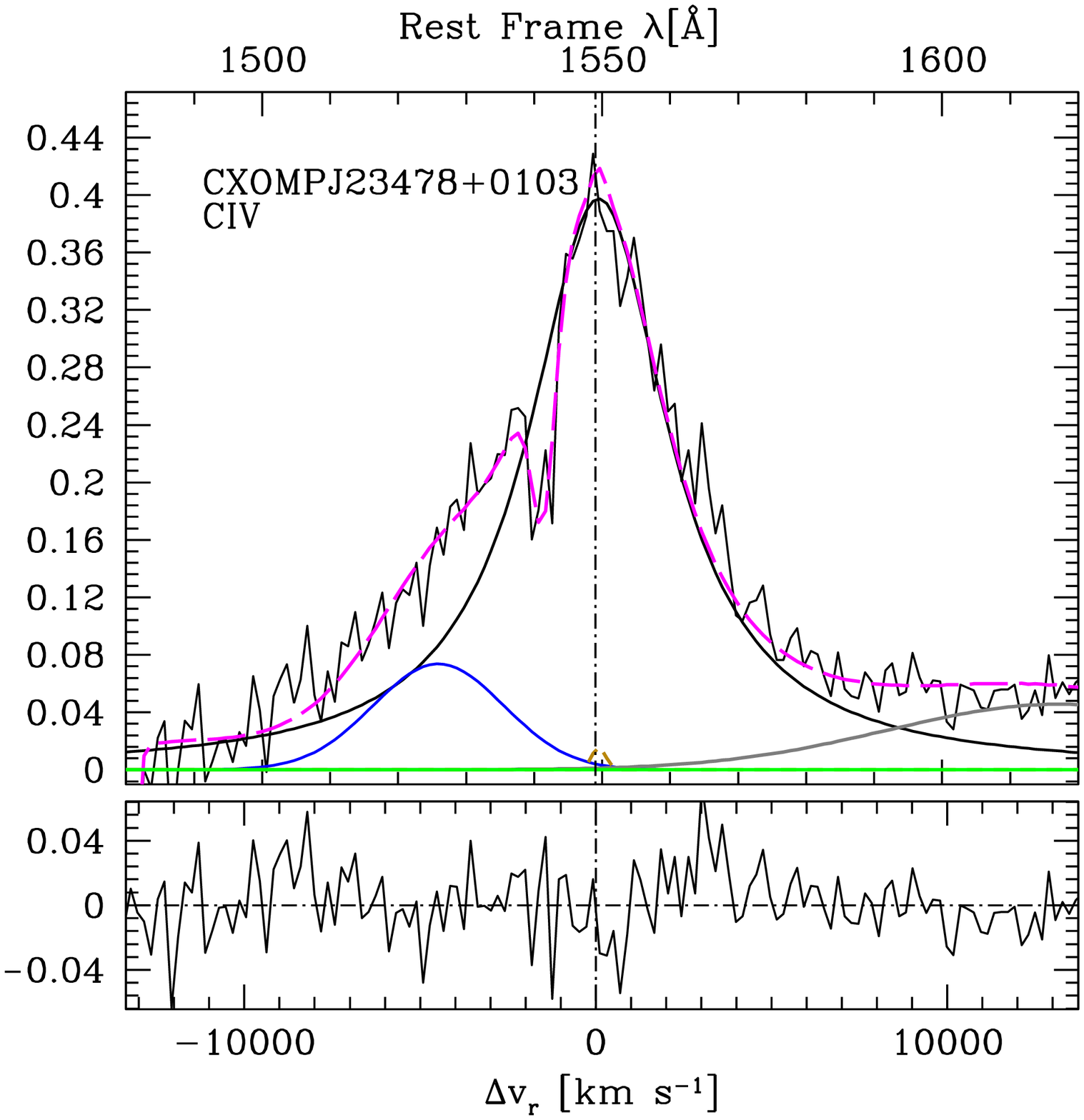}
\includegraphics[scale=0.2]{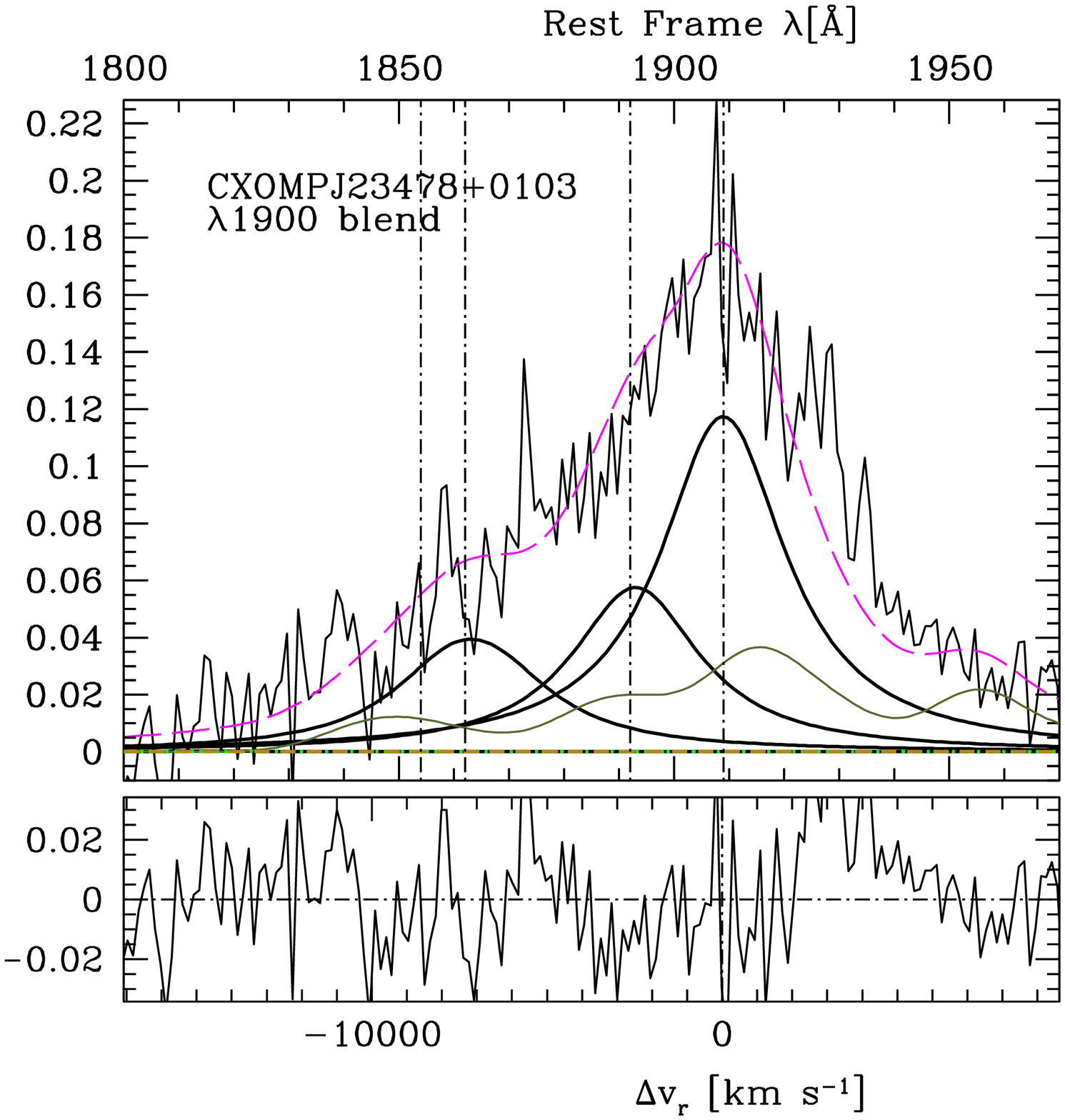}\\
\caption{(cont.)  }
\end{figure*}

\begin{figure*}
\includegraphics[scale=0.2]{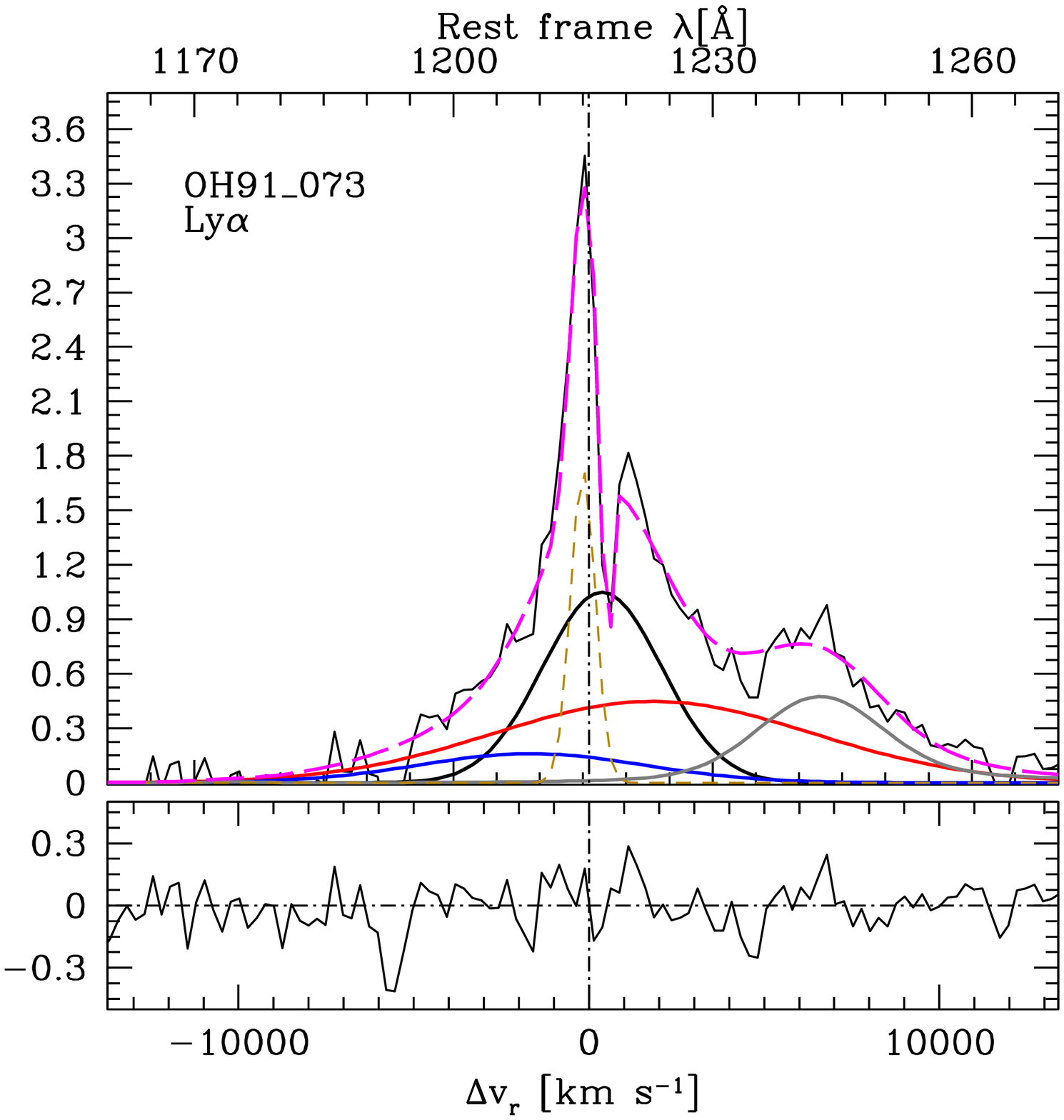}
\includegraphics[scale=0.2]{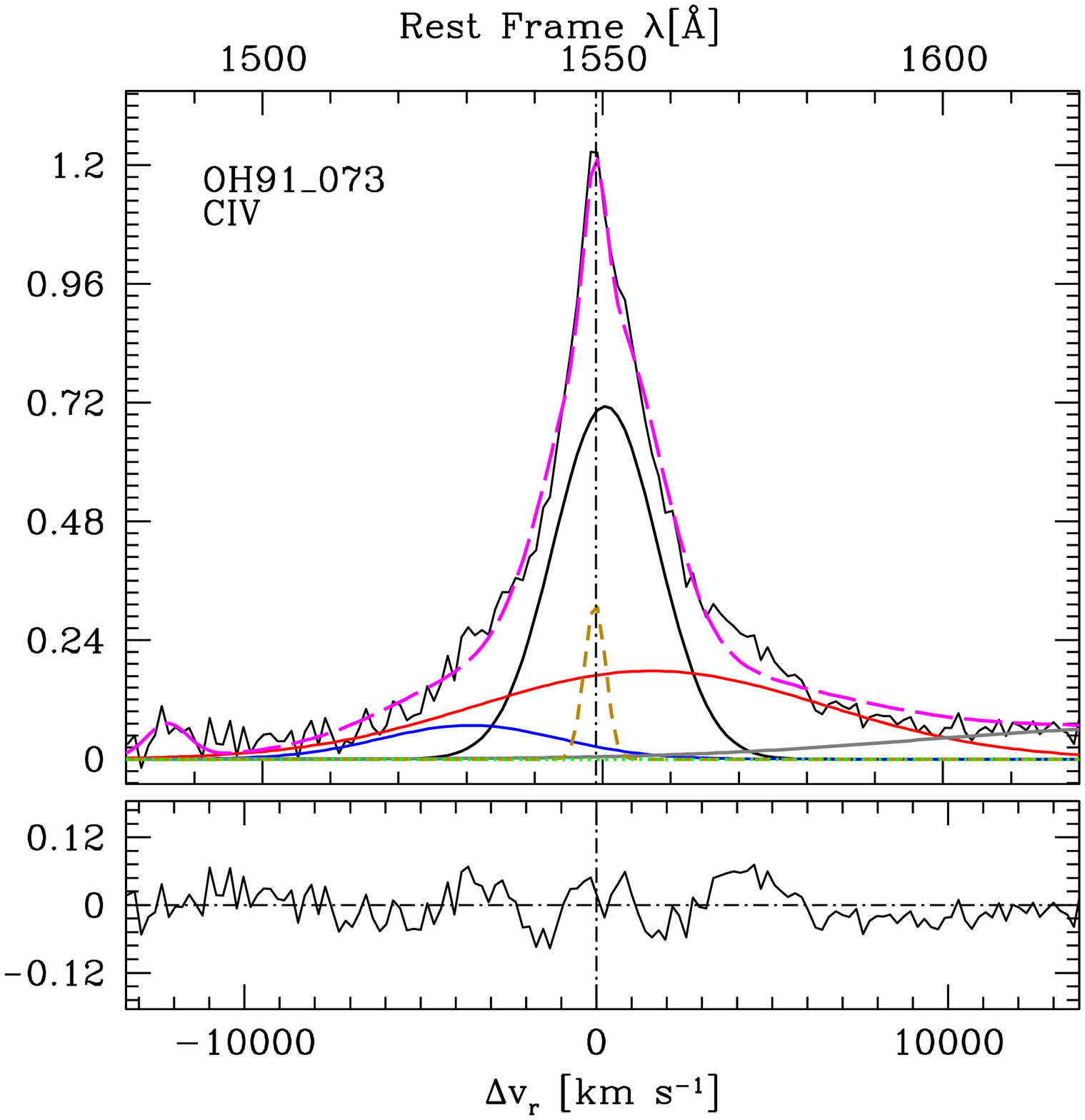}
\includegraphics[scale=0.2]{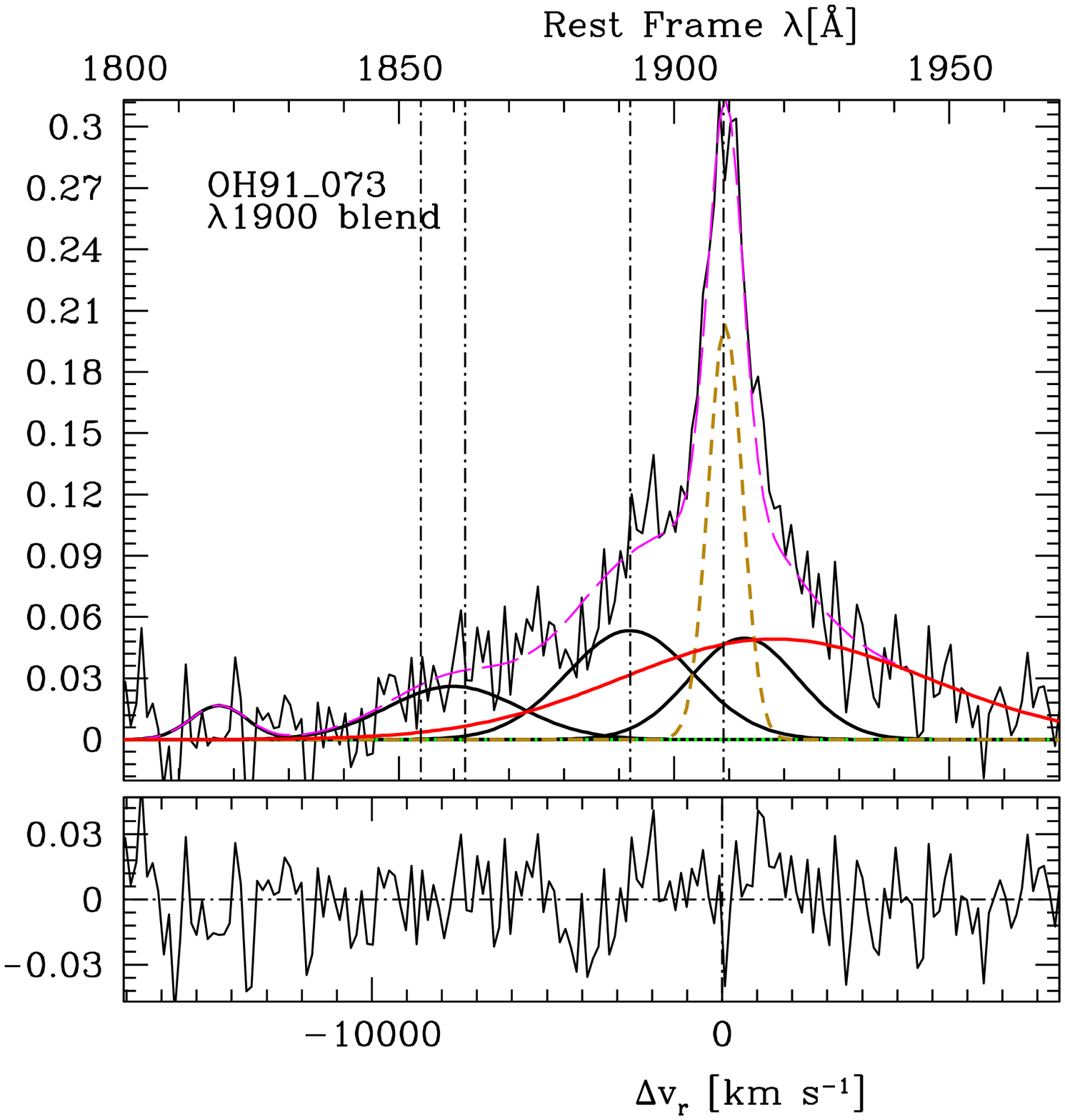}\\
\includegraphics[scale=0.2]{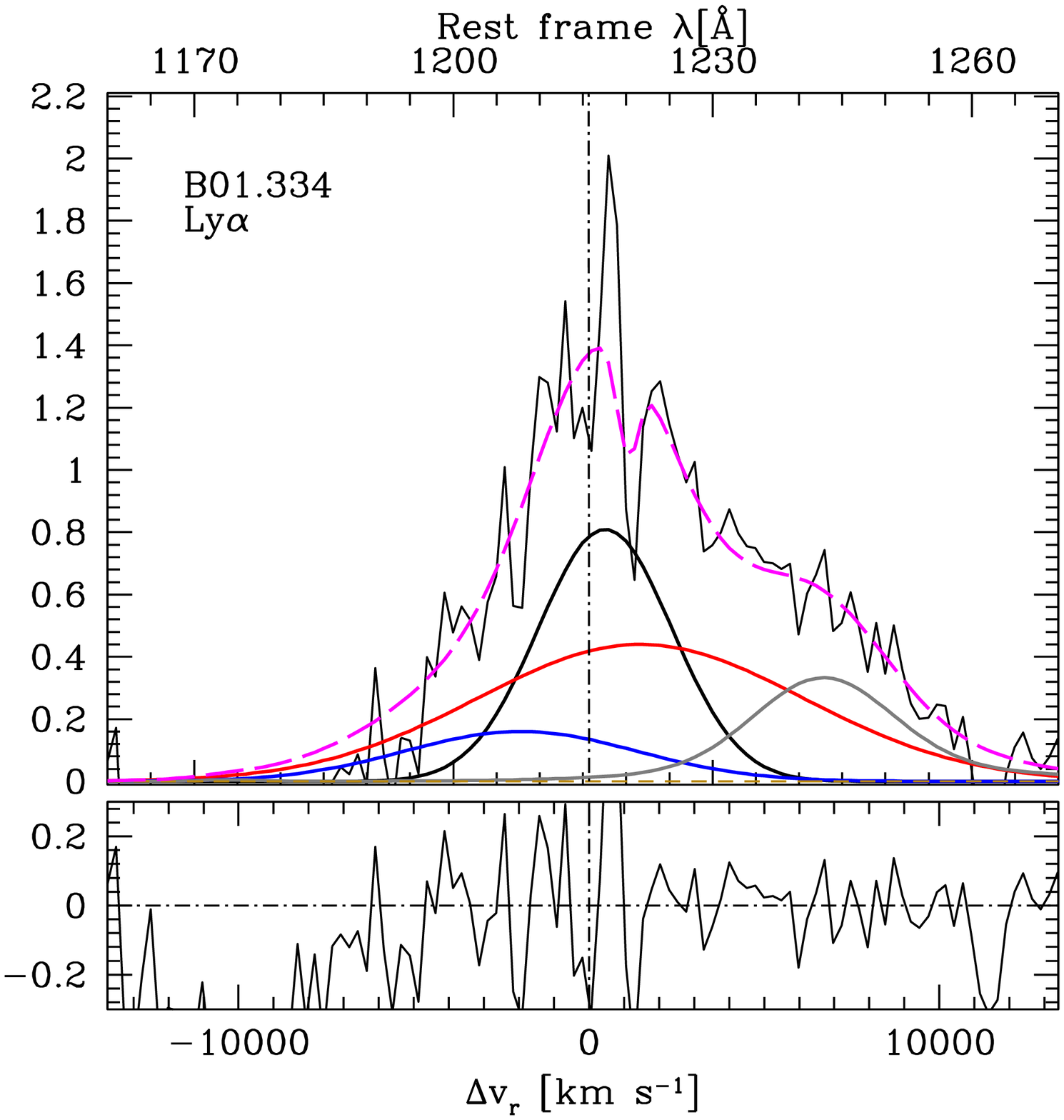}
\includegraphics[scale=0.2]{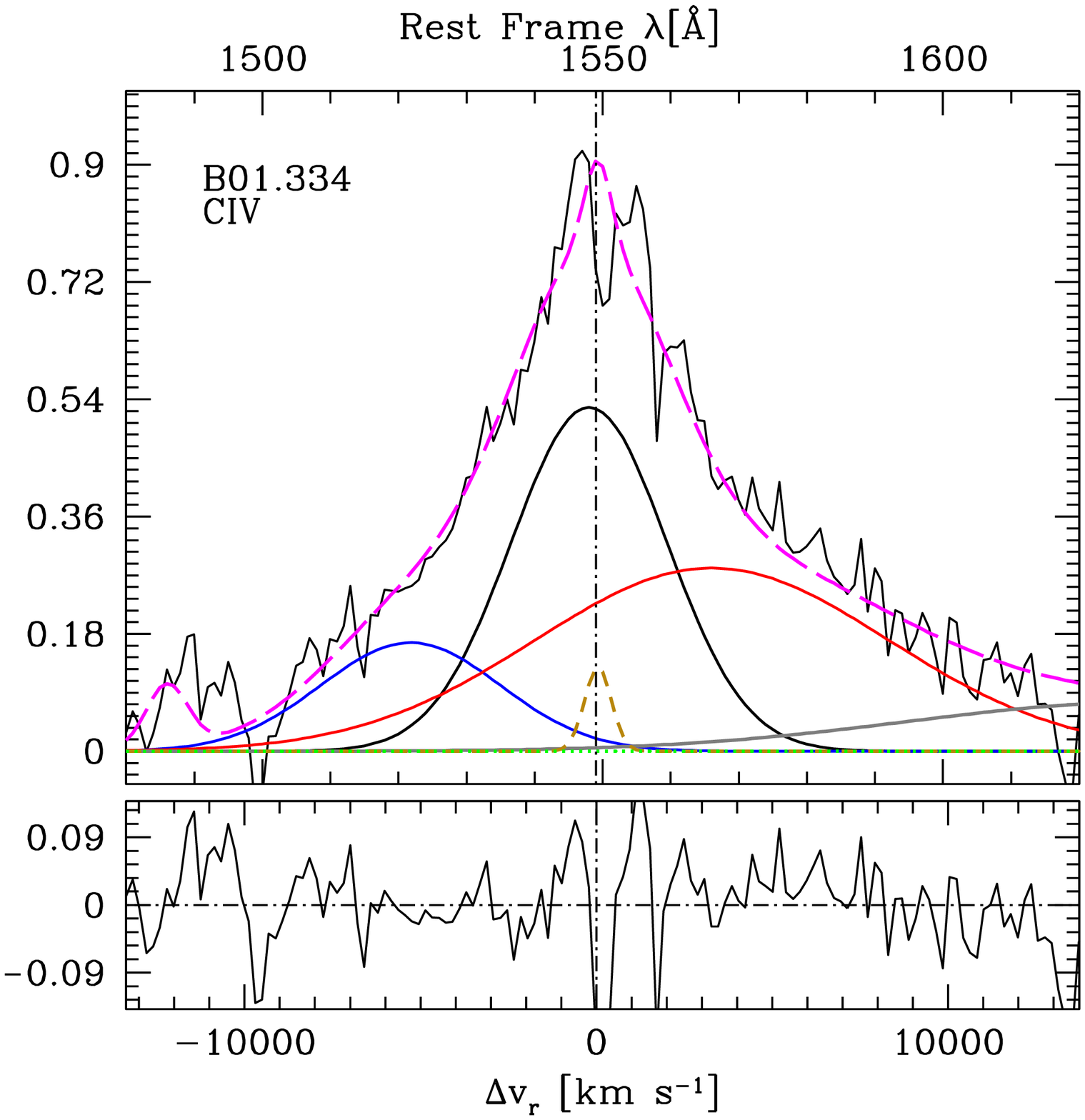}
\includegraphics[scale=0.2]{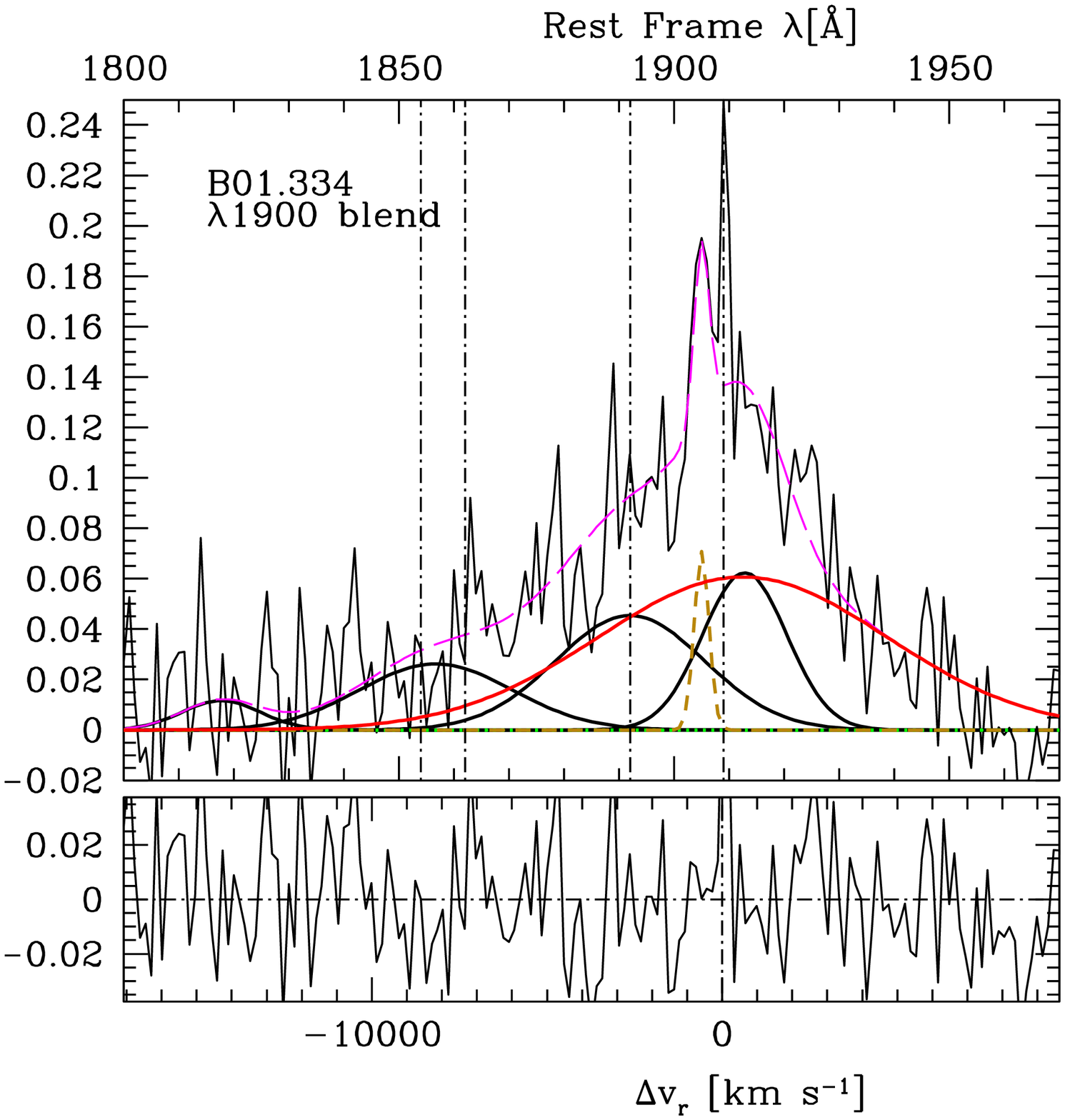}\\
\includegraphics[scale=0.2]{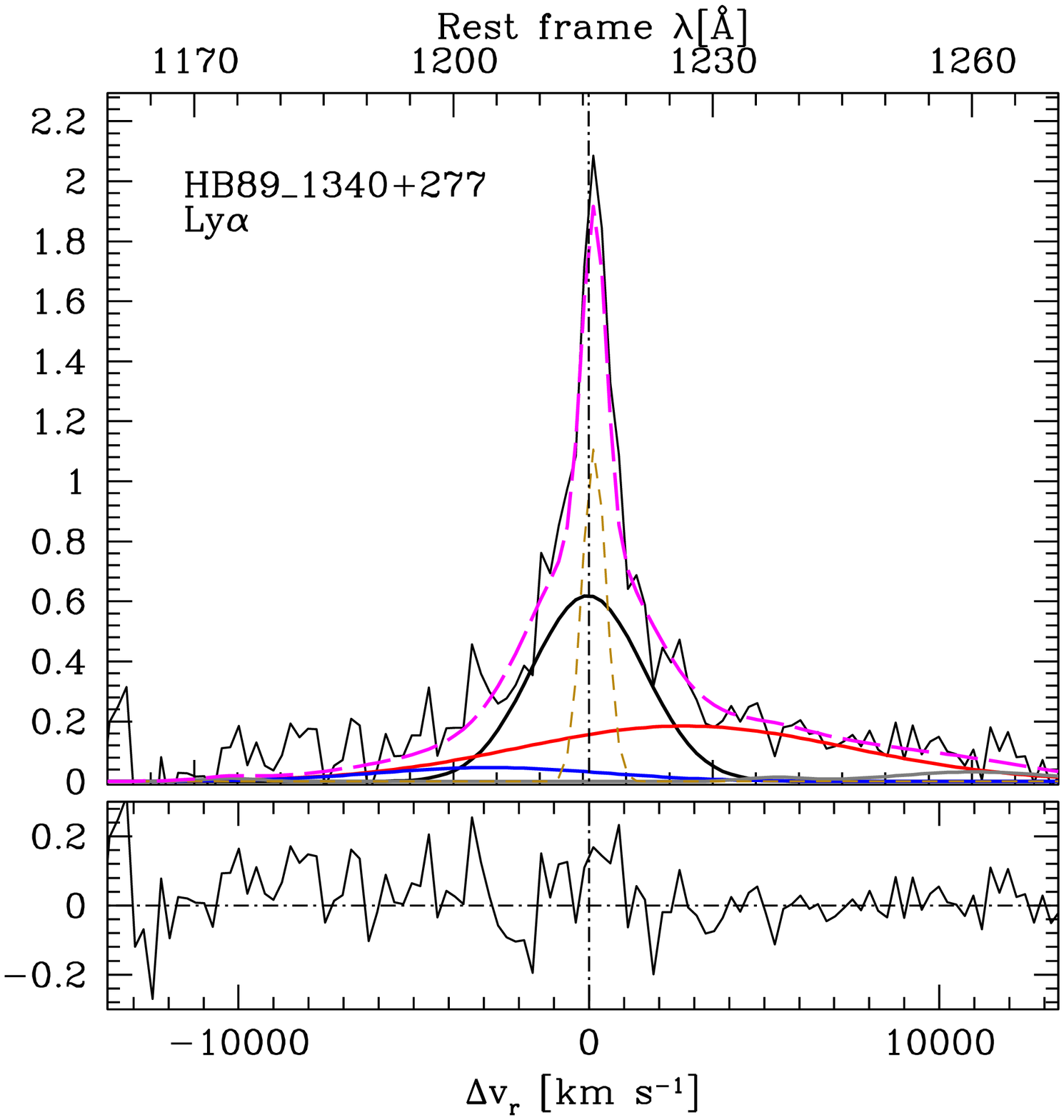}
\includegraphics[scale=0.2]{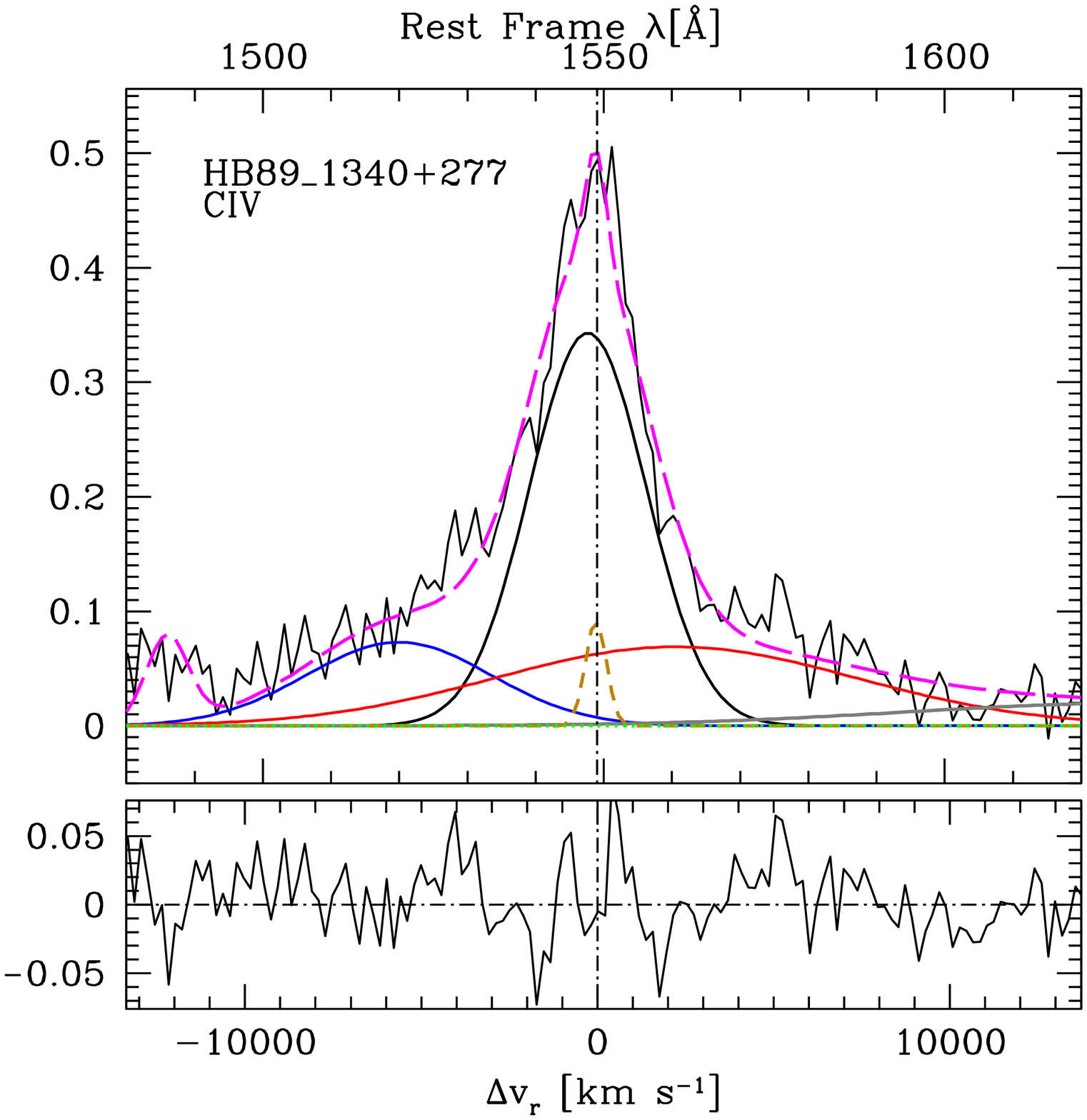}
\includegraphics[scale=0.2]{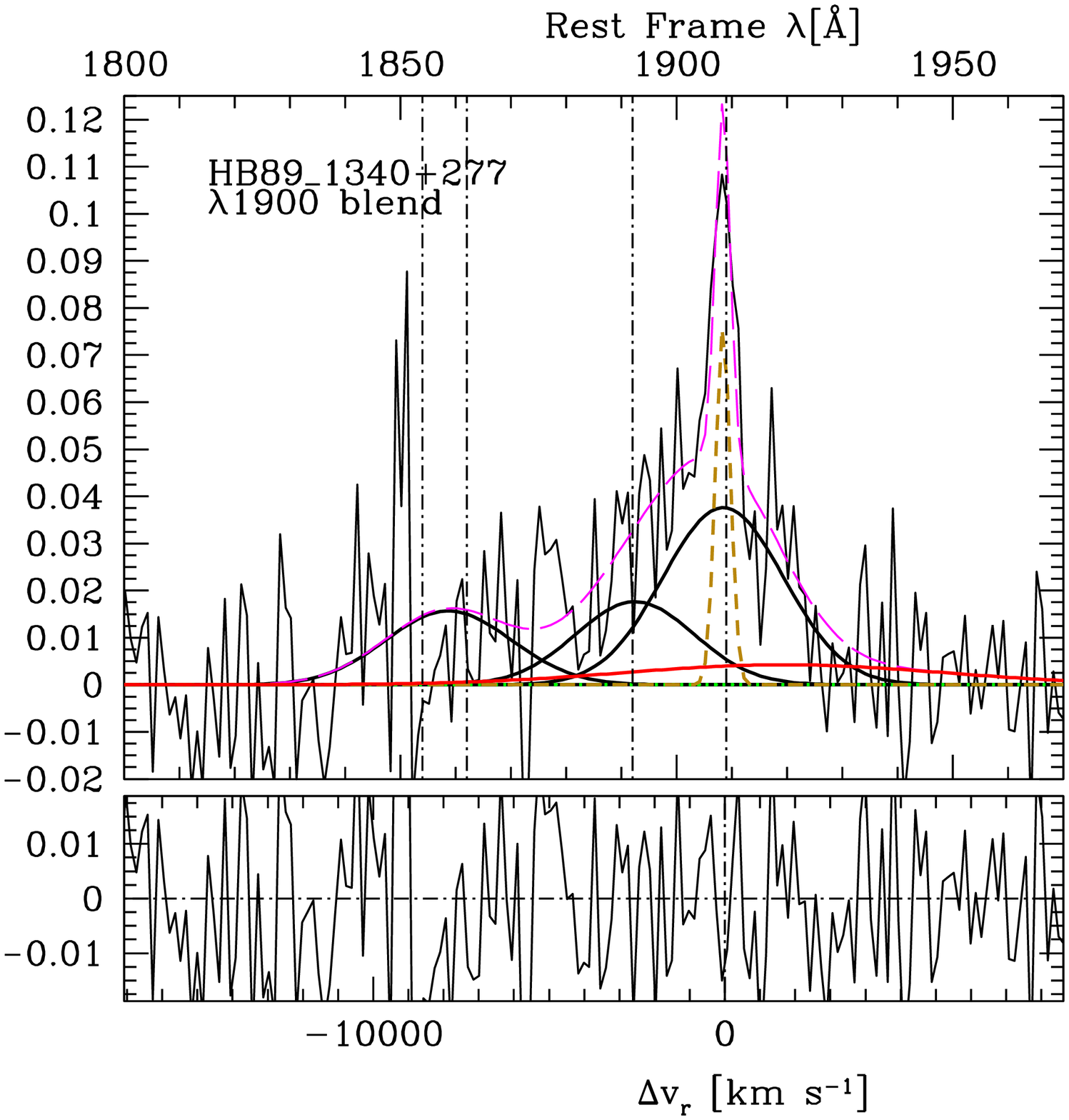}\\
\includegraphics[scale=0.2]{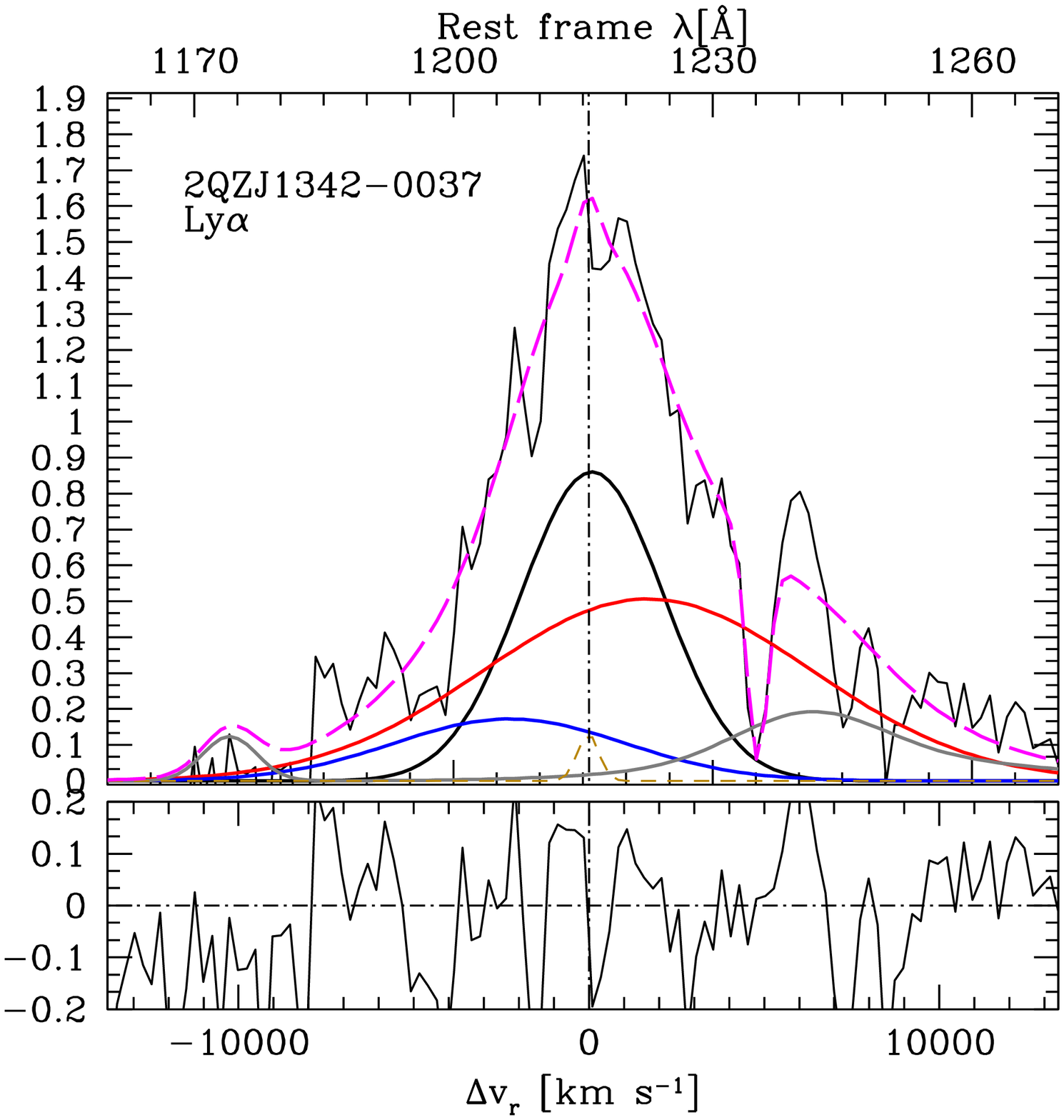}
\includegraphics[scale=0.2]{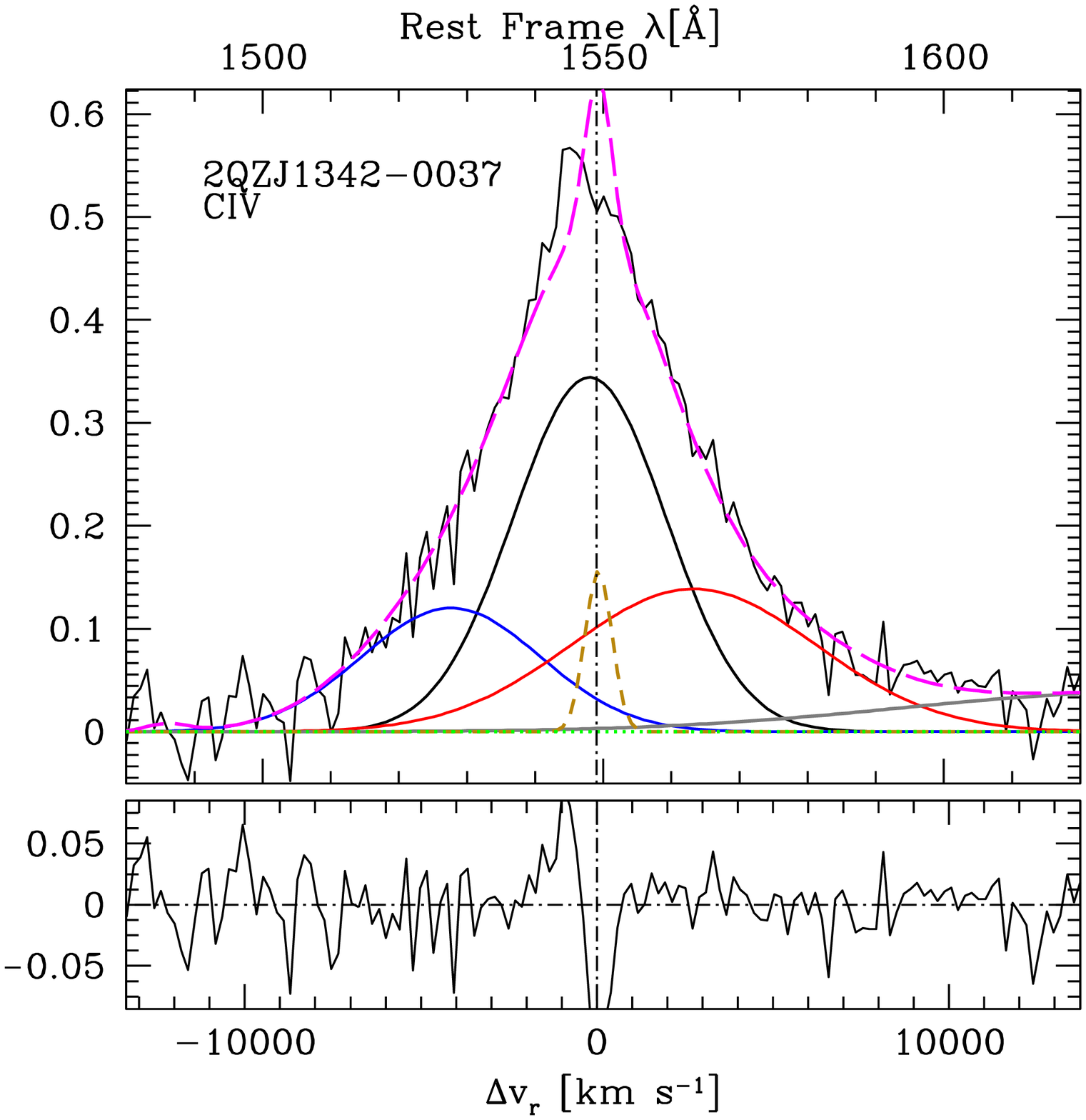}
\includegraphics[scale=0.2]{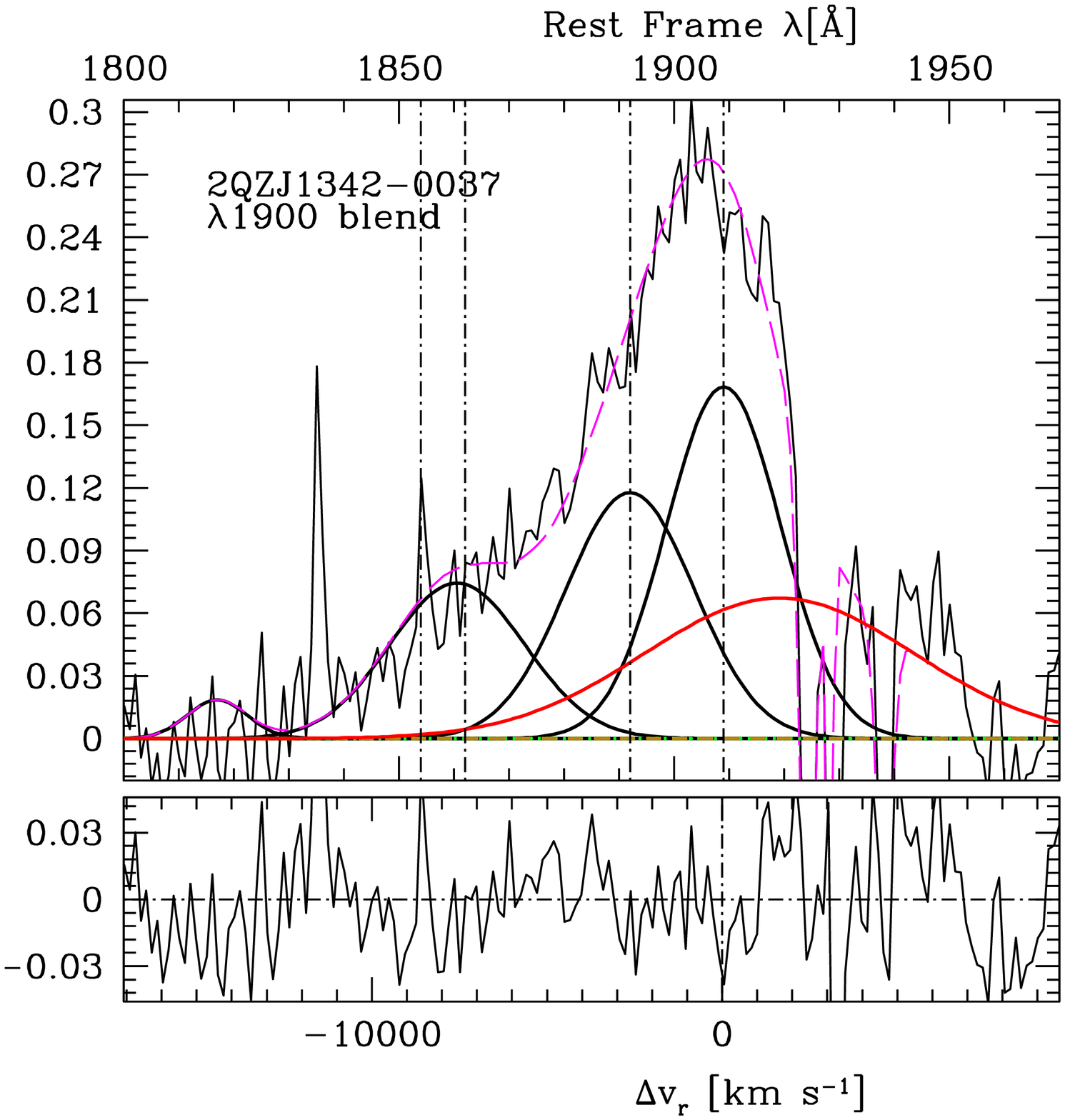}\\

\caption{Results of line fit analysis on the \lya\ (left panel) \civ\ (middle) and 1900 \AA\ emission 
features for Pop. B sources. Meaning of symbols and colors is as in the previous 
Figure. The VBC assumed to be present in the Pop. B emission line profiles is traced by a 
thick red line.  \label{fig:fitsb}}
\end{figure*}
\addtocounter{figure}{-1}

\newpage
\clearpage

\begin{figure*}
\includegraphics[scale=0.2]{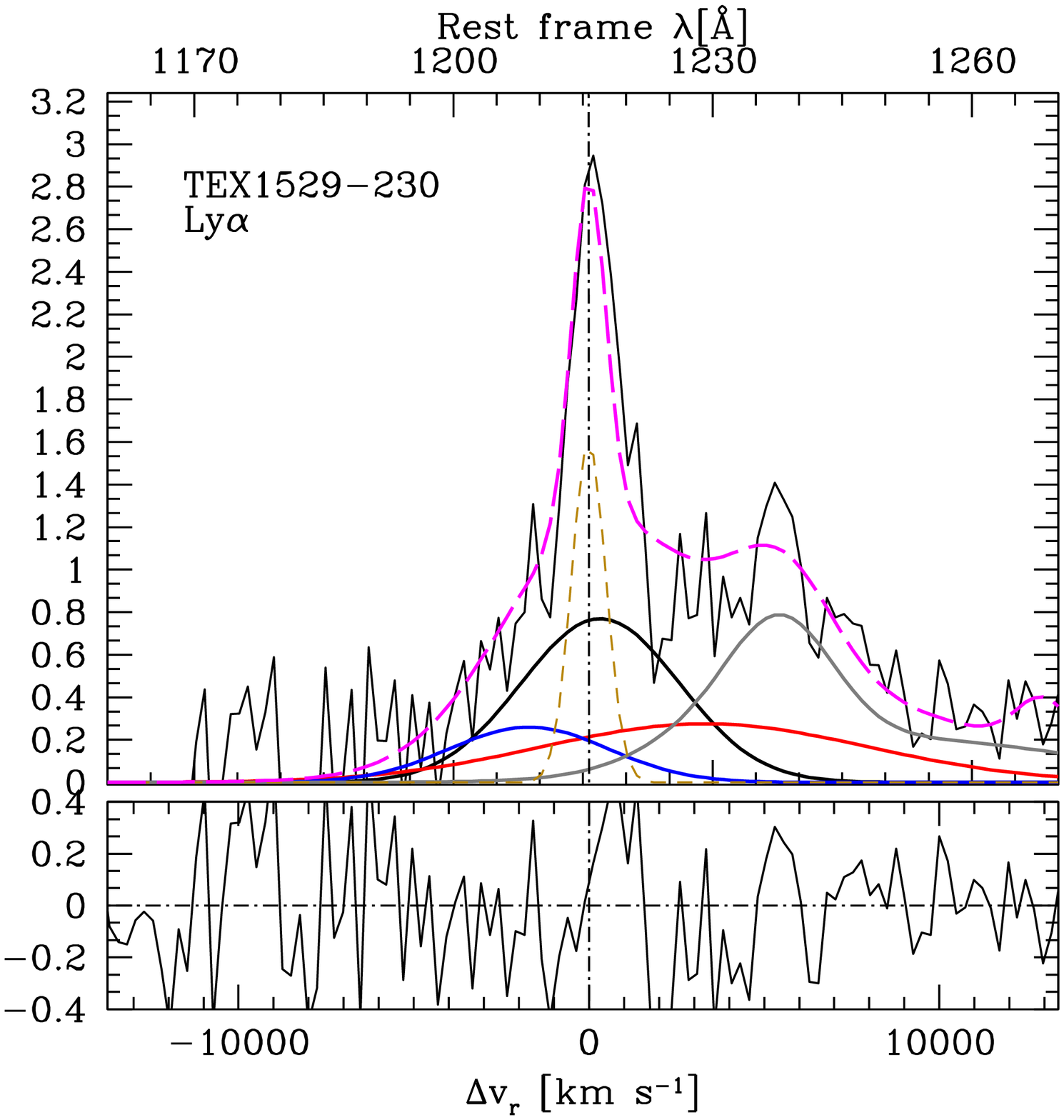}
\includegraphics[scale=0.2]{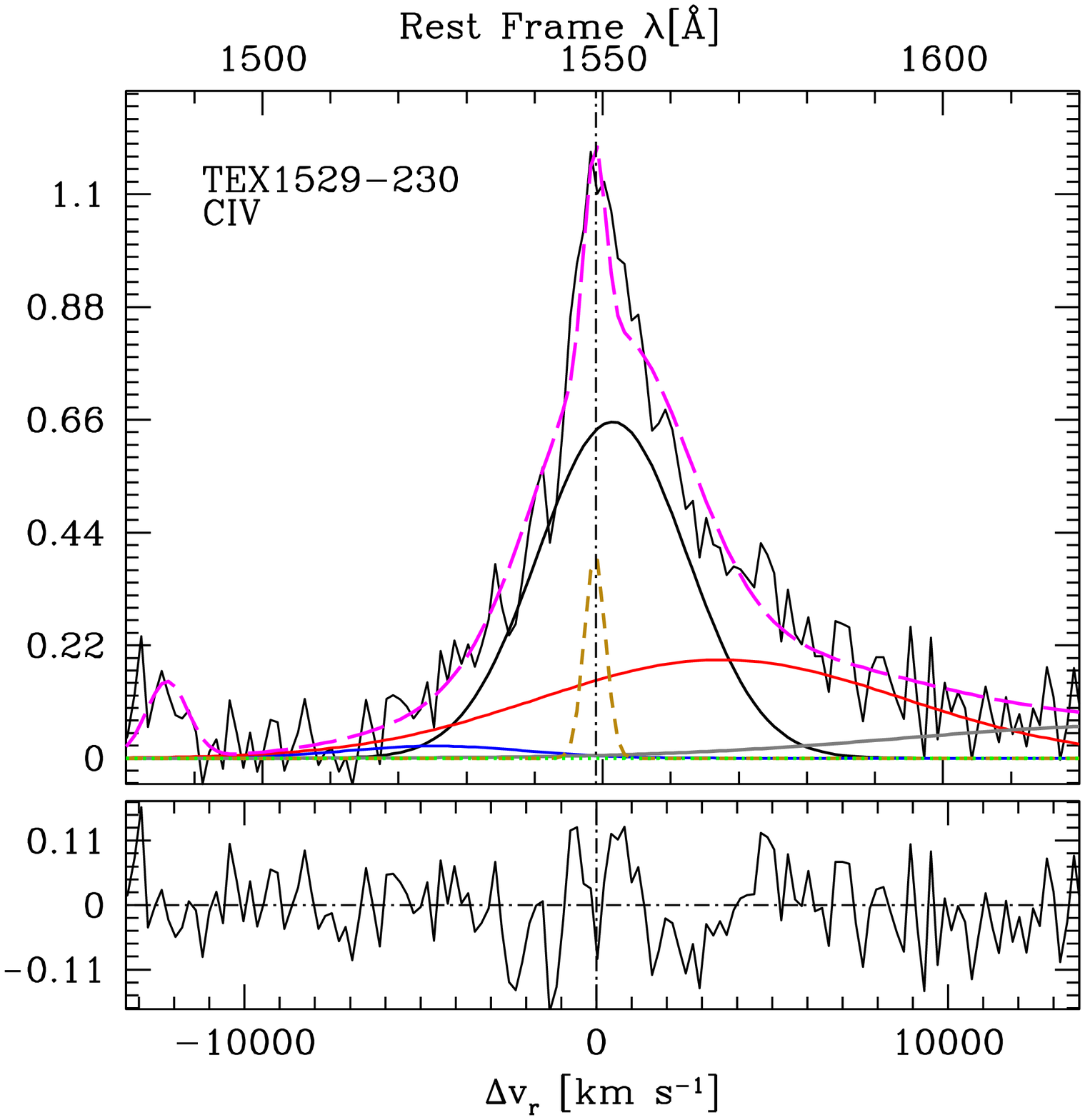}
\includegraphics[scale=0.2]{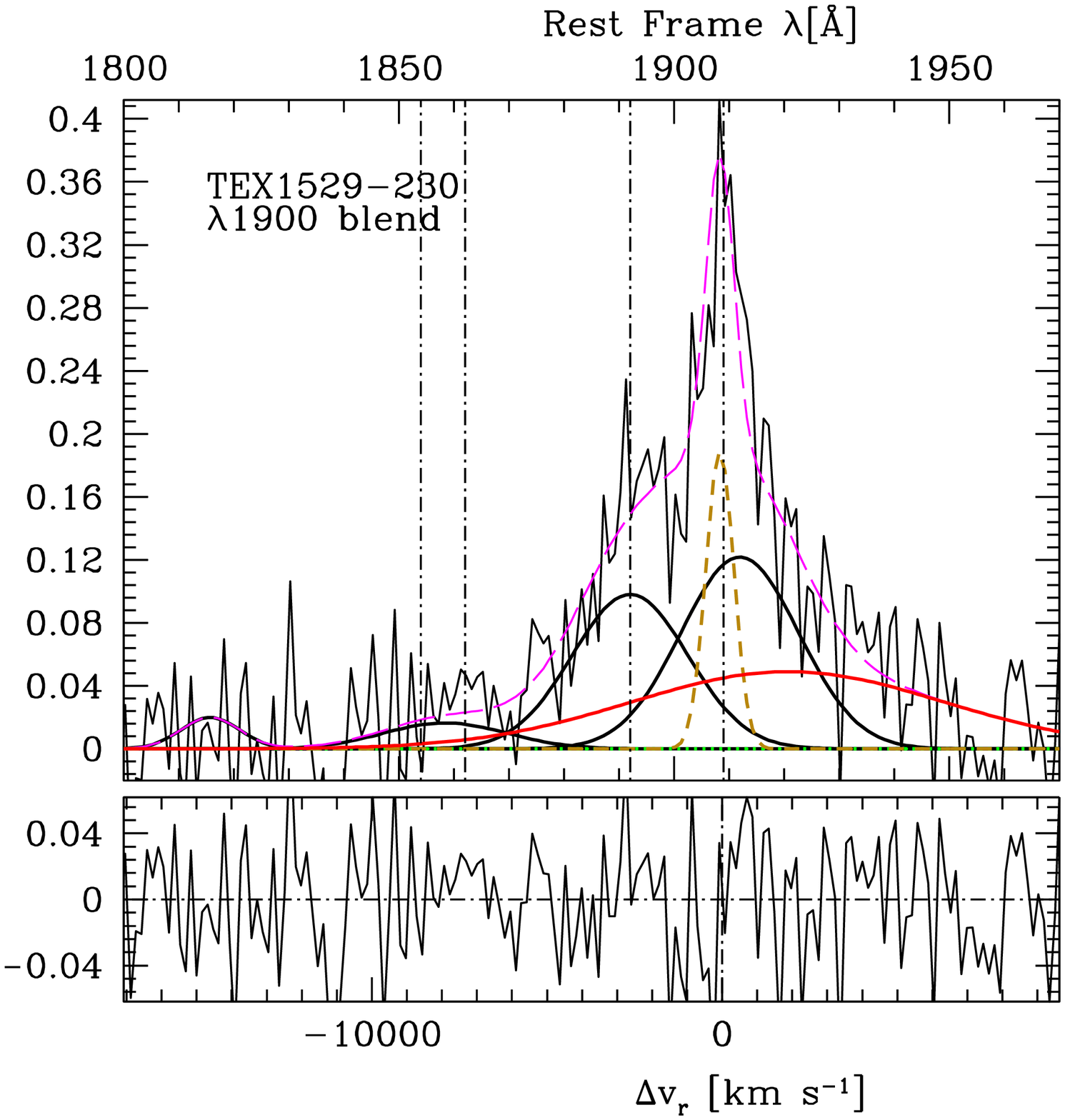}\\
\includegraphics[scale=0.2]{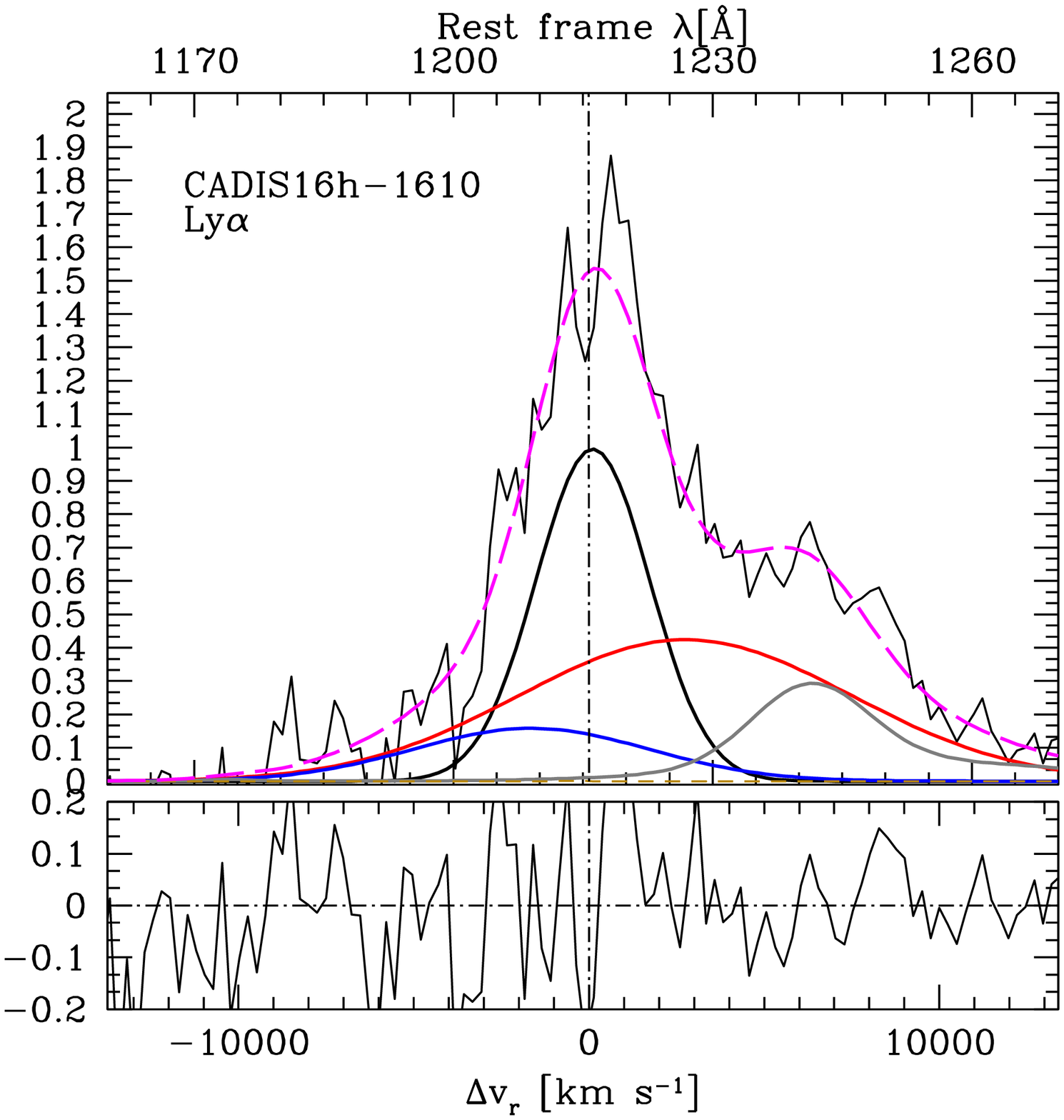}
\includegraphics[scale=0.2]{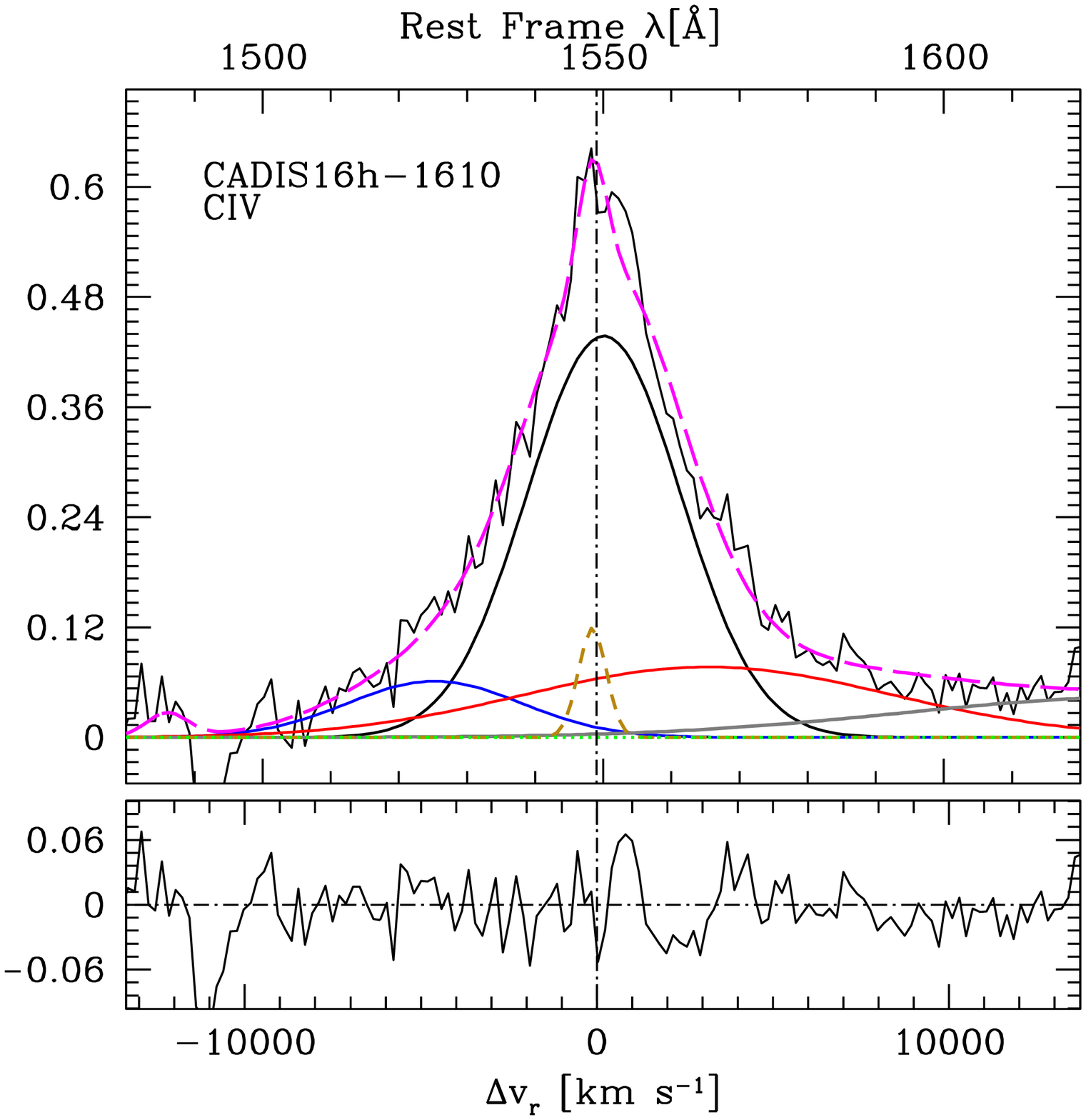}
\includegraphics[scale=0.2]{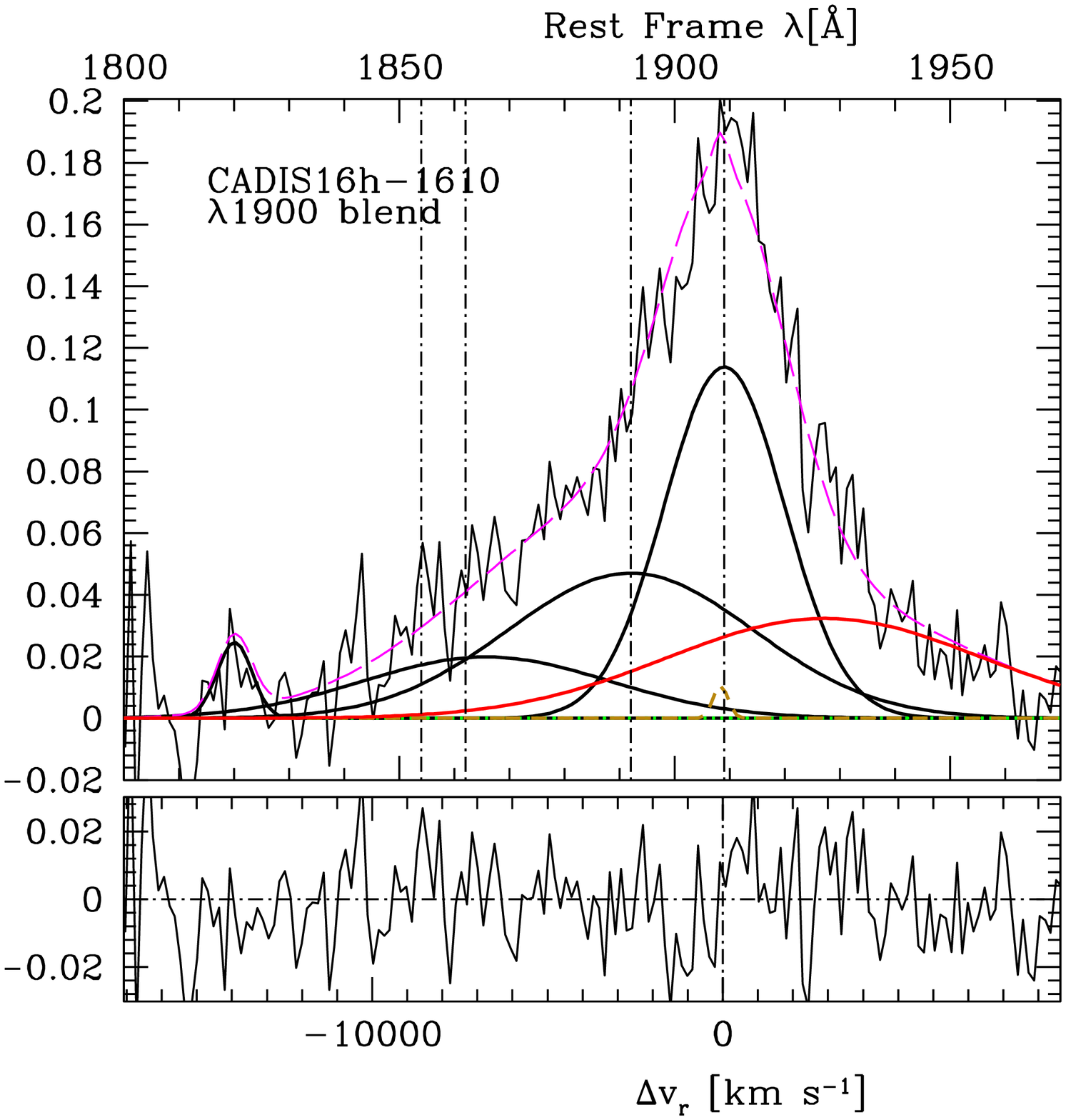}\\
\includegraphics[scale=0.2]{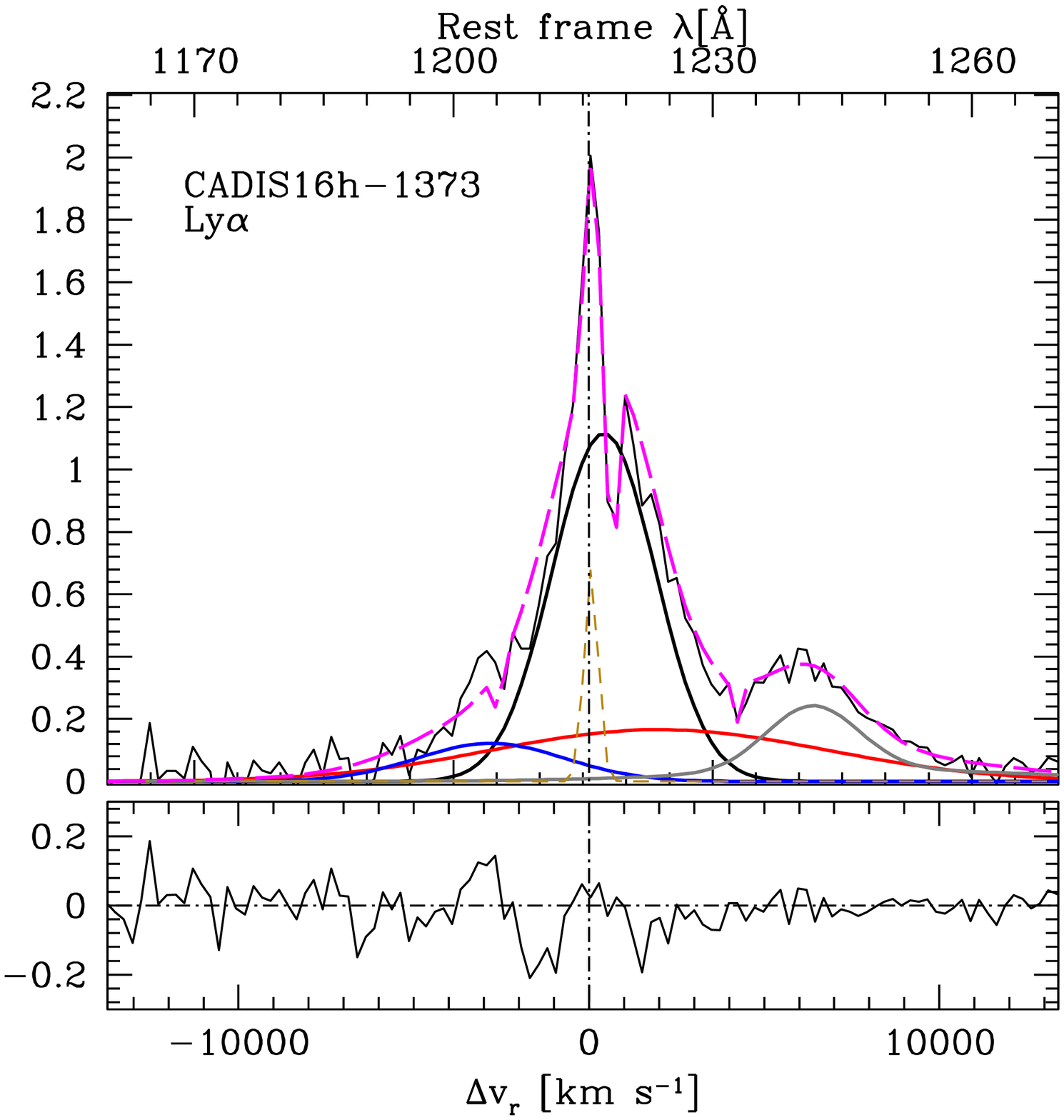}
\includegraphics[scale=0.2]{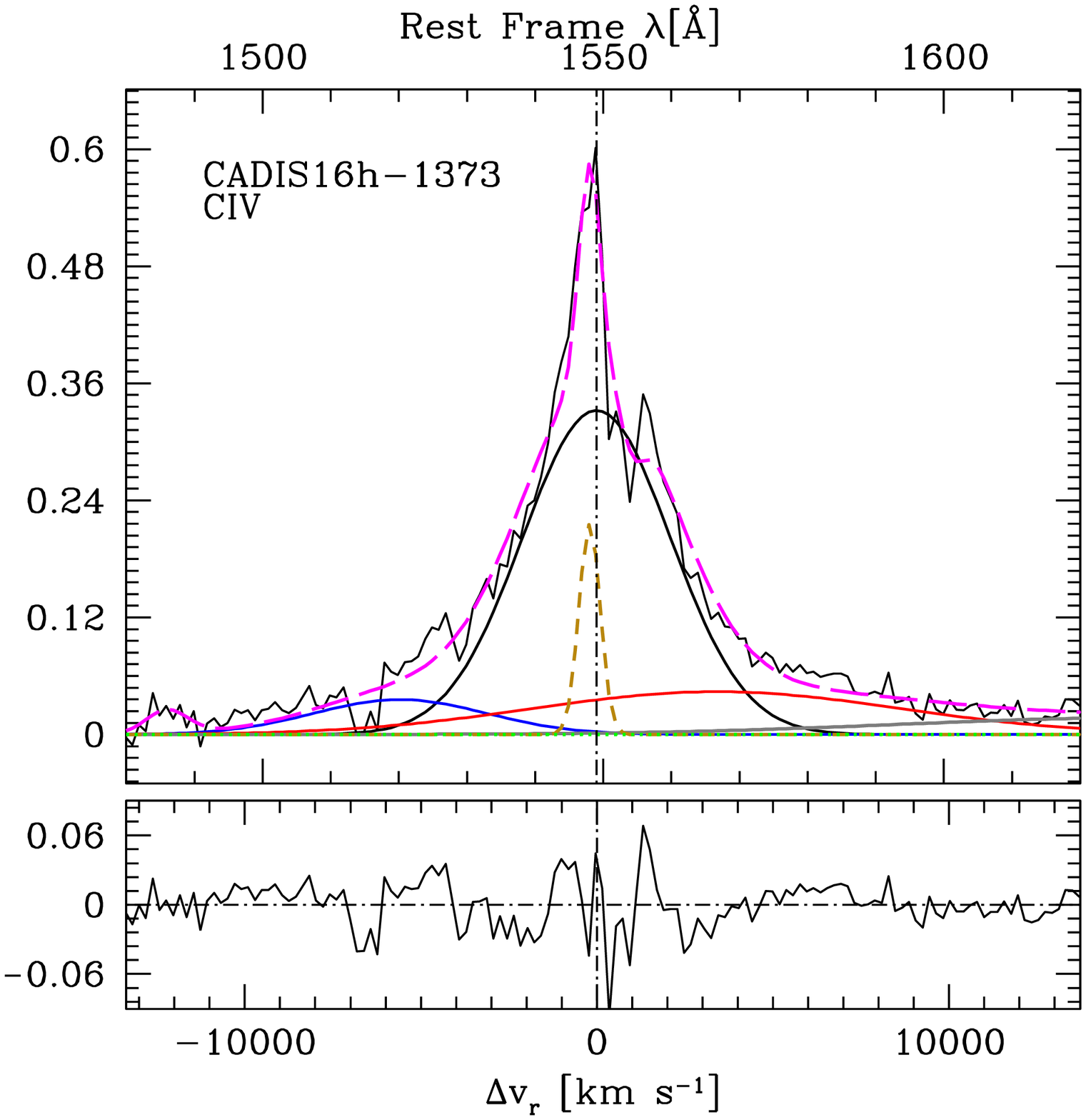}
\includegraphics[scale=0.2]{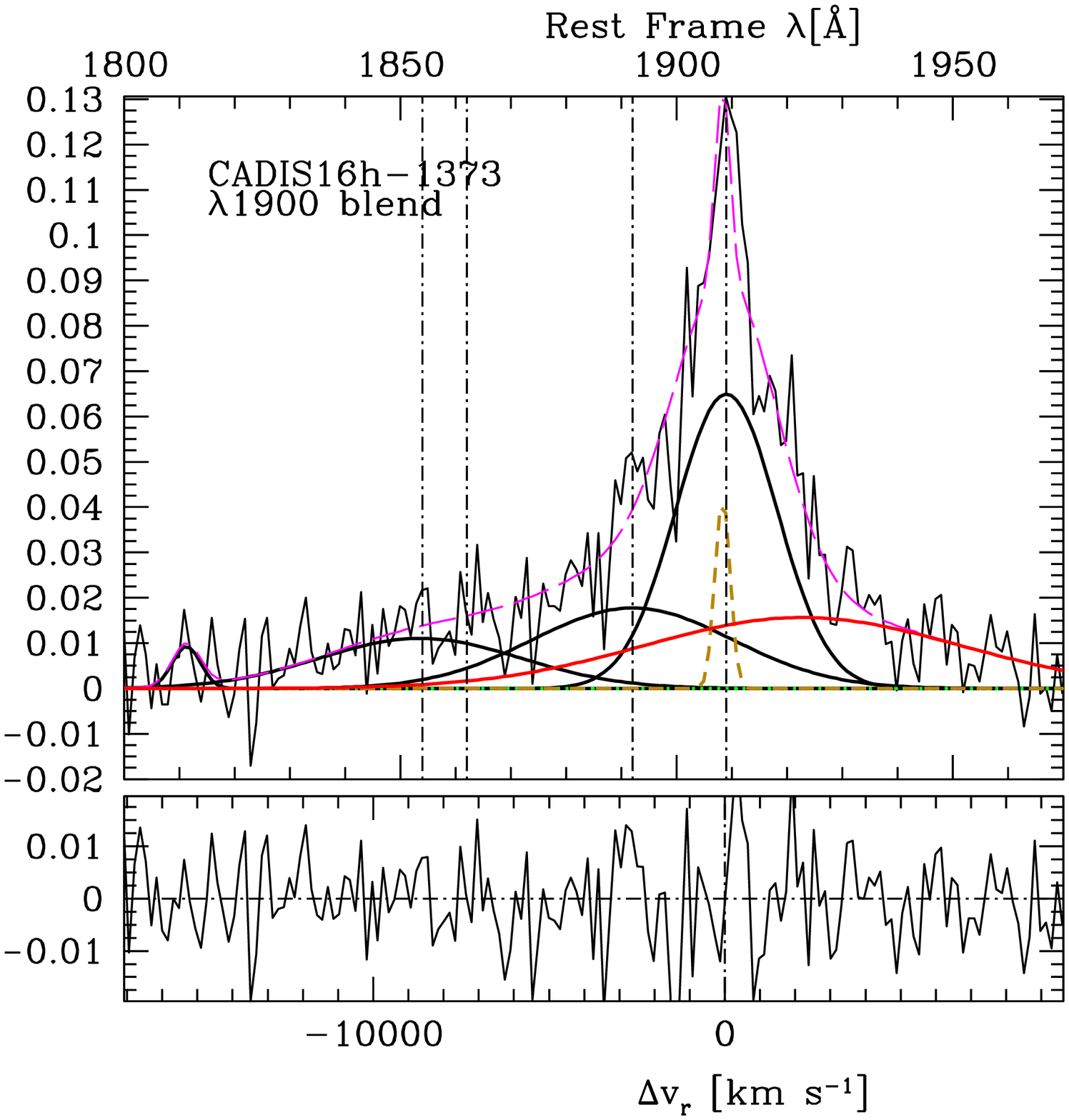}\\
\includegraphics[scale=0.2]{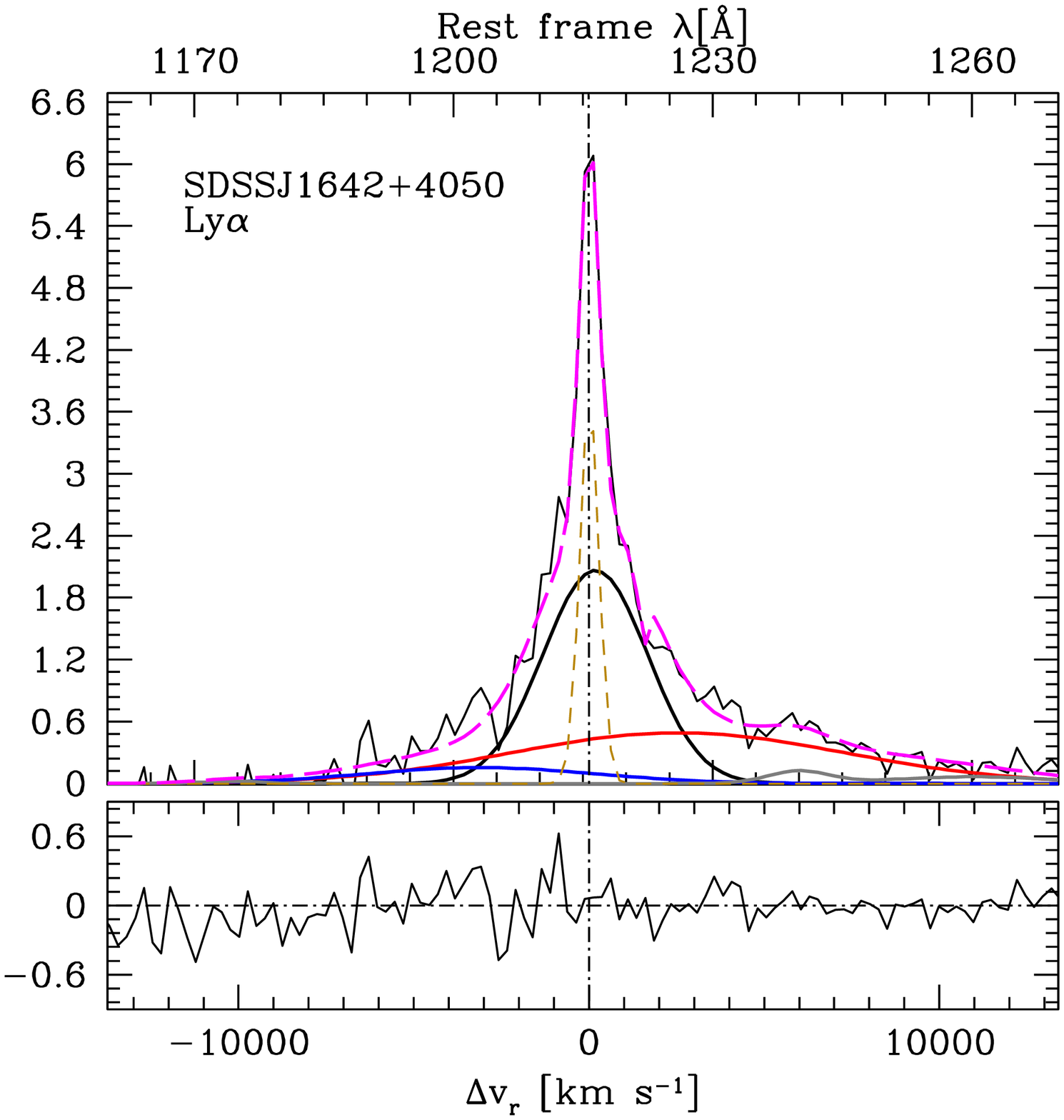}
\includegraphics[scale=0.2]{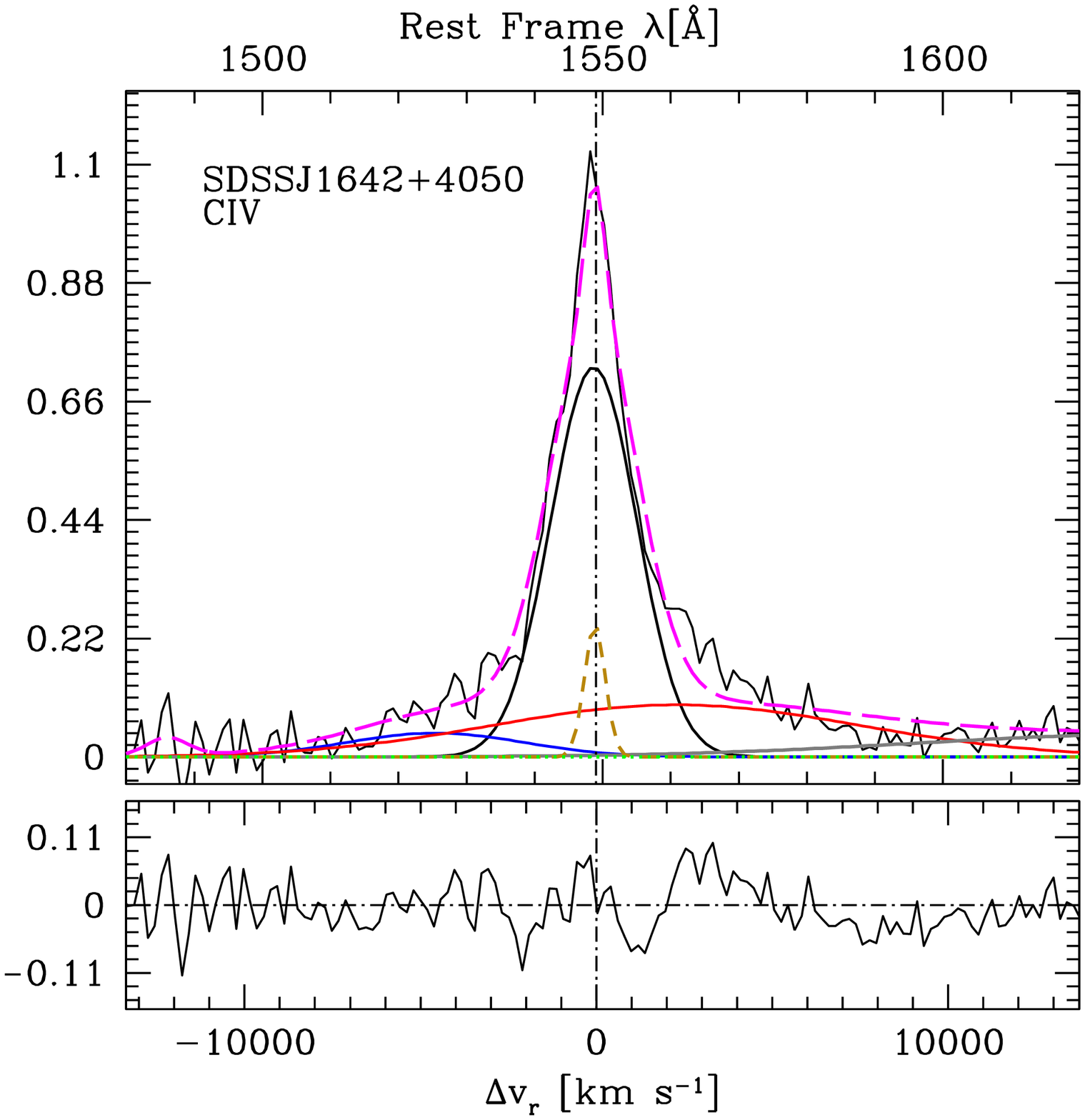}
\includegraphics[scale=0.2]{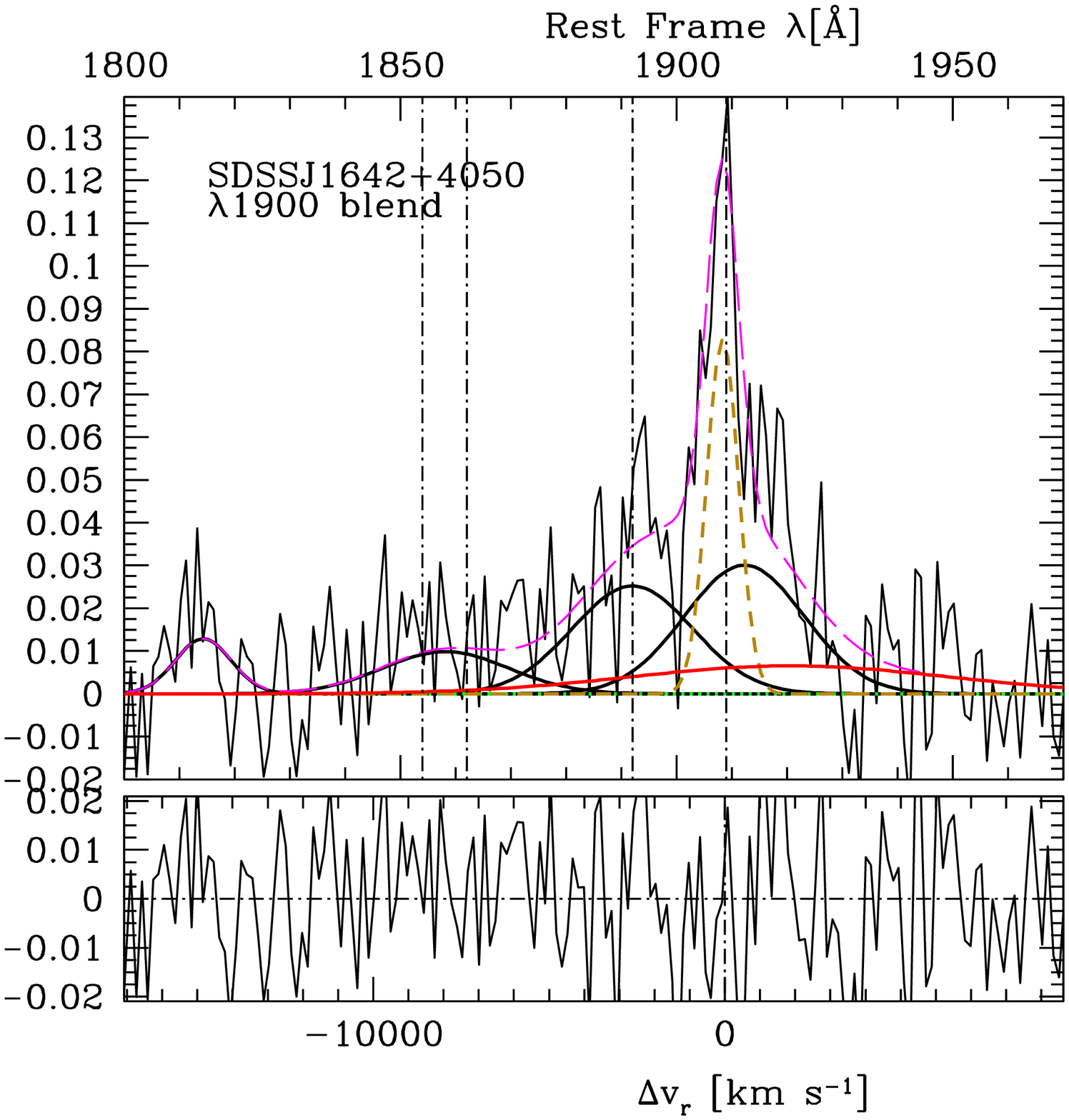}\\
\includegraphics[scale=0.2]{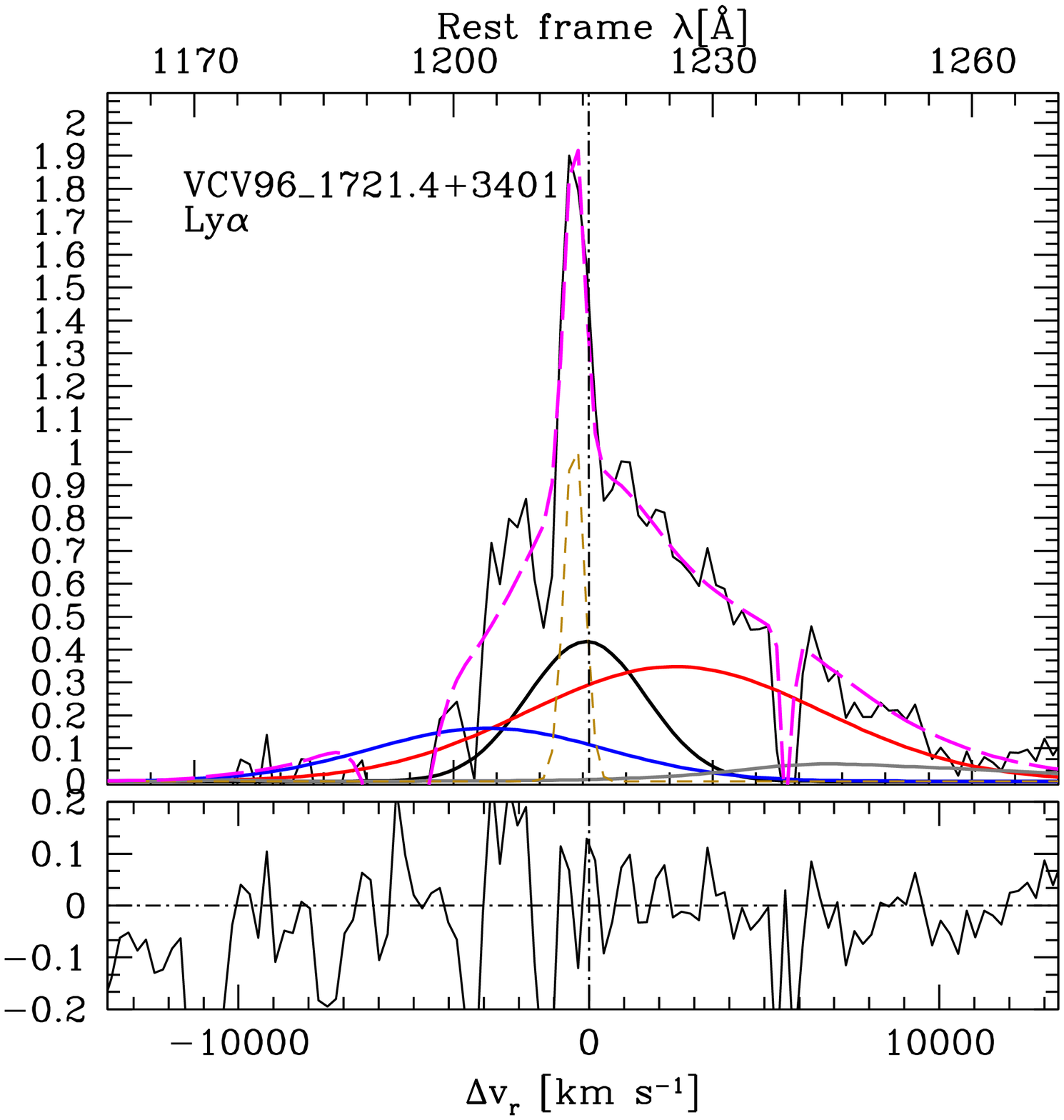}
\includegraphics[scale=0.2]{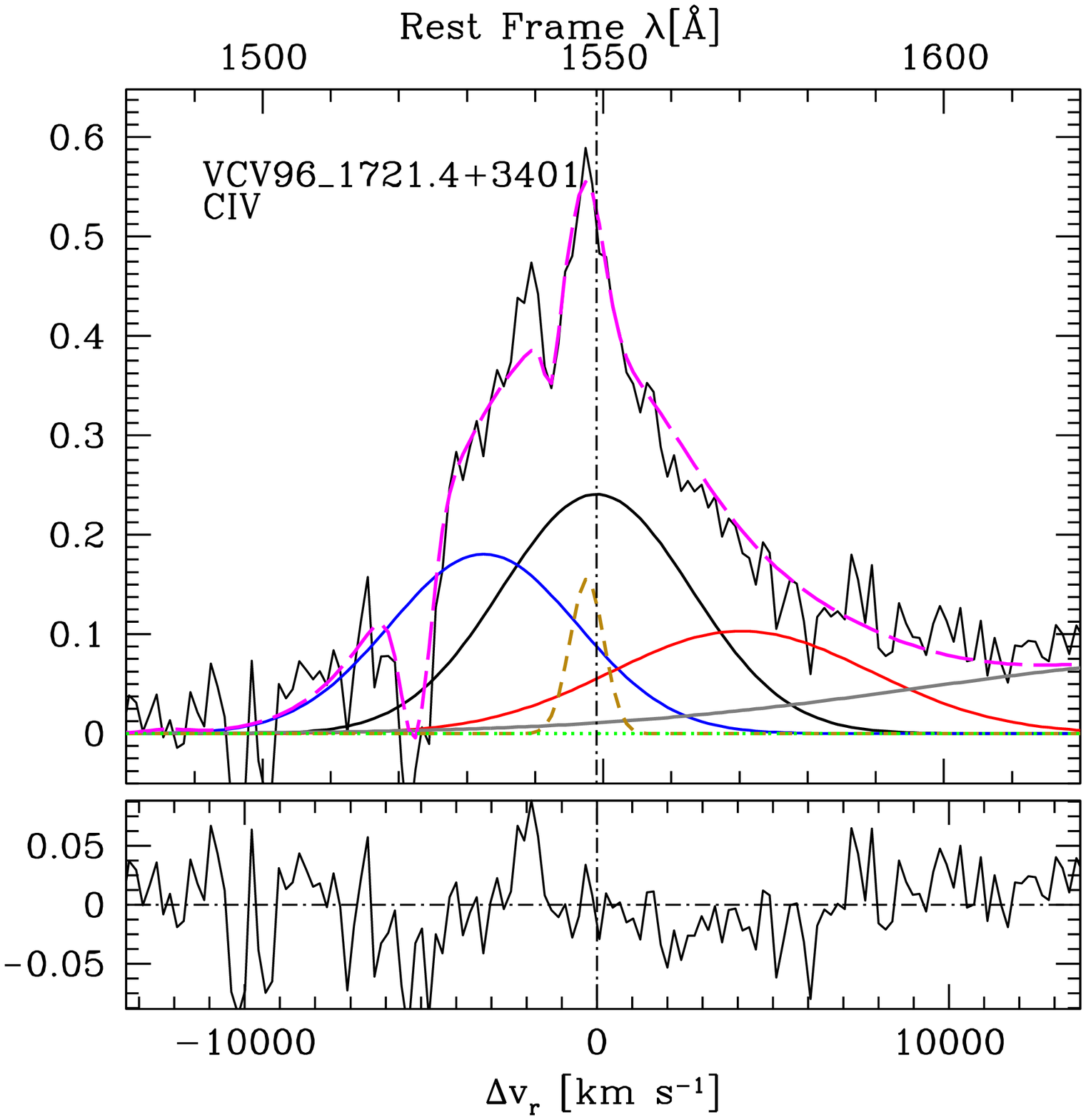}
\includegraphics[scale=0.2]{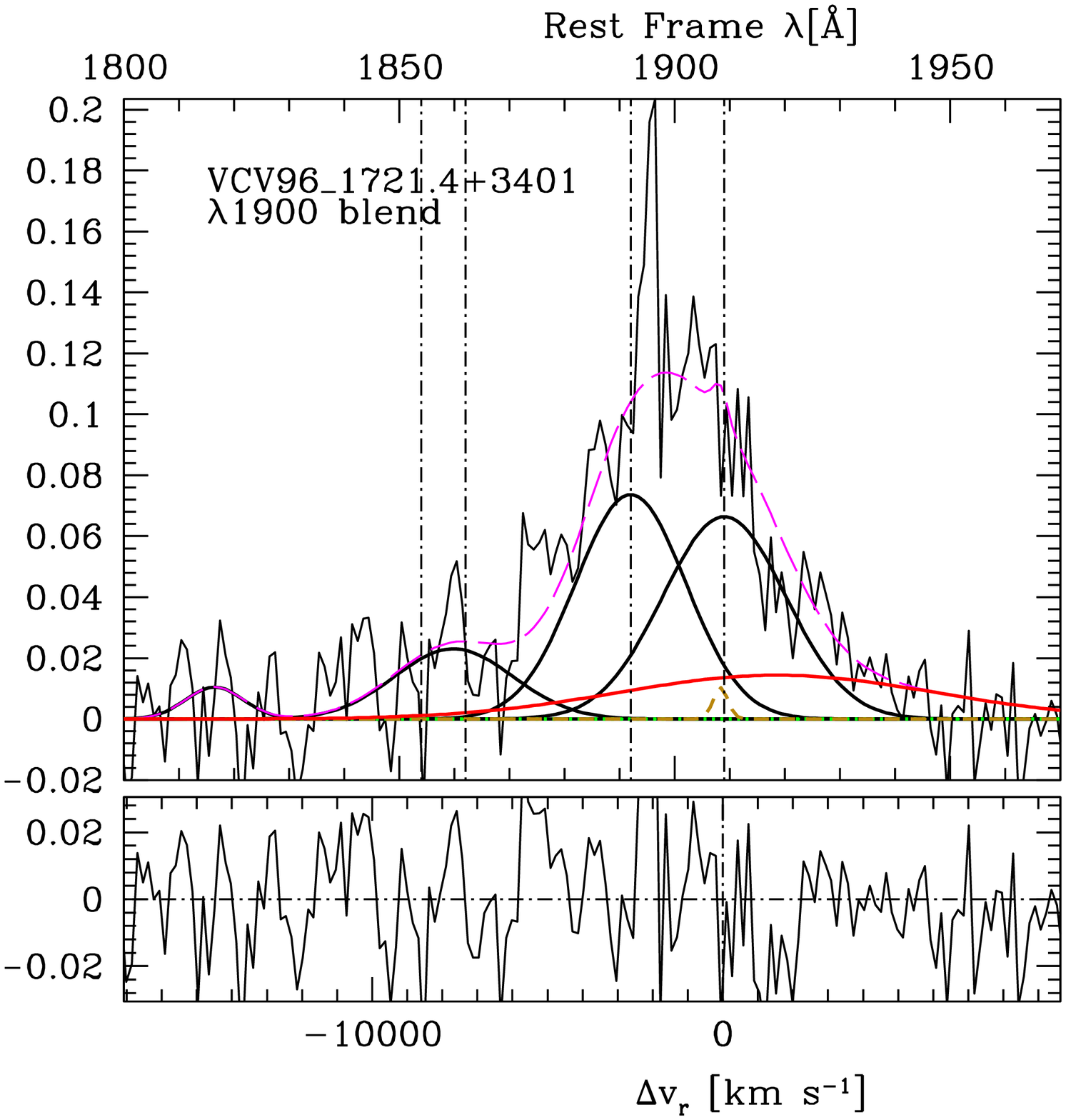}\\
\caption{(cont.) }
\end{figure*}

\newpage
\clearpage
\begin{figure}
\begin{center}
\includegraphics[scale=0.35]{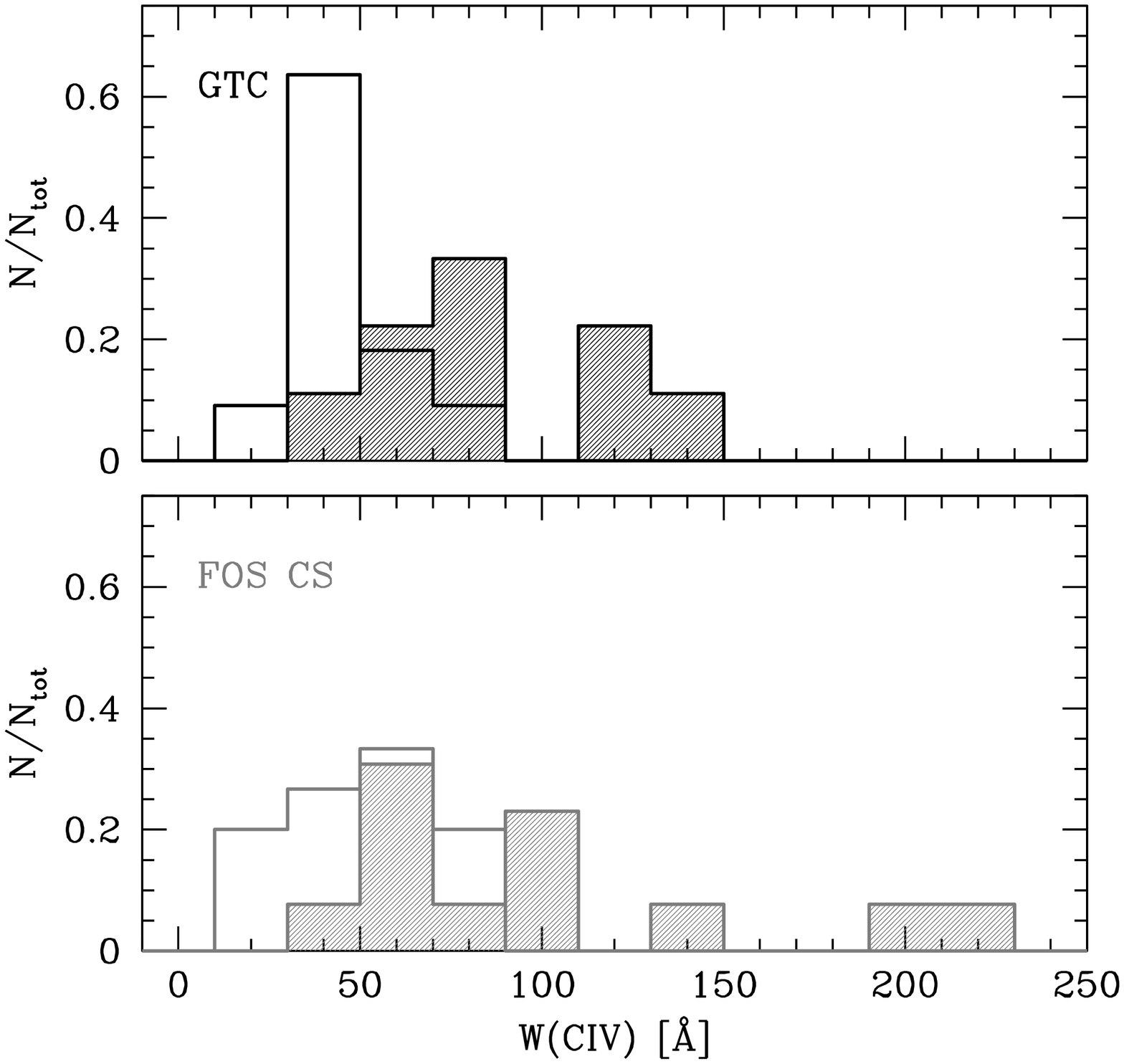}
\includegraphics[scale=0.35]{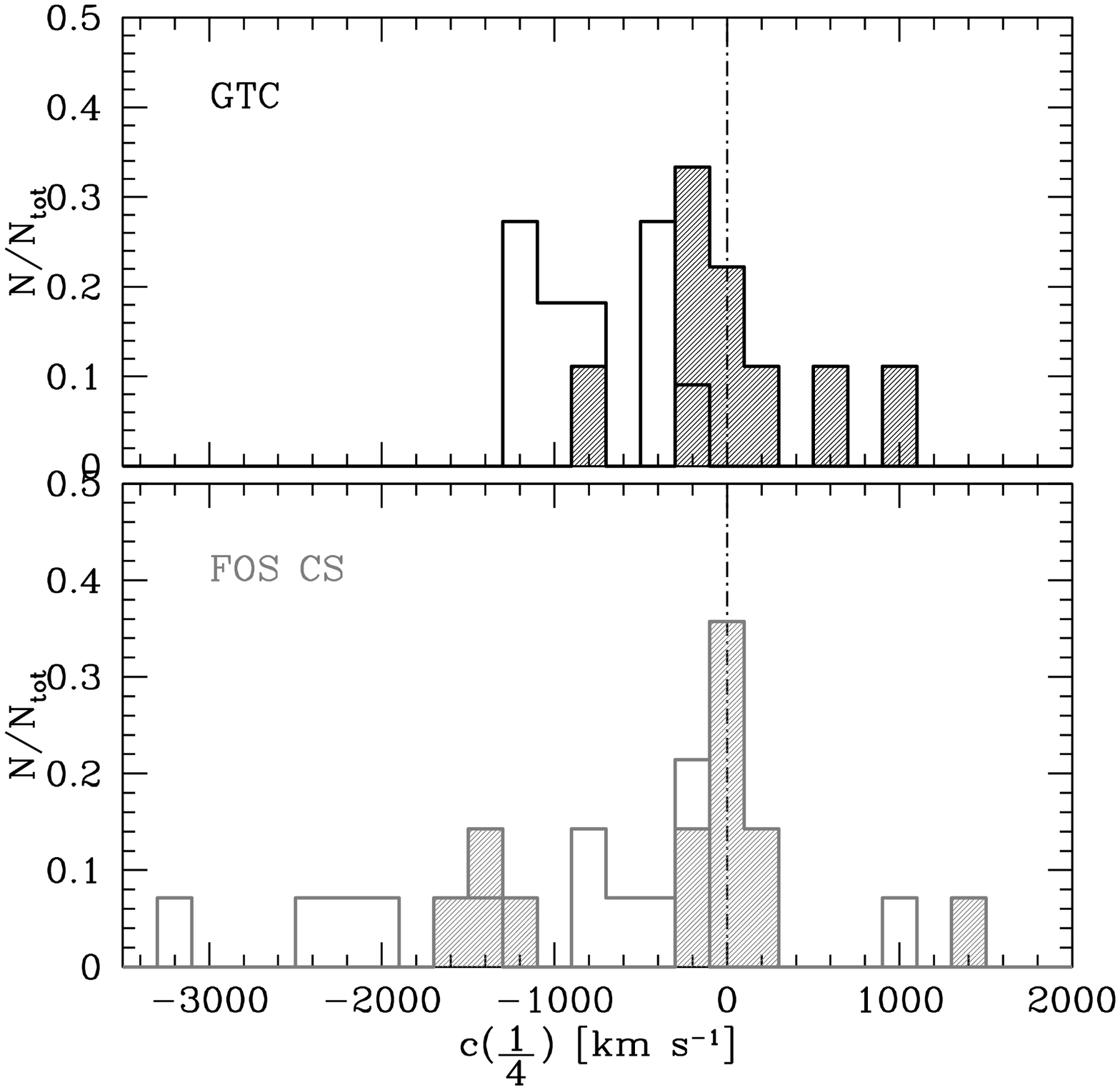}
\includegraphics[scale=0.35]{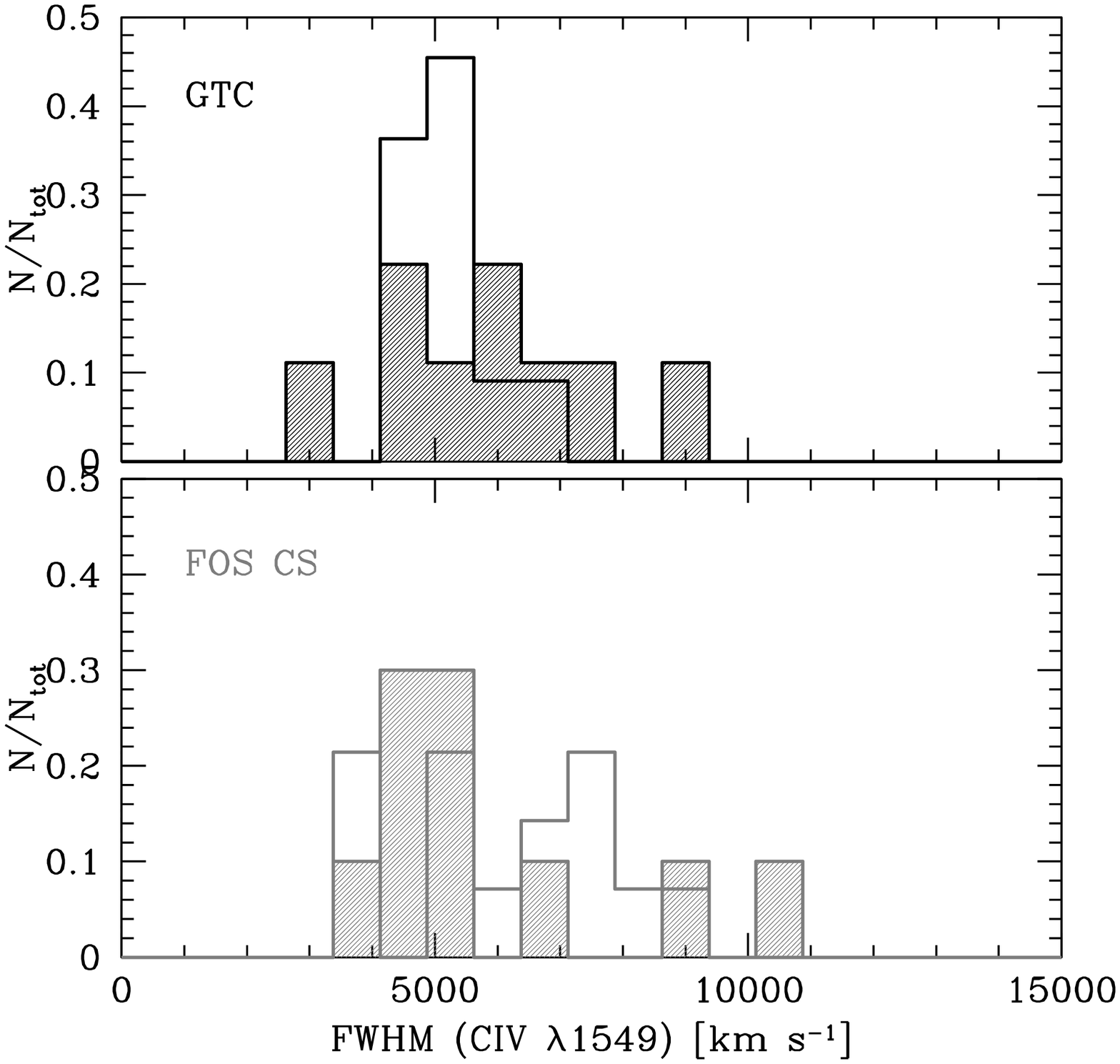}
\caption{Top to bottom: Distributions of W(\civ) for GTC (upper panel; black) and one realization 
of the FOS CS (grey). The shaded histogram refers to Pop. B sources. Middle: same, 
for the  distributions  of $c(\frac{1}{4})$; bottom: same for the distributions of 
FWHM(\civ). \label{fig:distributions}}
\end{center}
\end{figure}

\newpage
\clearpage

\begin{figure}
\includegraphics[scale=0.55]{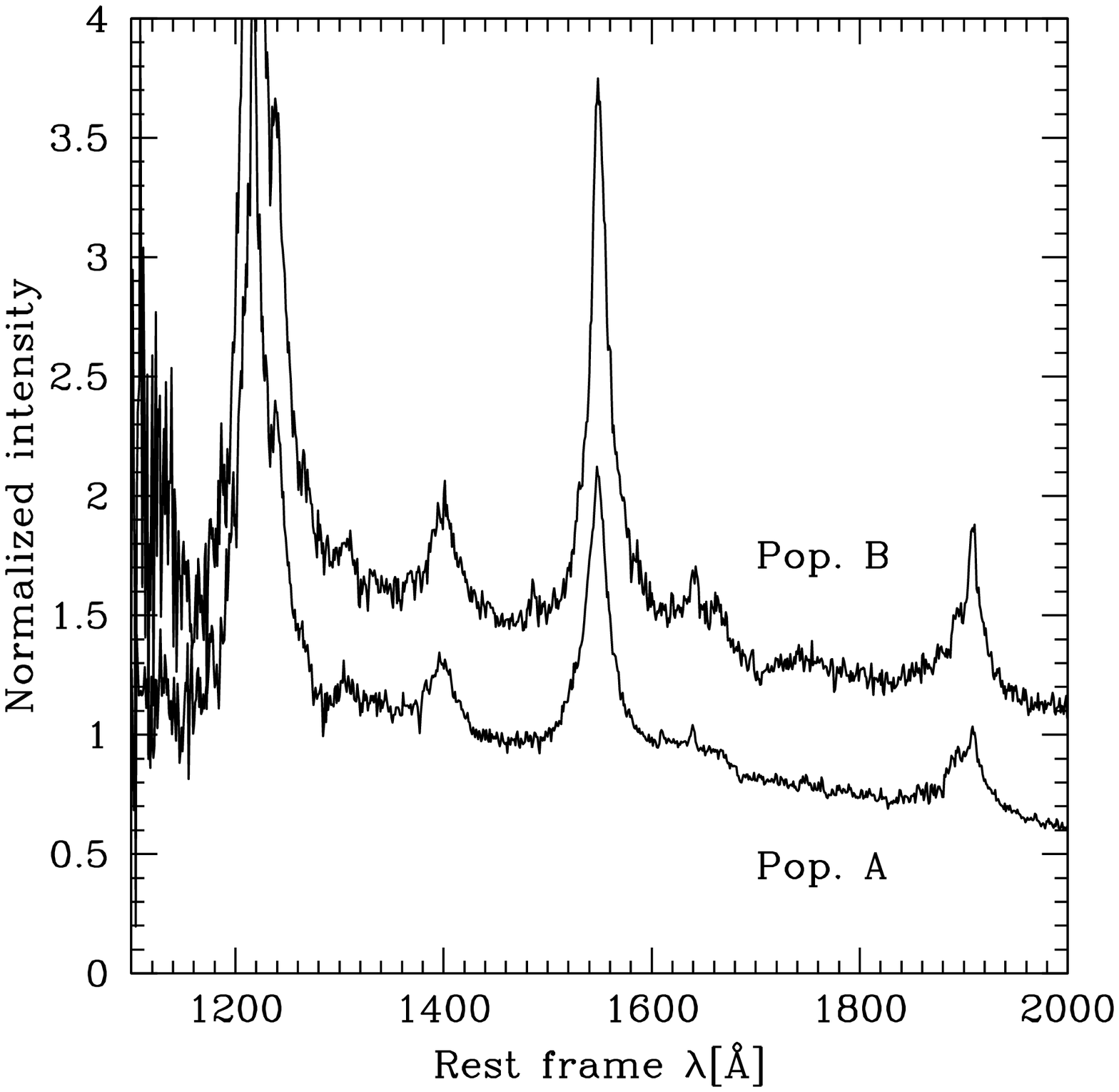}
\includegraphics[scale=0.55]{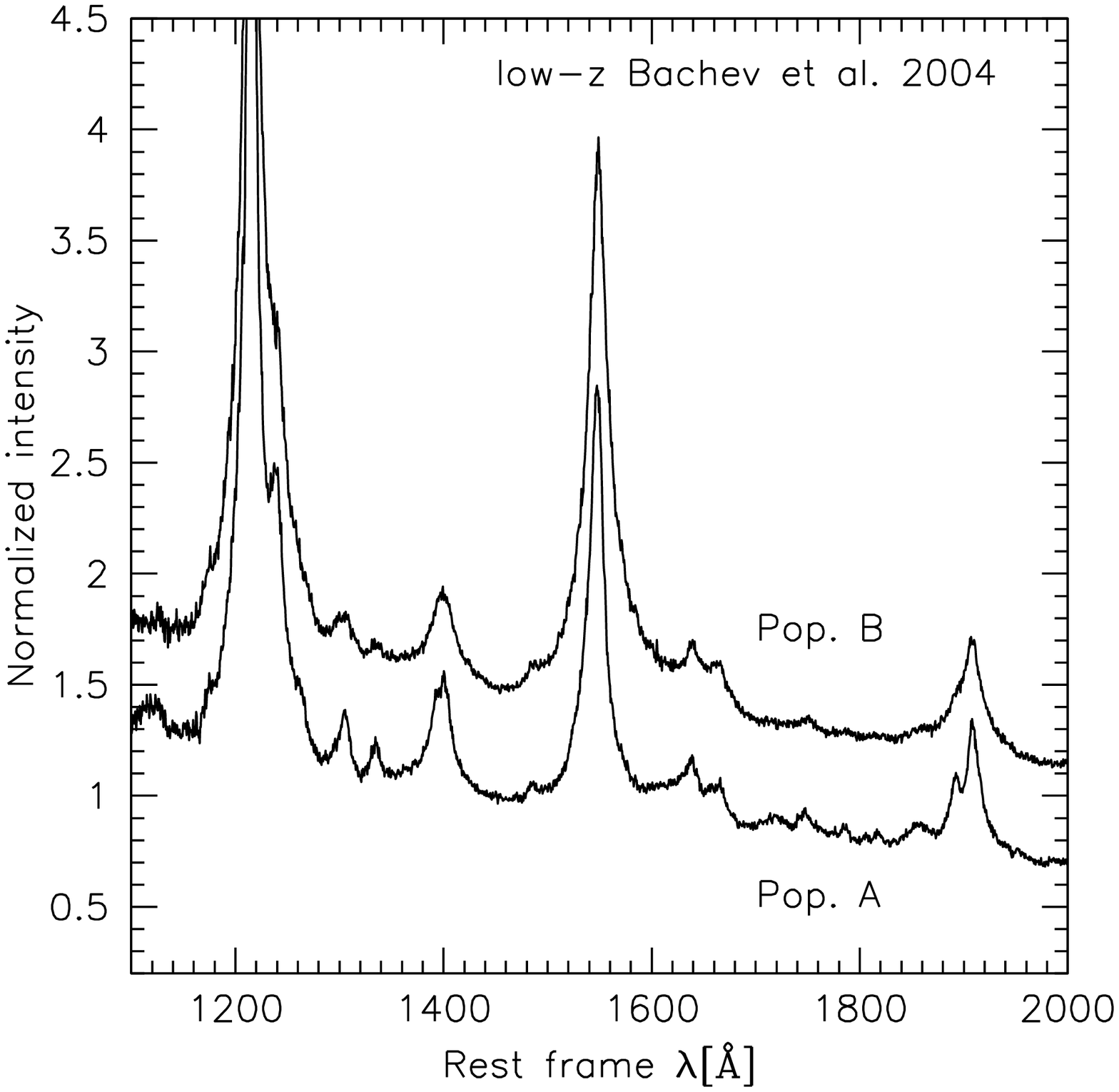}
\caption{Pop. A and B {median} spectra for the present GTC sample and for the 
low-$z$\ sample of \citet{bachevetal04} (bottom). \label{fig:popab}}
\end{figure}

\newpage
\clearpage
\begin{figure*}
\includegraphics[scale=0.2]{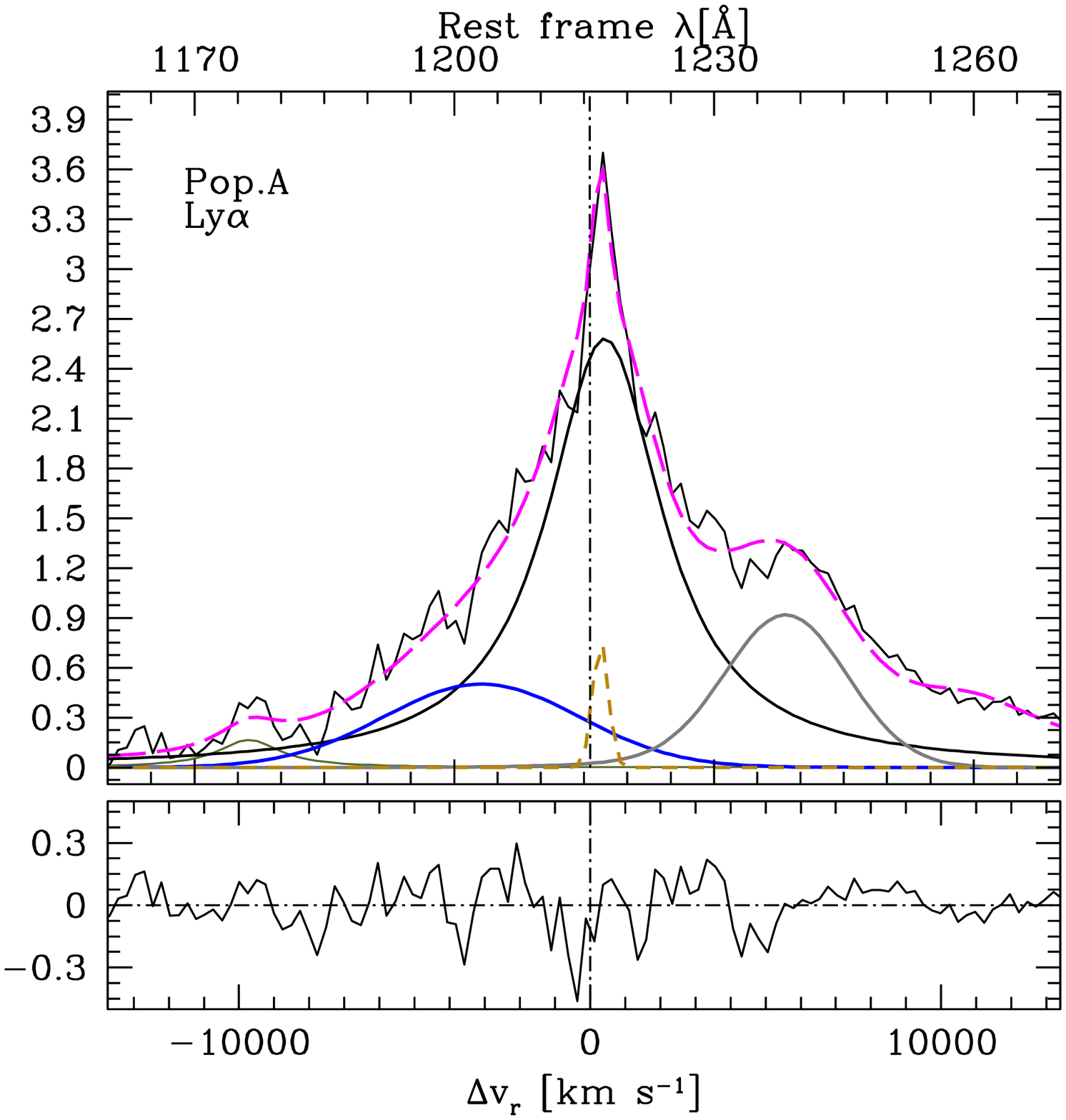}
\includegraphics[scale=0.2]{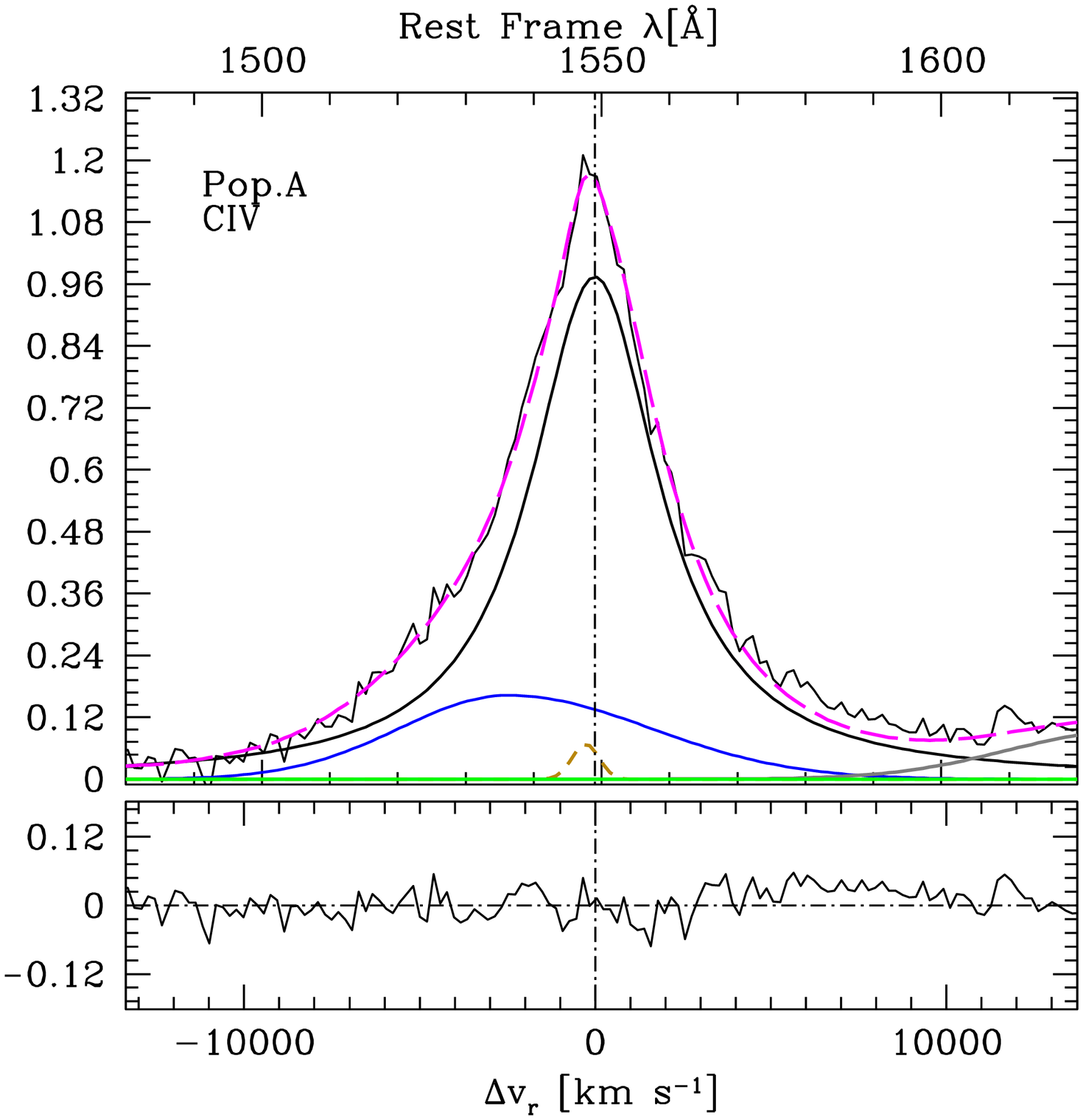}
\includegraphics[scale=0.2]{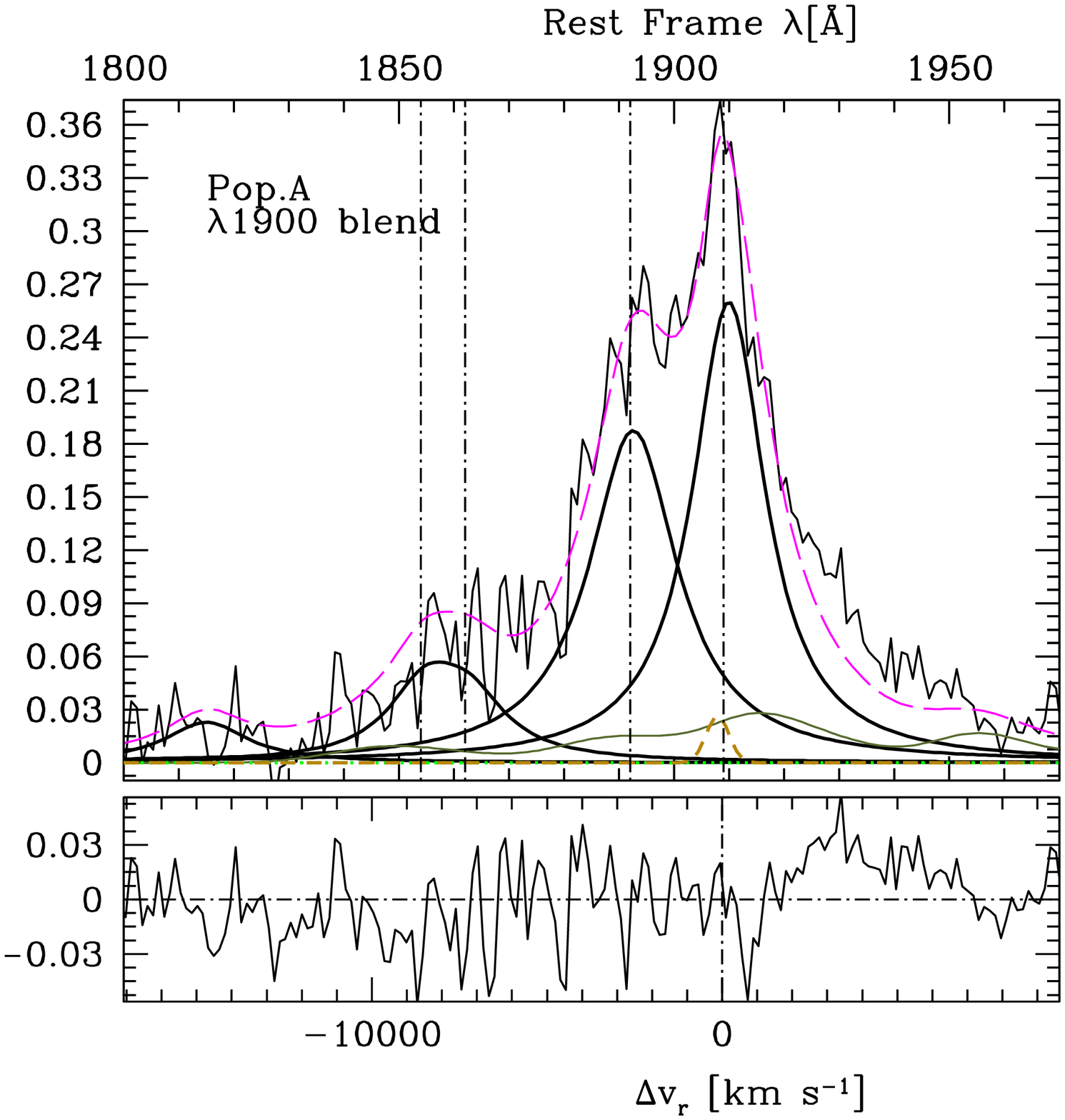}\\
\includegraphics[scale=0.2]{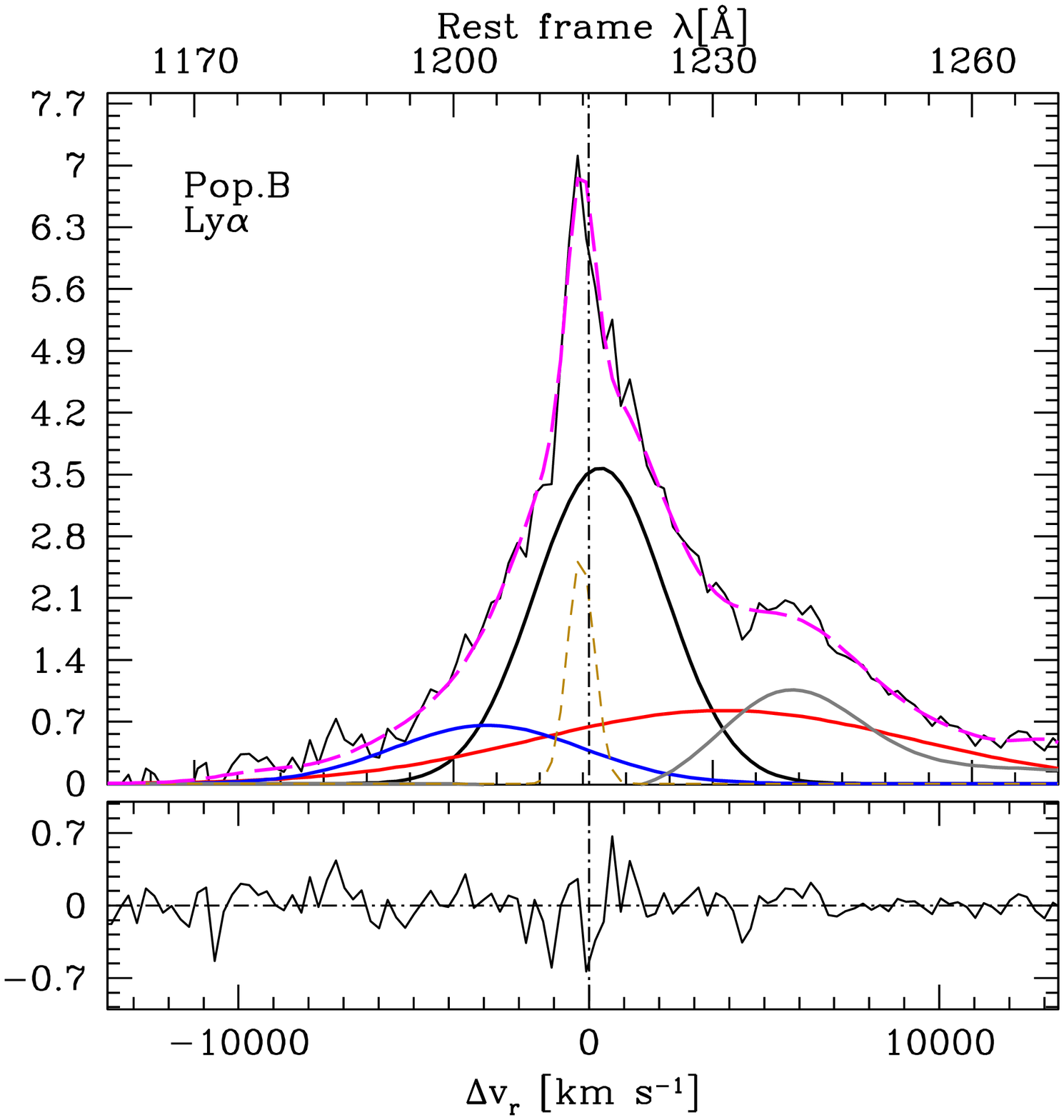}
\includegraphics[scale=0.2]{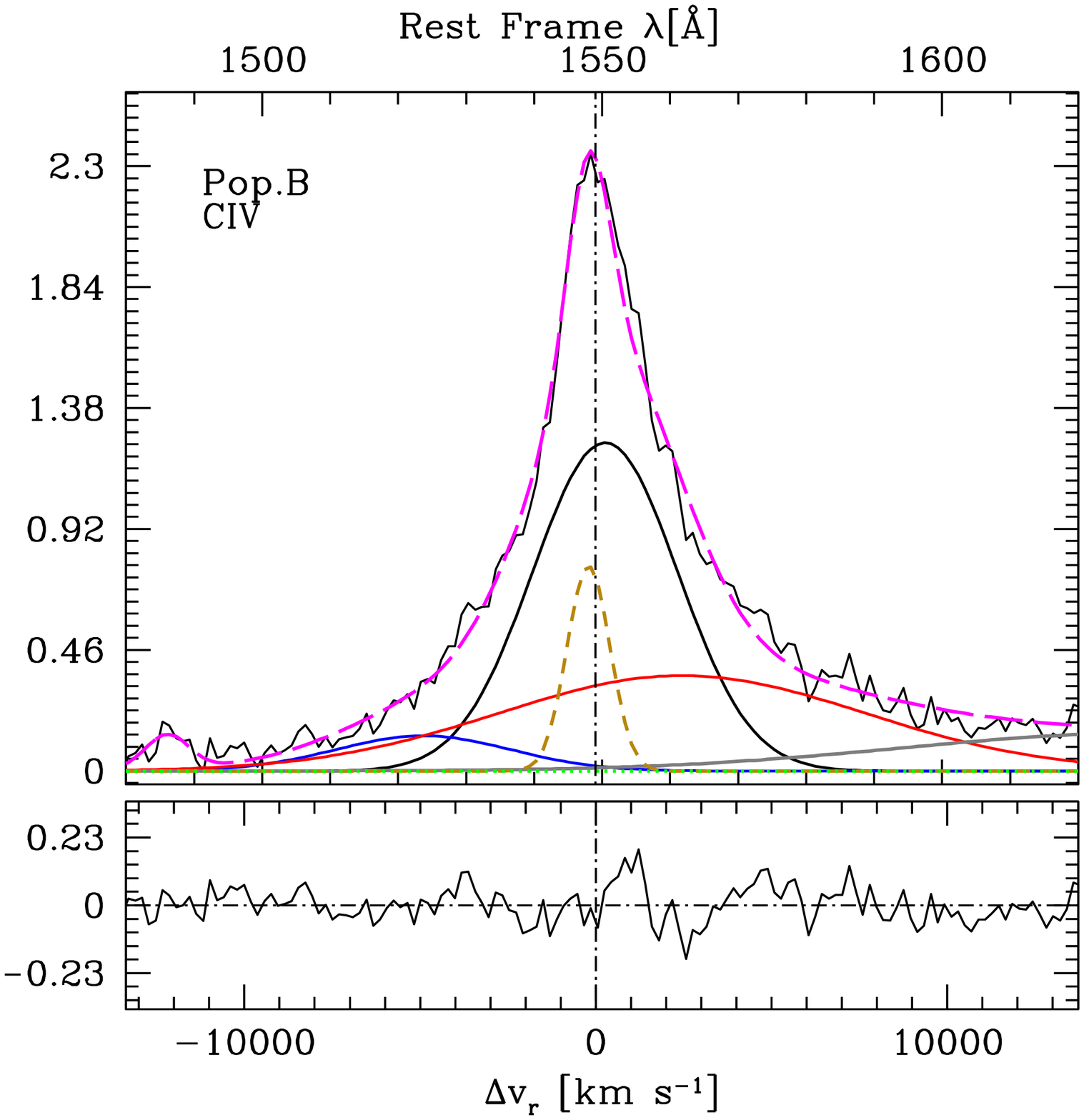}
\includegraphics[scale=0.2]{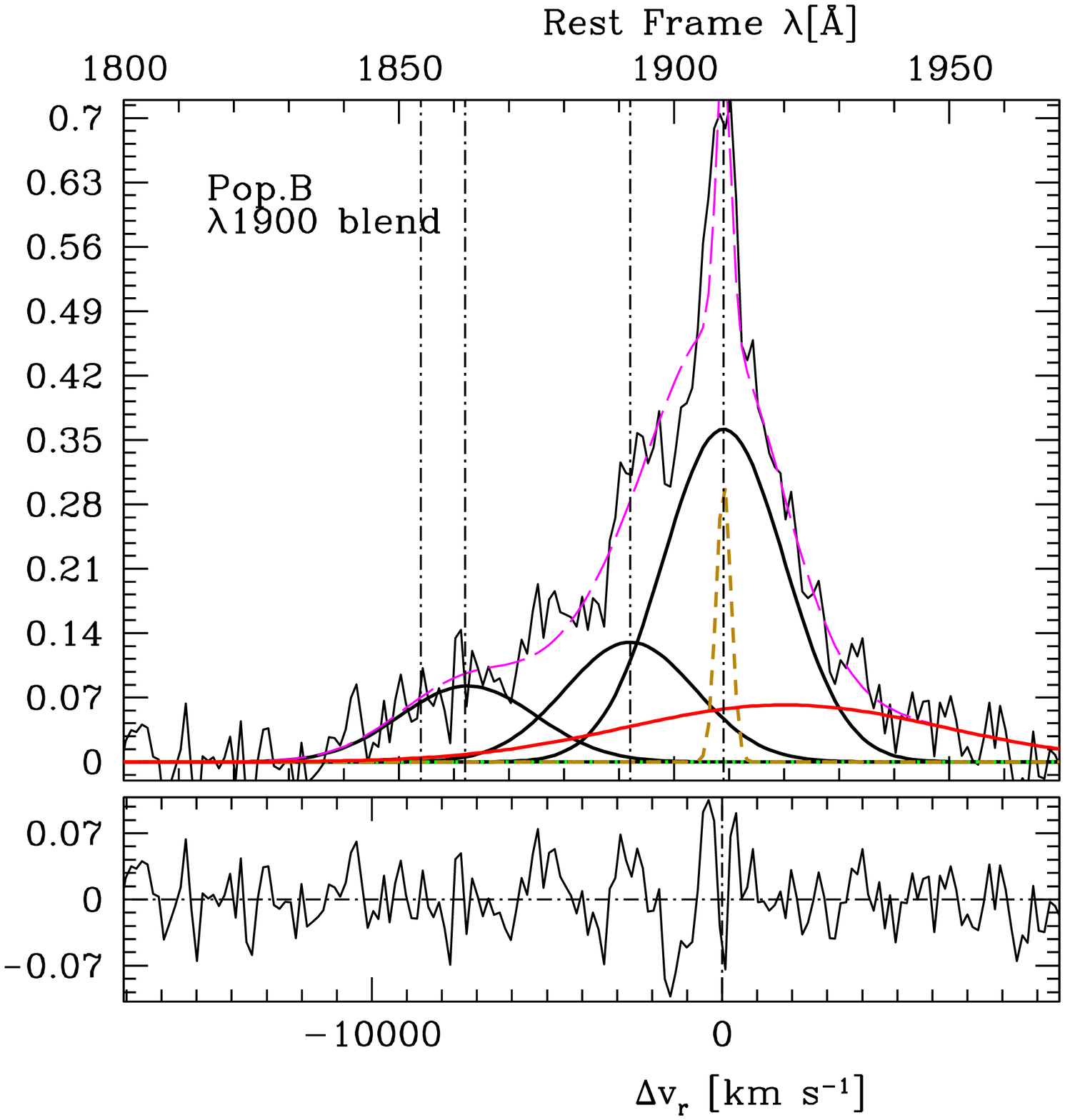}\\
\includegraphics[scale=0.2]{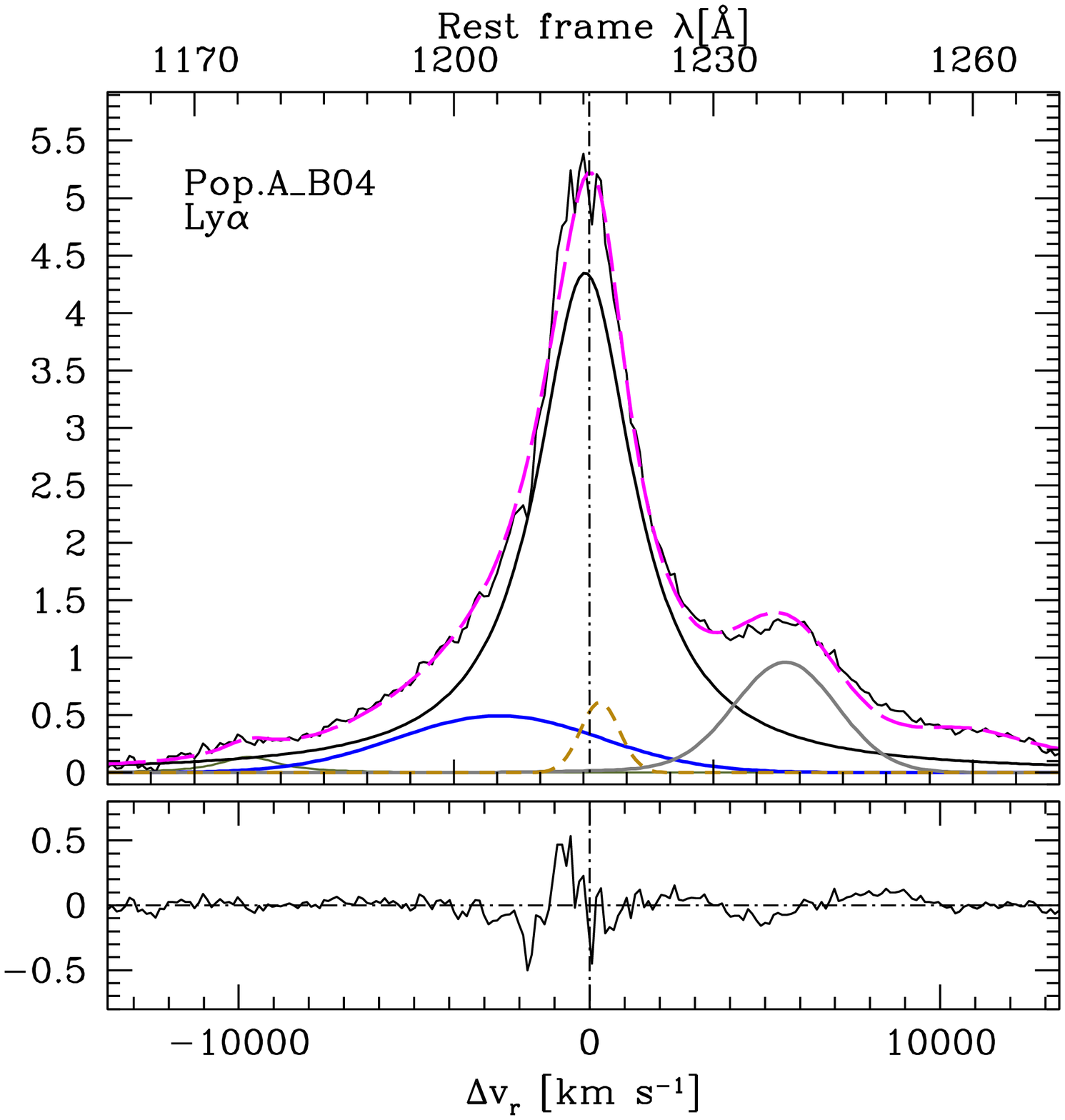}
\includegraphics[scale=0.2]{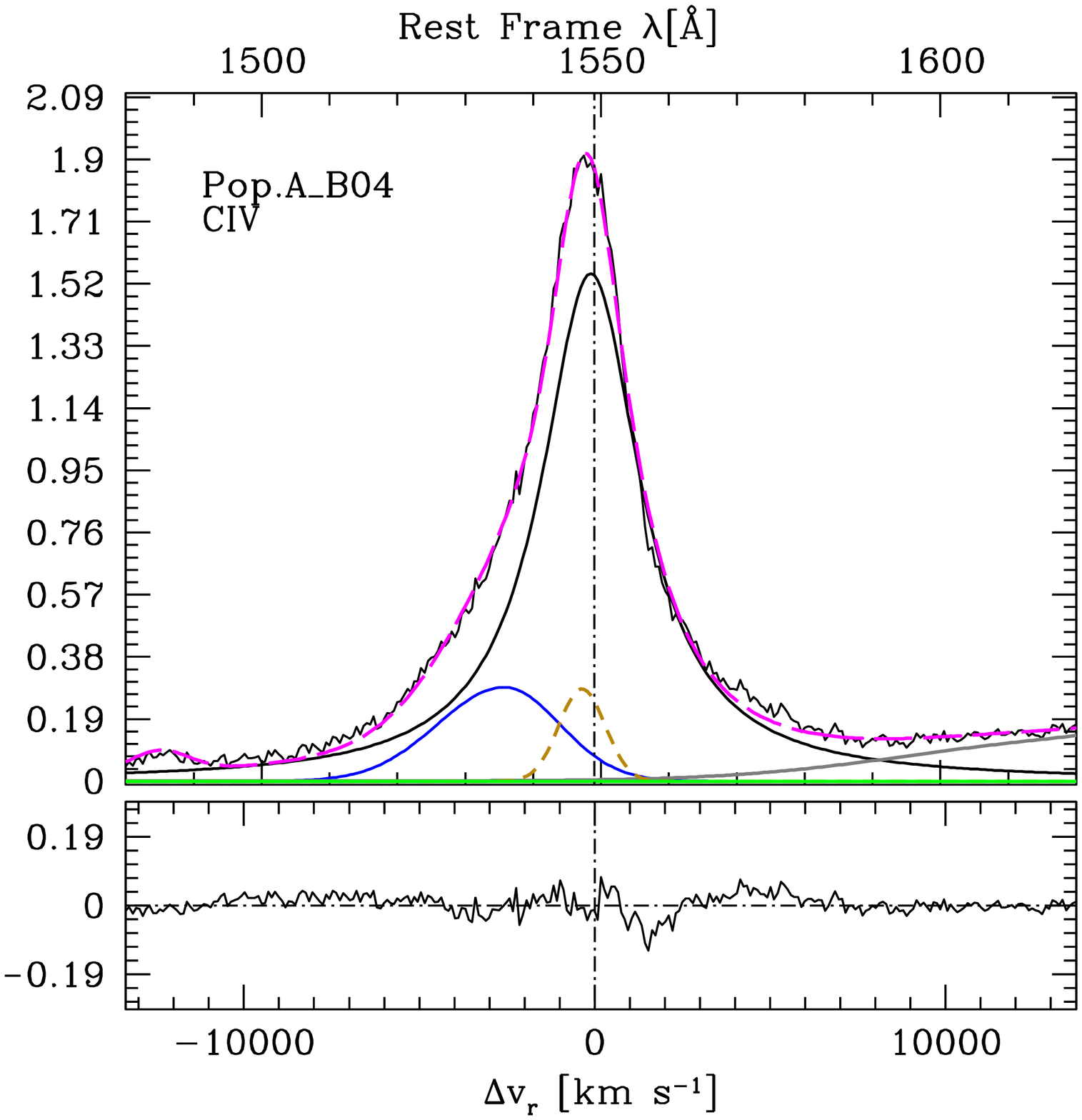}
\includegraphics[scale=0.2]{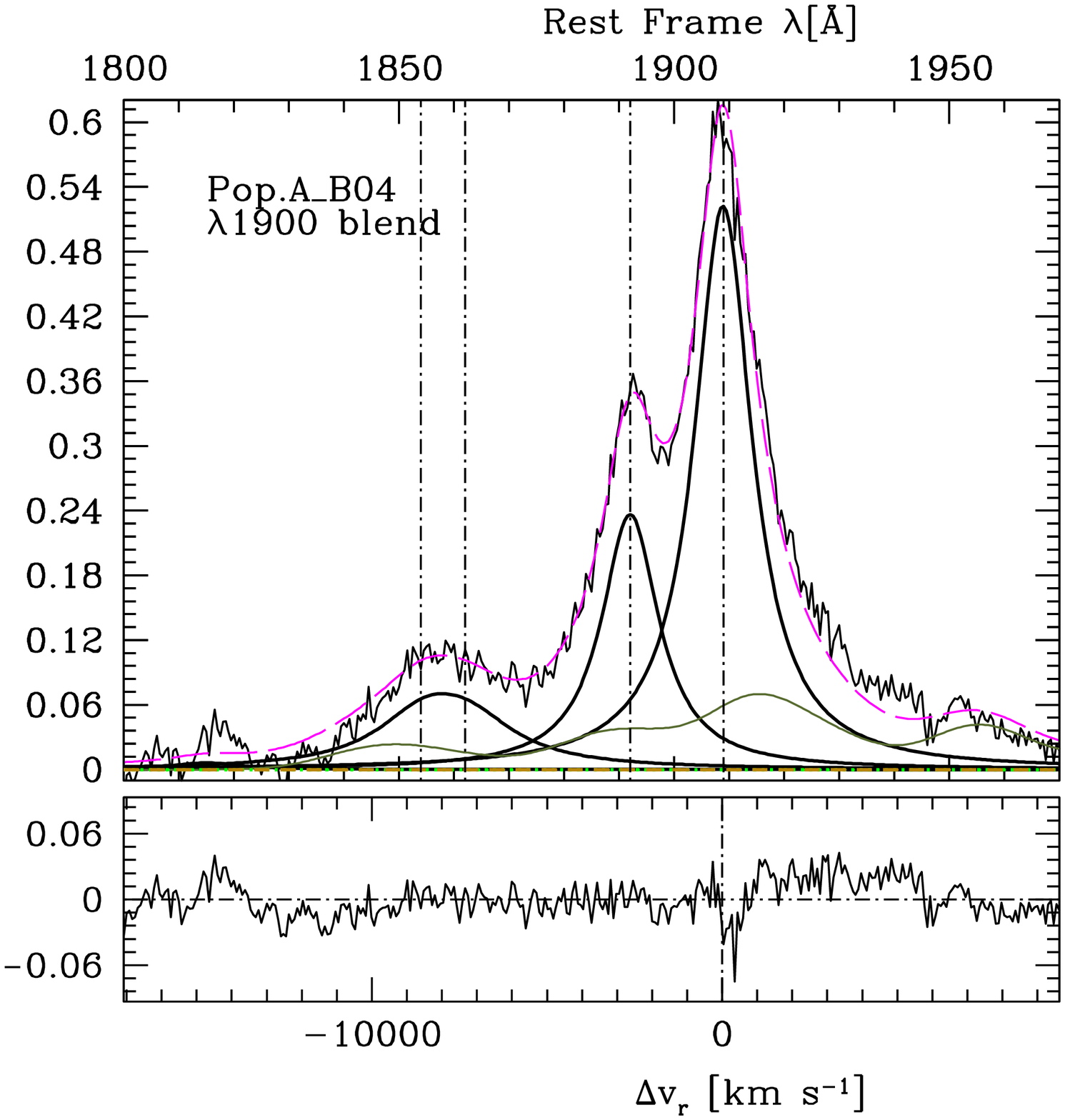}\\
\includegraphics[scale=0.2]{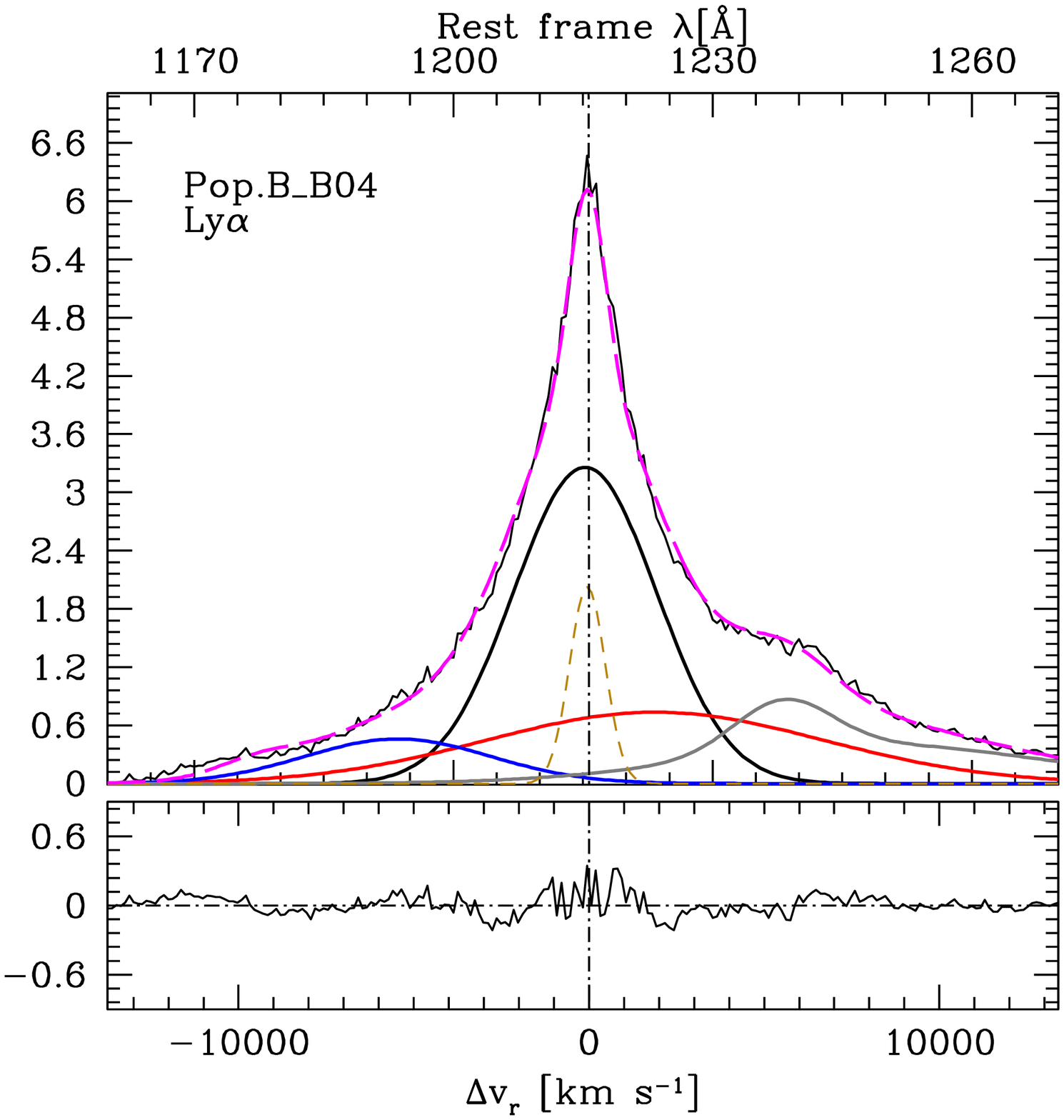}
\includegraphics[scale=0.2]{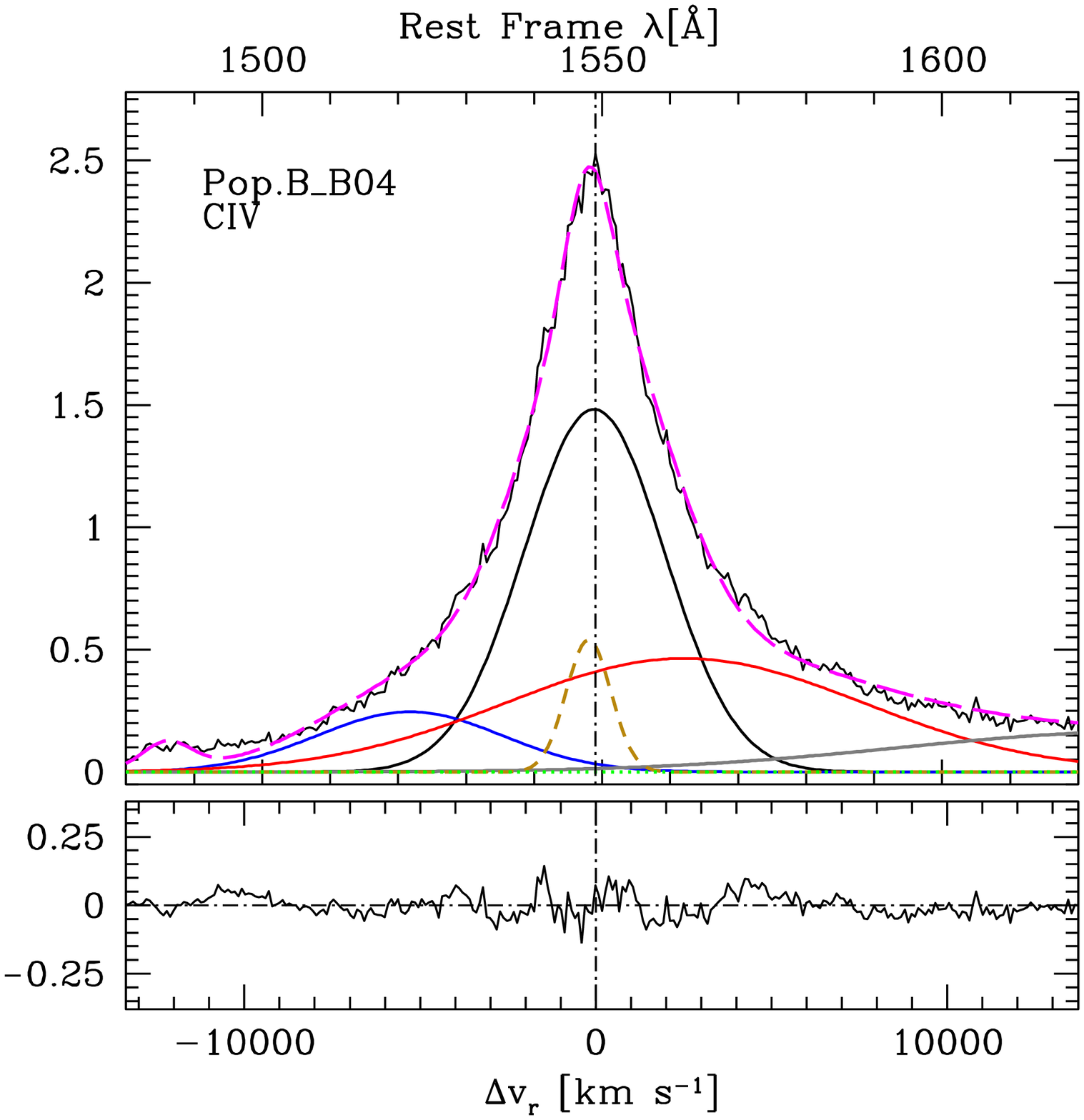}
\includegraphics[scale=0.2]{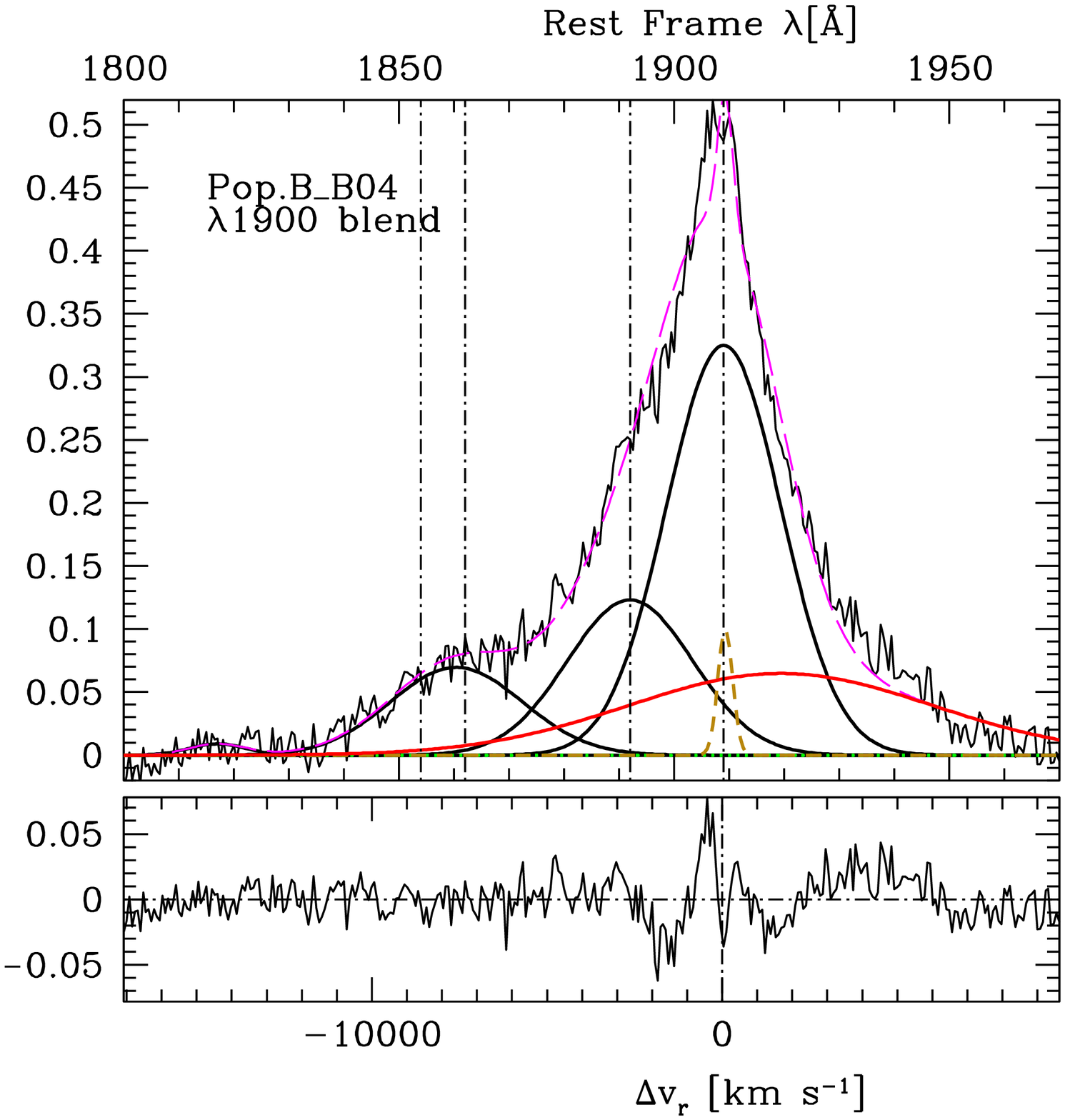}
\caption{Pop. A and B GTC median spectra emission line feature fits. The two lower panel show 
the same analysis for the \citet{bachevetal04} median composite. Meaning of colors and 
symbols is the same of Fig. \ref{fig:fitsa} and \ref{fig:fitsb}. \label{fig:popabfits} }
\end{figure*}

\newpage
\clearpage
\begin{figure*}
\begin{center}
\includegraphics[scale=0.75]{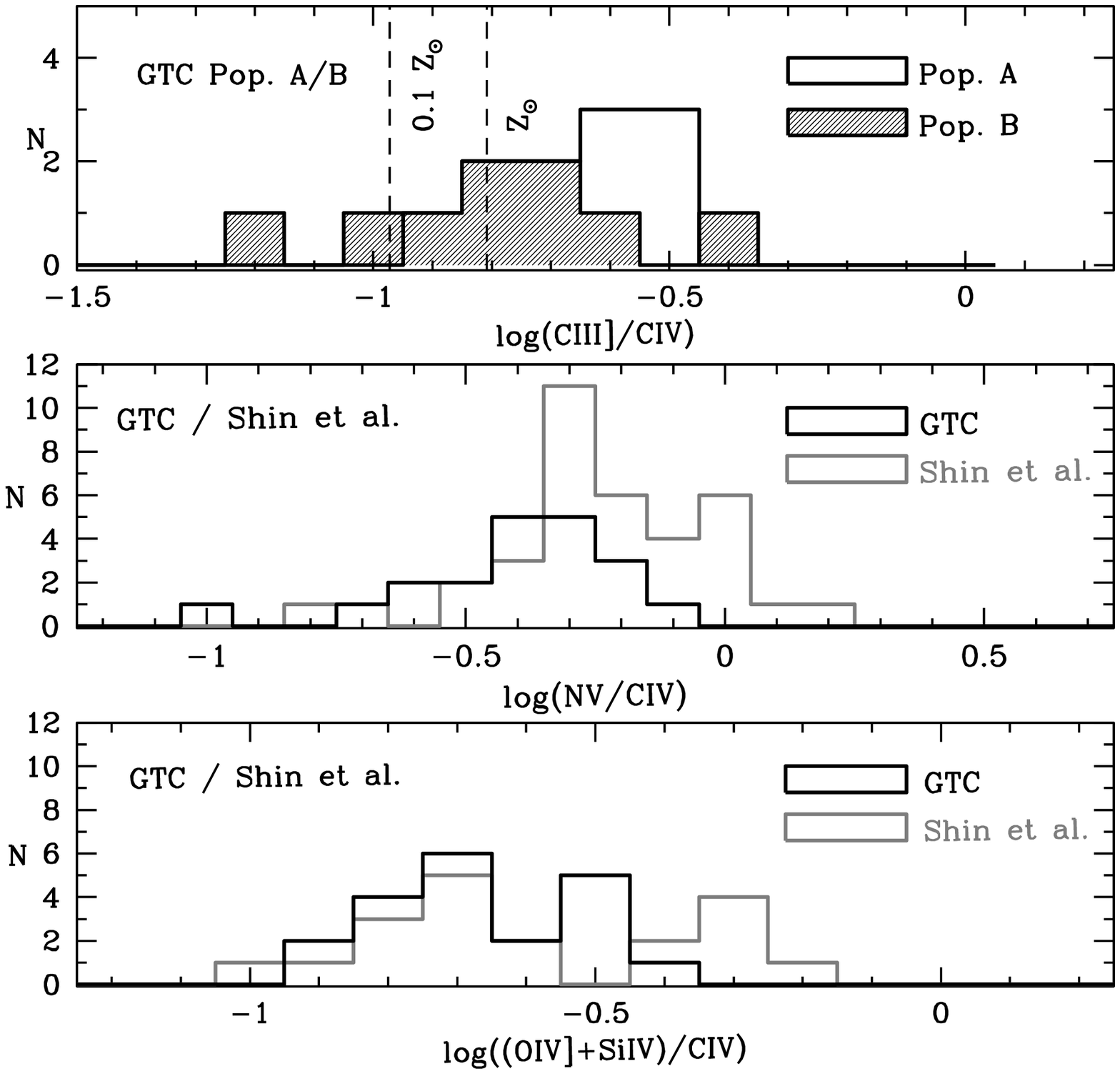}\\
\caption{Histograms showing the distributions of $Z$ sensitive ratios. 
From top to bottom: \ciii/\civ\ for the GTC sample (the dashed histogram is for Pop. B). 
Dot-dashed lines trace the expected ratios for $Z$ = 0.1 $Z_{\odot}$\ and $Z$ =   $Z_{\odot}$ 
for $\log$\nh = 10 and $\log U$ = --1.75. Middle: Distribution of the ratio \nv/\civ\ for 
the GTC (black lines) and the \citet{shinetal13} low-$z$ sample (grey lines), with the 
restriction $\log L \ge 46.2$ [\ergss]. Bottom: same for ratio (\oiv+\siiv)/\civ.
 \label{fig:metaldistr} }
\end{center}
\end{figure*}
\newpage
\clearpage


\begin{figure*}
\begin{center}
\includegraphics[scale=0.50]{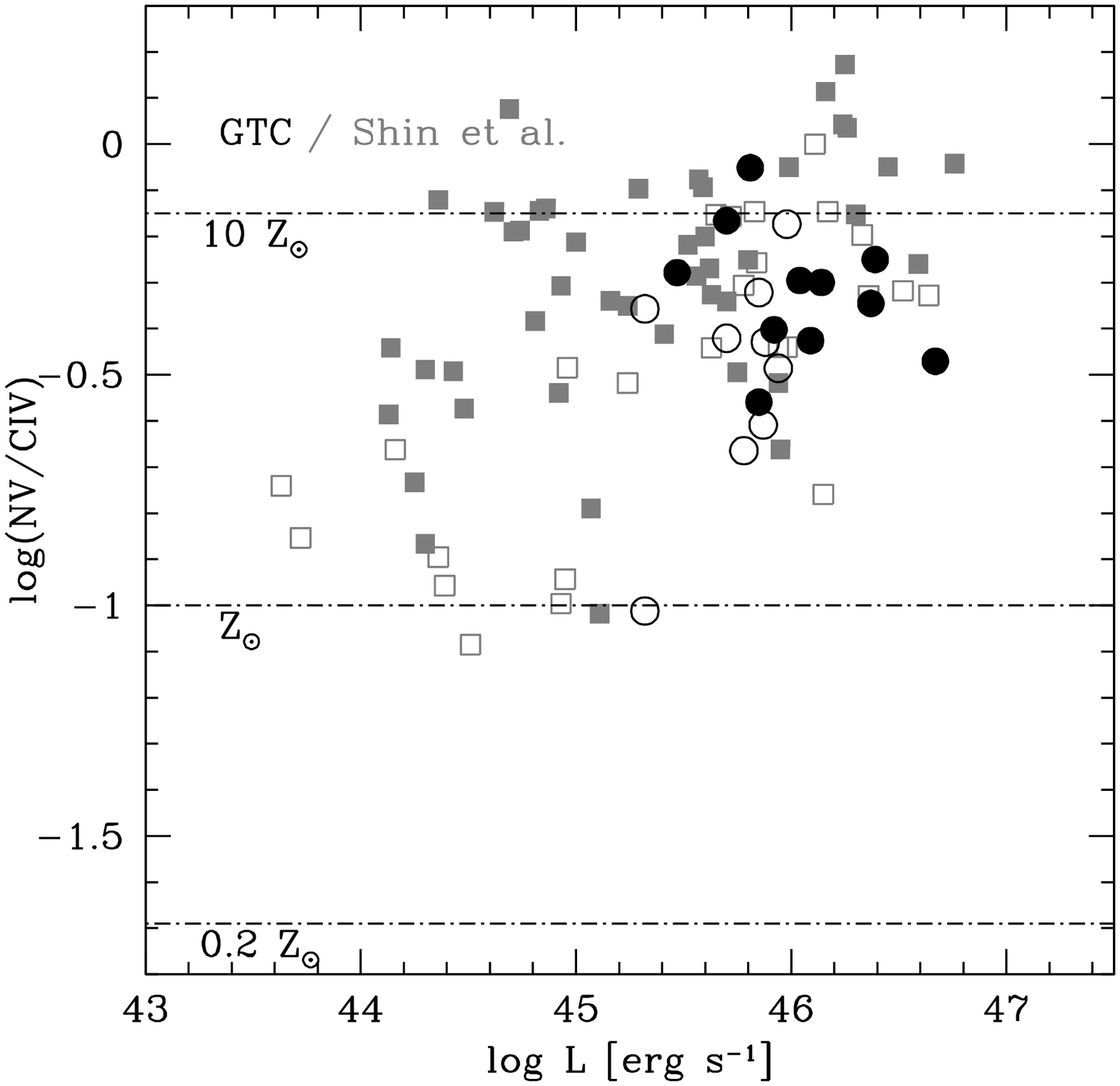}\\
\includegraphics[scale=0.50]{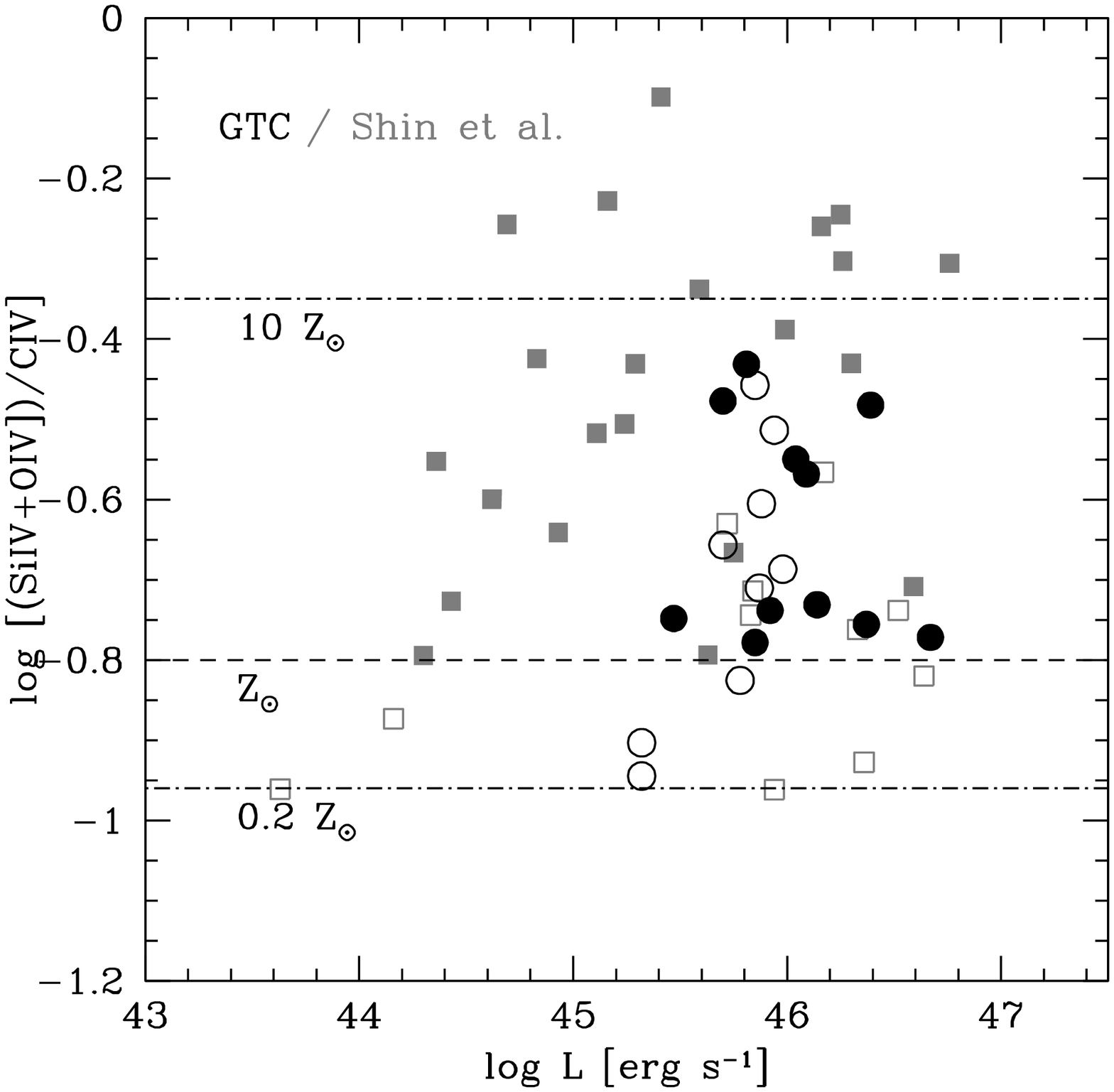}
\caption{Metallicity sensitive line intensity ratios as a function of bolometric 
luminosity for the sample of \citet{shinetal13} (grey squares) and the GTC sample (black circles). 
Open symbols are for Pop. B sources; filled symbols for  Pop. A. Ordinate of 
top panel is decimal logarithm of intensity ratio between \nv\ and \civ; 
ordinate of bottom panel: decimal logarithm of intensity ratio of the 1400\AA\ 
blend due to multiplets of SiIV and OIV] and \civ. The dashed lines show the 
value expected for 0.2, 1, and 10 times solar metallicity following the LOC 
models reported in \citet{nagaoetal06}. 
\label{fig:metals} }
\end{center}
\end{figure*}

\newpage
\clearpage

\begin{figure*}
\includegraphics[scale=0.75]{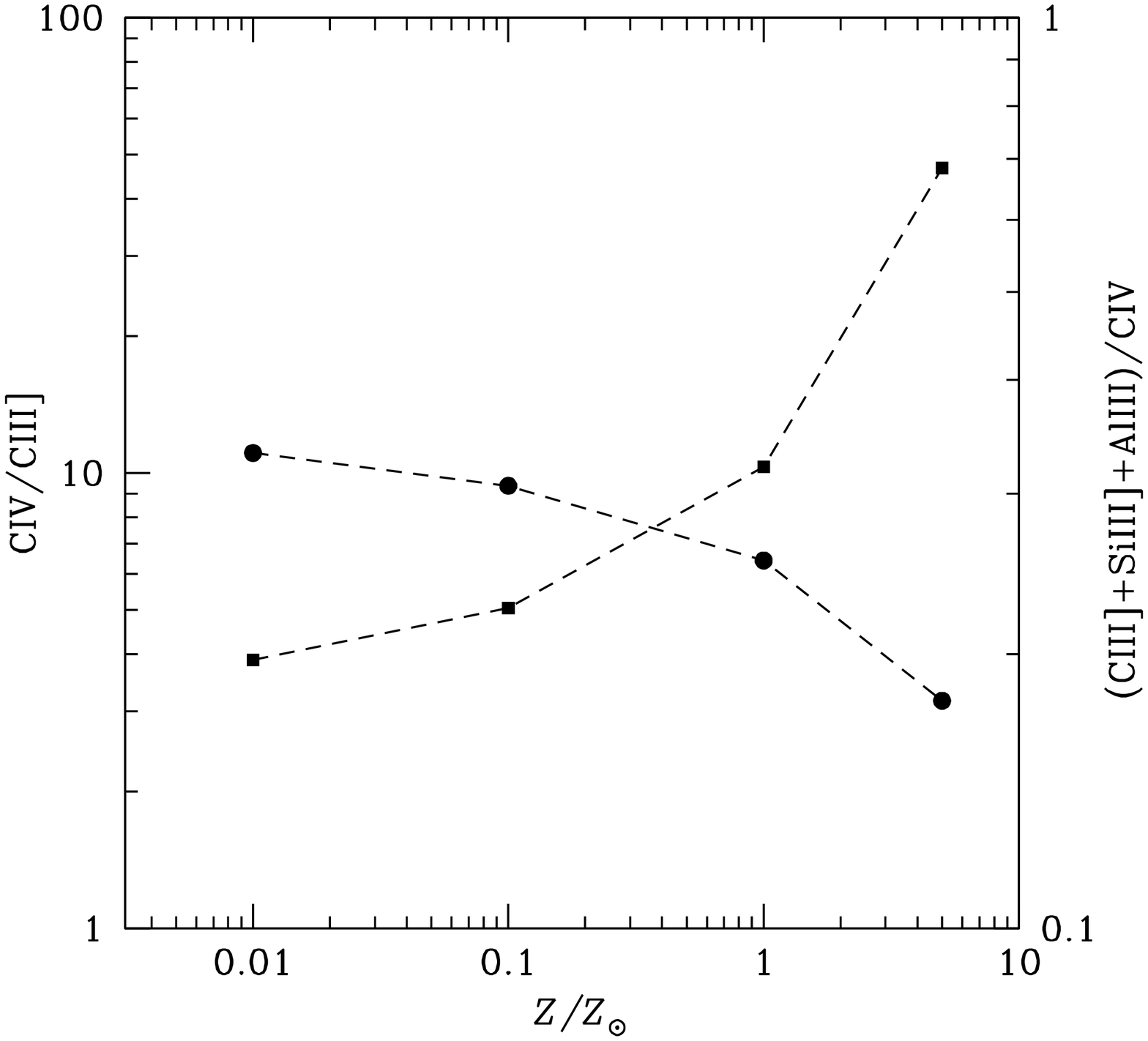}
\caption{Line intensity ratios \civ/\ciii\ (filled circles) and total 1900 blend/\civ\ (filled squares), for four values of $Z$, 0.01, 0.1, 1, 5 times solar, computed for $\log U $\  =  -1.75 and $\log$ \nh\ = 10. \label{fig:ratios}}
\end{figure*}

\newpage
\clearpage
\begin{figure*}
\includegraphics[scale=0.75]{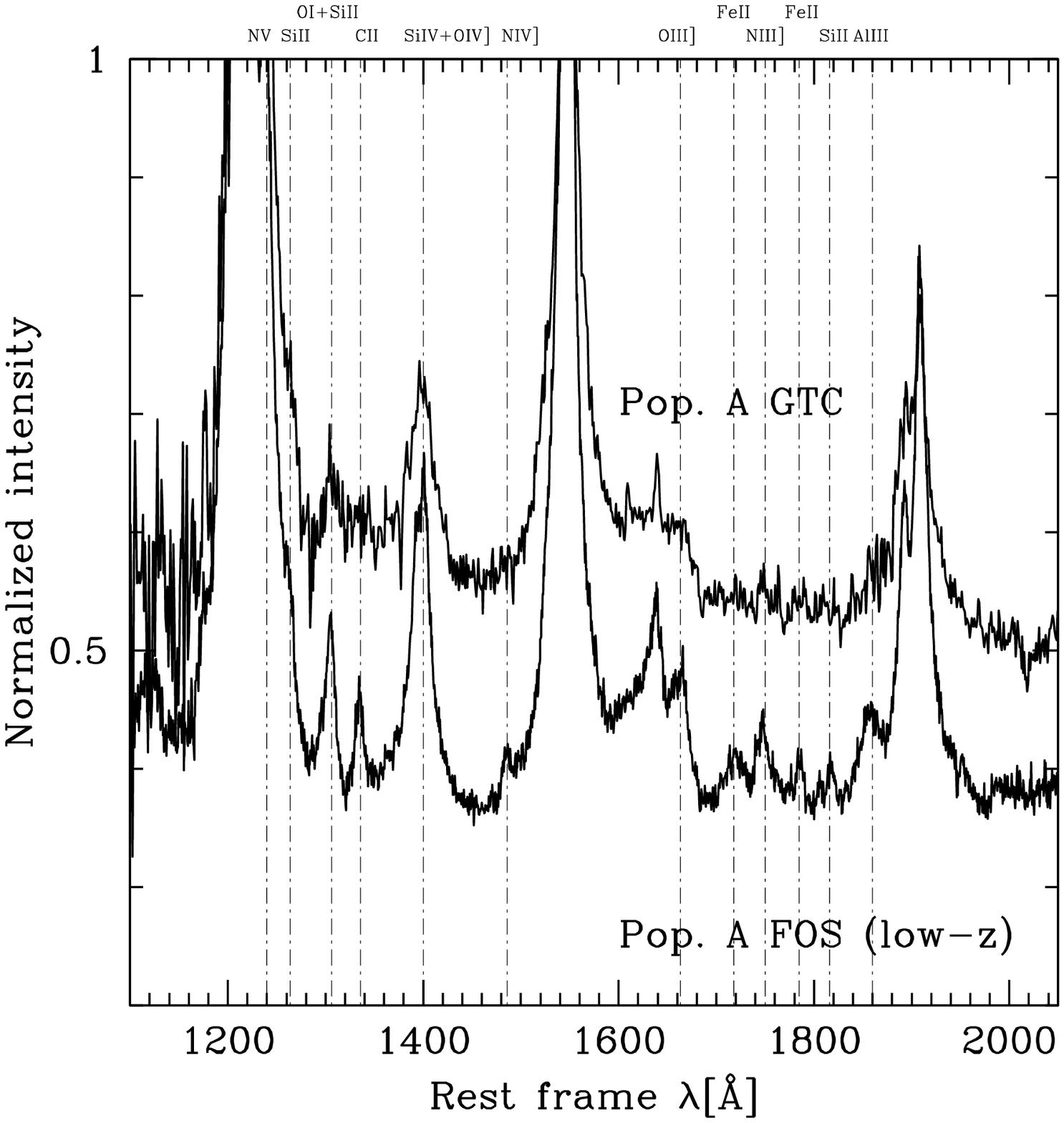}
\caption{Median spectra of Pop. A  for the GTC sample (top) and for the   spectra for the low-$z$\ sample of \citet{bachevetal04} (bottom). The continuum has been flattened and the intensity limit cut set   to emphasize faint features possibly related to differences in metal content. The displacement of the two spectra is due to an artificial offset.  \label{fig:compara}}
\end{figure*}

\newpage
\clearpage

\appendix
\section{Examining a reddened quasar (tip of iceberg?) observed by GTC}
\label{red}

Dust reddened and obscured quasars play an increasing role in the
relationship  between supermassive black hole growth and the
evolution of galaxies on cosmic scales. Recently a new co-evolution
scenario has emerged from the identification of a significant sample
of red QSOs \citep[e.g.][and references therein]{glikmanetal12}   where the
reddening is not linked to the central disk/torus phenomenon but is
associated with cold dust from the host galaxy that would also be
experiencing starburst processes. In this scenario red-QSOs
constitute a  population of young high accreting quasars probably involved
with a feedback mechanism. They could be precursors of the blue luminous
quasars that are revealed when the dusty cocoon is swept away.

\begin{figure}
\includegraphics*[width=8.75cm,angle=0]{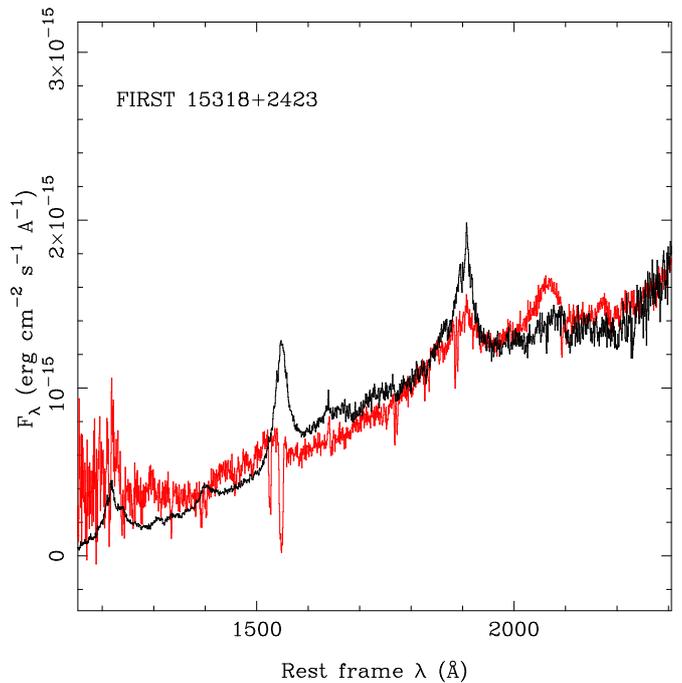}
\caption{Rest frame spectrum of FIRST J15318+2423 (red). Abscissa is rest frame wavelength, ordinate rest frame specific flux. The reddened Pop. A template is shown in black. See text for more details. \label{fig:qsored}}
\end{figure}

In our random selection of quasars from the \citet{veroncettyveron10} catalog
we included one of these red quasars. SDSS J153150.41+242317.7
(also known as FIRST 15318+2423 and F2M1531+2423) was at 
firstly identify as a red QSO by
\citet{glikmanetal07} from a sample of FIRST-2MASS candidates. They
obtained NIR spectroscopy and detected broad H$\alpha$\ and \hb\ emission
lines with $z \approx$\ 2.287. Our GTC spectrum is shown in Figure \ref{fig:qsored} where the
highly reddened continuum is obvious as well as the presence of \lya, \heiiuv and 
deep absorptions at the wavelengths of broad lines such as \civ\ and \ciii\
from which it can be classified as a mini-BAL. Taking into
account all these lines we obtained a redshift of $2.284$.

We have estimated the reddening of this object, parametrized by the color
excess $E(B -V)$,  by fitting its UV continuum with three quasar
templates,  excluding the regions of broad emission lines (\lya,
\civ, \ciii, Fe band). In order to redden the templates we used an SMC
extinction law \citep{gordonclayton98,gordonetal03} which appears
to be the most appropriate reddening law for modeling dust reddening in
quasars \citep{yorketal06}. We assume a $R_\mathrm{V}$\ coefficient of 3.07 for the
extinction law. Templates employed included two derived GTC composite
spectra corresponding to Pop. A and B, that provide
internal concordance since they were obtained from our quasar
sample with the same instrumental setup,  and also a third template
involving the composite FIRST Bright Quasar Survey spectrum (FBQS; \citealt{brothertonetal01}).

The three fits yield essentially the same value for the color excess.
In the first two cases, using the Pop. A \& B\ composites, we obtained $E(B - V) =
0.41\pm 0.01$\ while the FBQS template yields $E(B - V)$ = 0.40, all cases
with a high correlation coefficient ($\ga$ 0.97). Taking into account the
estimated extinction, FIRST15318 becomes the brightest quasar in our
sample with $M_B \approx -26.3$.  In Fig. \ref{fig:qsored}  we have plotted the fitted
reddened Pop A template over the observed spectrum. Attempts to
learn more about its nature are difficult using the present data.
Formal fitting of the red quasar continuum gives  a slightly better
solution as a reddened Pop A. spectrum but both Pop. A and
B fits yield similar results. Taking advantage of NIR
spectroscopic data \citep{glikmanetal07,glikmanetal12}  allow us to model the
H$\alpha$\ and \hb\ lines. The fit for H$\alpha$\ involves a
Lorentzian profile with FWHM (BC) $\approx$ 3400 \kms\ that identifies
FIRST15318 as a Pop. A quasar  in concordance with the UV continuum
we fitted.  In the \hb\ region, the spectrum is noisier but can be
clearly shows the strong \feii\ emission characteristic of a highly
accreting Pop. A sources (spectral type A2 or A3).

\end{document}